\documentclass[apj, twocolappendix, numberedappendix, appendixfloats]{emulateapj}
\usepackage[pdftex,backref,breaklinks,colorlinks,linkcolor=magenta,citecolor=blue]{hyperref}
\usepackage{graphics,graphicx}
\usepackage{subfloat}
\usepackage{color, xcolor}
\usepackage{multirow}
\usepackage{appendix}
\usepackage{amsfonts}
\usepackage{adjustbox}
\usepackage{rotating}
\usepackage{url}
\usepackage{hyperref}
\usepackage[pass,letterpaper]{geometry}
\usepackage[utf8]{inputenc}

\newcommand{\Msun}{$M_{\odot}$}
\newcommand{\Mbh}{$M_{\rm BH}$}
\newcommand{\Mnsc}{$M_{\rm NSC}$}
\newcommand{\Mstar}{$M_{\star}$}
\newcommand{\Mbulge}{$M_{\rm Bulge}$}
\newcommand{\Lbulge}{$L_{\rm Bulge}$}
\newcommand{\Lsun}{$L_\odot$}
\newcommand{\Lstar}{$L_{\star}$}
\newcommand{\betaz}{$\beta_{\rm z}$}

\renewcommand{\deg}{\ensuremath{^{\circ}}}
\newcommand{\ml}{\emph{M/L}}
\newcommand{\hst}{\emph{HST}}
\newcommand{\kms}{km~s$^{-1}$}
\newcommand{\se}{$\arcsec$}
\newcommand{\mm}{$\arcmin$}

\newcommand{\Lppc}{\Lsun/${\rm pc^2}$}
\newcommand{\Mppc}{\Msun/${\rm pc^2}$}

\shortauthors{NGUYEN ET AL.}
\begin{document}

\title{Nearby Early-Type Galactic Nuclei at High Resolution -- Dynamical Black Hole and Nuclear Star Cluster Mass Measurements}

\author{\mbox{Dieu D. Nguyen\altaffilmark{1}}}
\author{\mbox{Anil C. Seth\altaffilmark{1}}}
\author{\mbox{Nadine Neumayer\altaffilmark{2}}}
\author{\mbox{Sebastian Kamann\altaffilmark{3}}}
\author{\mbox{Karina T. Voggel\altaffilmark{1}}}
\author{\mbox{Michele Cappellari\altaffilmark{4}}} 
\author{\mbox{Arianna Picotti\altaffilmark{2}}}
\author{\mbox{Phuong M. Nguyen\altaffilmark{5}}}
\author{\mbox{Torsten B\"oker\altaffilmark{6}}}
\author{\mbox{Victor Debattista\altaffilmark{7}}}
\author{\mbox{Nelson Caldwell\altaffilmark{8}}}
\author{\mbox{Richard McDermid\altaffilmark{9}}}
\author{\mbox{Nathan Bastian\altaffilmark{3}}}
\author{\mbox{Christopher C. Ahn\altaffilmark{1}}}
\author{\mbox{Renuka Pechetti\altaffilmark{1}}}

\altaffiltext{1}{Department of Physics and Astronomy, University of Utah, 115 South 1400 East, Salt Lake City, UT 84112, USA\\
dieu.nguyen@utah.edu, aseth@astro.utah.edu, kvoggel@astro.utah.edu, renuka.pechetti@utah.edu, chris.ahn43@gmail.com}
\altaffiltext{2}{Max Planck Institut f\"ur Astronomie (MPIA), K\"onigstuhl 17, 69121 Heidelberg, Germany\\
neumayer@mpia.de, picotti@mpia.de}
\altaffiltext{3}{Astrophysics Research Institute, Liverpool John Moores University, 146 Brownlow Hill, Liverpool L3 5RF, UK\\
s.kamann@ljmu.ac.uk, N.J.Bastian@ljmu.ac.uk}
\altaffiltext{4}{Sub-department of Astrophysics, Department of Physics, University of Oxford, Denys Wilkinson Building, Keble Road, Oxford OX1 3RH, UK\\
michele.cappellari@physics.ox.ac.uk}
\altaffiltext{5}{Department~of~Physics, Quy~Nhon~University, 170~An~Duong~Vuong, Quy~Nhon, Vietnam\\
nguyenthiminhphuong@qnu.edu.vn}
\altaffiltext{6}{European Space Agency, c/o STScI, 3700 San Martin Drive, Baltimore, MD 21218, USA\\
tboeker@cosmos.esa.int}
\altaffiltext{7}{School of Physical Sciences and Computing, University of Central Lancashire, Preston, Lancashire, PR1 2HE, UK\\
vpdebattista@uclan.ac.uk}
\altaffiltext{8}{Harvard-Smithsonian Center for Astrophysics, Harvard University, 60 Garden St, Cambridge, MA 02138, USA\\
caldwell@cfa.harvard.edu}
\altaffiltext{9}{Department of Physics and Astronomy, Macquarie University, NSW 2109, Australia\\
richard.mcdermid@mq.edu.au}

%%%%%%%%%%%%%%%%%%%%%%%%%%%%%%%%%%%%%%%%%%%%%%%%%%%%%%%%%%%%%%%%%%%%
% WRITE THE ABSTRACT HERE
%%%%%%%%%%%%%%%%%%%%%%%%%%%%%%%%%%%%%%%%%%%%%%%%%%%%%%%%%%%%%%%%%%%%
\begin{abstract}\label{sec:abst}

  We present a detailed study of the nuclear star clusters (NSCs) and massive black holes (BHs) of four of the nearest low-mass early-type galaxies: M32, NGC~205, NGC~5102, and NGC~5206.  We measure dynamical masses of both the BHs and NSCs in these galaxies using Gemini/NIFS or VLT/SINFONI stellar kinematics, \emph{Hubble Space Telescope} (\hst) imaging, and Jeans Anisotropic Models.  We detect massive BHs in M32, NGC~5102, and NGC~5206, while in NGC~205, we find only an upper limit.  These BH mass estimates are consistent with previous measurements in M32 and NGC~205, while those in NGC~5102 \& NGC~5206 are estimated for the first time, and both found to be $<$$10^6$\Msun. This adds to just a handful of galaxies with dynamically measured sub-million \Msun~central BHs.   Combining these BH detections with our recent work on NGC~404's BH, we find that 80\% (4/5) of nearby, low-mass ($10^9-10^{10}$\Msun; $\sigma_{\star}\sim20-70$~\kms) early-type galaxies host BHs.   Such a high occupation fraction suggests the BH seeds formed in the early epoch of cosmic assembly likely resulted in abundant seeds, favoring a low-mass seed mechanism of the remnants, most likely from the first generation of massive stars.  We find dynamical masses of the NSCs ranging from $2-73\times10^6$\Msun~and compare these masses to scaling relations for NSCs based primarily on photometric mass estimates. Color gradients suggest younger stellar populations lie at the centers of the NSCs in three of the four galaxies (NGC~205, NGC~5102, and NGC~5206), while the morphology of two are complex and are best-fit with multiple morphological components (NGC~5102 and NGC~5206).  The NSC kinematics show they are rotating, especially in M32 and NGC~5102 ($V/\sigma_{\star}\sim0.7$).  


\end{abstract}
\maketitle

%%%%%%%%%%%%%%%%%%%%%%%%%%%%%%%%%%%%%%%%%%%%%%%%%%%%%%%%%%%%%%%%%%%%%
% INTRODUCTION SECTION
%%%%%%%%%%%%%%%%%%%%%%%%%%%%%%%%%%%%%%%%%%%%%%%%%%%%%%%%%%%%%%%%%%%%%%
\section{Introduction}\label{sec:intro}

Supermassive black holes (SMBHs) appear to be ubiquitous features of the centers of massive galaxies.    This is inferred in nearby galaxies, both from their dynamical detection \citep[see review by][]{Kormendy13}, and from the presence of accretion signatures \citep[e.g.,][]{Ho09}.     Furthermore, the mass density of these BHs in nearby massive galaxies is compatible with the inferred mass accretion of active galactic nuclei (AGN) in the distant universe \citep[e.g.,][]{Marconi04}.    However, the presence of central BHs is not well constrained in lower mass galaxies, particularly below stellar masses of $\sim$10$^{10}$~\Msun~\citep[e.g.,][]{Greene12, Miller15}.    These lower mass galaxy nuclei are almost universally populated by bright, compact nuclear star clusters (NSCs) \citep[e.g.,][]{Georgiev14, Denbrok14} with half-light/effective radius of $r_{\rm eff} \sim 3$~pc.    These NSCs are known to co-exist in some cases with massive BHs, including in the Milky Way \citep[MW][]{Seth08a, Graham09, Neumayer12}.

Finding and weighing central BHs in lower mass galaxies is challenging due to the difficulty of dynamically detecting the low-mass ($\lesssim$$10^6$\Msun) BHs they host.    However, it is a key measurement for addressing several related science topics.    First, low-mass galaxies are abundant \citep[e.g.,][]{Blanton05}, and thus if they commonly host BHs, these will dominate the number density (but not mass density) of BHs in the local universe.    This has consequences for a wide range of studies, from the expected rate of tidal disruptions \citep{Kochanek16}, to the number of BHs we expect to find in stripped galaxy nuclei \citep[e.g.,][]{Pfeffer14, Ahn17}.   Second, the number of low-mass galaxies with central BHs is currently most feasible way to probe the unknown formation mechanism of massive BHs.     More specifically, if massive BHs form from the direct collapse of relatively massive $\sim$10$^5$\Msun~seed black holes scenario \citep[e.g.,][]{Lodato06, Bonoli14}, few would be expected to inhabit low-mass galaxies, if, on the other hand, they form from the remnants of Population~III stars, a much higher ``occupation fraction'' is expected in low-mass galaxies \citep{Volonteri08,  vanWassenhove10, Volonteri10}.   This work is complementary to work that is starting to be undertaken at higher redshifts to probe accreting black holes at the earliest epochs \citep[e.g.,][]{Weigel15, Natarajan17, Volonteri17}.    Tidal disruptions are also starting to probe the BH population of low mass galaxies \citep[e.g.,][]{Wevers17, Law-Smith17}, and may eventually provide constraints on occupation fraction \citep[e.g.,][]{Stone16}.    Third, dynamical mass measurements of SMBHs ($M_{\rm SMBH}$~$\sim10^6-10^{10}$\Msun)  have shown that their masses scale with the properties of their host galaxies such as their bulge luminosity, bulge mass, and velocity dispersion \citep{Kormendy13, McConnell13, Graham15, Saglia16}.  These scaling relations have been used to suggest that galaxies and their central SMBHs coevolve, likely due to feedback from the AGN radiation onto the surrounding gas \citep[see review by][]{Fabian12}.

The black hole--galaxy relations are especially tight for massive early-type galaxies (ETGs), while later-type, lower dispersion galaxies show a much larger scatter \citep{Greene16, Lasker16}.  However, the lack of measurements at the low-mass, low-dispersion end, especially in ETGs means that our knowledge of how BHs populate host galaxies is very incomplete.

The population of known BHs in low-mass galaxies was recently compiled by \citet{Reines15}.  Of the BHs known in galaxies with stellar masses $<$10$^{10}$\Msun, most have been found by the detection of optical broad-line emission, with their masses inferred from the velocity widths of their broad-lines \citep[e.g.,][]{Greene07, Dong12, Reines13}.  Many of these galaxies host BHs with inferred masses below 10$^6$\Msun, especially those with stellar mass $\lesssim$$3 \times 10^9$\Msun.  Broad-line emission is however found in only a tiny fraction ($<$1\%) of low-mass galaxies; other accretion signatures that are also useful for identifying BHs in low-mass galaxies include narrow-line emission \citep[e.g.,][]{Moran14}, coronal emission in the mid-infrared \citep[e.g.,][]{Satyapal09}, tidal-disruption events \citep[e.g.,][]{Maksym13}, and hard X-ray emission \citep[e.g.,][]{Miller15, She17}.

The current record holder for the lowest mass central BH in a late-type galaxy (LTG) is the broad-line AGN in RGG~118 with $M_\star \sim 2 \times 10^9$~\Msun~\citep[\Mbh~=~$5\times10^4$\Msun;][]{Baldassare15}.  The lowest-mass systems known to host central massive BHs are ultracompact dwarfs, which are likely stripped galaxy nuclei \citep{Seth14, Ahn17}.  Apart from these systems, the lowest mass BH-hosting galaxies are the $M_\star \sim 3 \times 10^8$~\Msun~early-type dwarf in Abell~1795, where a central BH is suggested by the detection of a tidal-disruption event \citep{Maksym13}, and the similar mass dwarf elliptical galaxy Pox~52, which hosts a broad-line AGN \citep{Barth04, Thornton08}.  Currently, there are only a few galaxies with sub-million solar mass {\em dynamical} BH mass estimates, including NGC~4395 \citep[\Mbh~=~$4^{+8}_{-3}\times10^5$\Msun;][]{denBrok15}, NGC~404 \citep[\Mbh~$\lesssim1.5\times10^5$\Msun;][]{Nguyen17}, and NGC~4414 \citep[\Mbh~$\lesssim1.6\times10^6$\Msun;][]{Thater17}. 
 
Unlike BHs, the morphologies, stellar populations, and kinematics of NSCs provide an observable record for understanding mass accretion into the central parsecs of galaxies.  The NSCs' stellar mass accretion can be due to (1) the migration of massive star clusters formed at larger radii that then fall into the galactic center via dynamical friction \citep[e.g.,][]{Lotz01, Antonini13, Guillard16} or (2) the {\it in-situ} formation of stars from gas that falls into the nucleus \citep[e.g.,][]{Seth06, Antonini15b}.  Observations suggest the formation of NSCs is ongoing, as they typically have multiple populations  \citep[e.g.,][]{Rossa06, Nguyen17}.  Most studies on NSCs in ETGs have focused on galaxies in nearby galaxy clusters \citep{Walcher06}; spectroscopic and photometric studies suggest that typically the NSCs in these ETGs are younger than their surrounding galaxy, especially in galaxies $\lesssim2 \times 10^9$~M$_\odot$ \citep[e.g.,][]{Cote06, Paudel11, Krajnovic17, Spengler17}.  There is also a clear change in the scaling relations between NSC and galaxy luminosities and masses at about this mass, with the scaling being shallower at lower masses and steeper at higher masses \citep{Scott13, Denbrok14, Spengler17}.  This change in addition to the flattening of more luminous NSCs \citep{Spengler17} suggests a possible change in the formation mechanism of NSCs from migration to {\it in-situ} formation. However, the NSC scaling relations for ETGs are based almost entirely of photometric estimates, with a significant sample of dynamical measurements available only for nearby late-type spirals \citep{Walcher05} and only two in ETG FCC~277  \citep{Lyubenova13} and NGC~404 \citep{Nguyen17}.

The relationship between NSCs and BHs is not well understood. The sample of objects with both a detected NSC and BH is quite limited \citep{Seth08a, Graham09, Neumayer12, Georgiev16}, but even from this data it is clear that low-mass galaxy nuclei are typically dominated by NSCs, while high mass galaxy nuclei are dominated by SMBHs.  This transition could be due to a number of mechanisms, including the tidal disruption of infalling clusters  \citep{Ferrarese06, Antonini15a}, the growth of BH seeds in more massive galaxies by tidal disruption of stars \citep{Stone17}, or differences in feedback from star formation vs.~AGN accretion \citep{Nayakshin09b}.  In addition, the formation of NSCs and BHs could be explicitly linked with NSCs creating the needed initial seed BHs during formation \citep[e.g.,][]{PortegiesZwart04b}, or strong gas inflow creating the NSC and BH simultaneously \citep{Hopkins10a}.   However, the existence of a BH around which a NSC appears to be currently forming in the nearby galaxy Henize 2-10 suggests that BHs can form independently of NSCs \citep{Nguyen14}. 

This work presents a dynamical study of the BHs and NSCs of four nearby low-mass ETGs including the two brightest companions of Andromeda,  M32 and NGC~205, and two companions of Cen~A, NGC~5102 and NGC~5206.    Of these four galaxies, previous BH mass estimates exist for two; the detection of $(2.5\pm0.5)\times10^6$\Msun~BH in M32 \citep{Verolme02,  vandenBosch10}, and the upper limit of $3.8\times$$10^4$\Msun~(at 3$\sigma$) in NGC~205 \citep{Valluri05}.      In this paper, we use adaptive optics integral field spectroscopic measurements, combined with dynamical modeling using carefully constructed mass models to constrain the BH and NSC masses in all four galaxies, including the first published constraints on the BHs of NGC~5102 and NGC~5206.
 
This paper is organized into nine sections.     In Section~\ref{sec:data}, we describe the observations and data reduction.     The four galaxies' properties are presented in Section~\ref{sec:hosts}, and we construct luminosity and mass models in Section~\ref{sec:sbfs}.   We present kinematics measurements of their nuclei in Section~\ref{sec:kine}.    We model these kinematics using Jeans models and present the BH constraints and uncertainties in Section~\ref{sec:jeans}.   In section~\ref{sec:nscs} we analyze the properties of the NSCs and present our measurements of their masses.     We discuss our results in Section~\ref{sec:disc}, and conclude in Section~\ref{sec:concl_future}.    

%%%%%%%%%%%%%%%%%%%%%%%%%%%%%%%%%%%%%%%%%%%%%%%%%%%%%%%%%%%%%%%%%%%%%
% DATA SECTION 
%%%%%%%%%%%%%%%%%%%%%%%%%%%%%%%%%%%%%%%%%%%%%%%%%%%%%%%%%%%%%%%%%%%%%%
\section{Data and Data Reduction}\label{sec:data}
%%%%%%%%%%%
\subsection{HST Imaging}\label{ssec:hst}
The \hst/WFPC2 PC and ACS/HRC imaging we use in this work is summarized in Table~\ref{hst_data_tab}.   The nuclei of M32 and NGC~5206 are observed in the WFPC2 PC chip in F555W and F814W filters.    For NGC~205, we use the central ACS/HRC F555W and F814W images.  For NGC~5102, we use more imaging of WFPC2 including two central saturated F450W and F560W, unsaturated F547M, and H$\alpha$ emission F656N.    

%%%%%%%%%%%%%%%%%%%%%% Table 1 of HST WFPC2 data for 5 galaxies %%%%%%%%%%%%%%%%%%%%%%%%%%
\begin{table*}[ht]
\caption{\hst/WFPC2 PC/WF3 and ACS HRC Imaging}
\hspace{-4mm}
\begin{tabular}{ccccccccccccccc} 
\hline \hline
Object&$\alpha$(J2000)&$\delta$(J2000)&Camera&Aperture&   UT Date   &   PID &  Filter &Exptime& Pixelscale &Zeropoint$\tablenotemark{a}$&A$_{\rm \lambda}\tablenotemark{b}$\\
           &    (h~m~s)        &(\deg~\mm~\se)&            &              &                    &          &           &     (s)    &   (\se/pix)  &                     (mag)        &      (mag)  \\
  (1)    &           (2)          &           (3)          &  (4)      &    (5)      &       (6)        &    (7)  &    (8)   &     (9)    &      (10)     &                       (11)         &       (12)    \\	
\hline   
M32   &   00:42:38.37    &    40:51:49.3    &WFPC2&PC1-FIX&1994 Dec 26& 5236 &F555W& $4\times24$&0.0445&                   24.664       &    0.047     \\
          &                          &                         &WFPC2&PC1-FIX&1994 Dec 26& 5236 &F814W& $4\times24$&0.0445&                   23.758       &    0.026     \\
\hline
NGC~205 &  00:40:22.0  &     41:41:07.1   &    ACS/HRC  &  HRC  &2002 Sep 08 & 9448 &F555W&$4\times640$&0.0300&                    25.262      &    0.047    \\
          &                          &                         &    ACS/HRC &   HRC  &2002 Sep 08 & 9448 &F814W&$8\times305$&0.0300&                    24.861      &    0.026    \\
\hline
NGC~5102& 13:21:55.96&   -36:38:13.0    &WFPC2&PC1-FIX&1994 Sep 02&5400 &F547M& $8\times70$  &0.0445&                    23.781       &   0.050    \\
           &                         &                         &WFPC2&PC1-FIX&1994 Sep 02&5400 &F450W$^{\star}$&$11\times590$&0.0445&    24.106       &   0.058     \\
           &                         &                         &WFPC2&PC1-FIX&1994 Sep 02&5400 &F569W$^{\star}$&$11\times590$&0.0445&    24.460       &   0.043     \\
           &                         &                         &WFPC2&PC1-FIX&2001 May 27& 8591&F656N& $5\times340$ &0.0445&                   19.683       &   0.035     \\                      
 \hline
NGC~5206&13:33:43.92 &   -48:09:05.0    &WFPC2&PC1-FIX&1996 May 11& 6814&F555W& $6\times350$ &0.0445&                   24.664       &   0.047    \\
          &                          &                         &WFPC2&PC1-FIX&1996 May 11& 6814&F814W& $6\times295$ &0.0445&                   23.758       &   0.026     \\
 \hline
\end{tabular}
\tablenotemark{}
\tablecomments{\normalsize   Column 1: galaxy name. Columns 2 and 3: position (R.A. and Dec.) of the galaxy from \hst/HLA data.   Columns 4 and 5: the camera and the aperture in which the data were taken. Column 6: date when the observations were performed. Column 7: the principle investigator identification numbers. Column 8: filter. Column 9: the exposure times of the observations. Column 10: the pixel-scale of each camera.  Columns 11 and 12: the photometric zero point and extinction value in each filter.  The superscript $^{\star}$ indicates the nucleus was saturated in these data.}
\footnotetext[1]{\normalsize The photometric zero points were based on Vega System.}
\footnotetext[2]{\normalsize The extinction values A$_{\lambda}$ were obtained from \citet{Schlafly11} with interstellar extinction law from UV to near-infrared \citep[NIR;][]{Cardelli89}.}
\label{hst_data_tab}
\end{table*}
%%%%%%%%%%%%%%%%%%%%%% %%%%%%%%%%%%%%%%%%%%%%%%%%%%%%%%%%%%%%%%%%%%%%%

For the ACS/HRC data we downloaded reduced, drizzled images from the \hst/Hubble Legacy Archive (HLA). However, because the HLA images of the WFPC2 PC chip have a pixel-scale that is downgraded by 10\%, we re-reduce images, which are downloaded from the Mikulski Archive for Space Telescope (MAST), using \texttt{Astrodrizzle} \citep{Avila12} to a final pixel scale of 0$\farcs$0445.   For all images, we constrain the sky background by comparing them to ground-based data (see Section~\ref{ssec:1d}).   

We use the centroid positions of the nuclei to align images from all \hst~filters to the F814W (in M32, NGC~205, and NGC~5206) or F547M (in NGC~5102) data.    The astrometric alignment of  the Gemini/NIFS or VLT/SINFONI spectroscopic data (Section~\ref{ssec:spec}) was then also tied to the same images, providing a common reference frame for all images used in this study.

%%%%%%%%%%%
\subsection{Integral-Field Spectroscopic Data}\label{ssec:spec} 

\subsubsection{Gemini/NIFS Spectroscopy} \label{sssec:nifs}
 M32 and NGC~205 were observed with Gemini/NIFS using the Altair tip-tilt laser guide star system.   The M32 data were previously presented in \citet{Seth10b}.  The observational information can be found in Table~\ref{spec_tab}.     We use these data to derive stellar kinematics from the CO band-head absorption in Section~\ref{sec:kine}. 
 
 The data were reduced using the \texttt{IRAF} pipeline modified to propagate the error spectrum, for details see \citet{Seth10a}. The telluric calibration was done with A0V stars, for M32, HIP 116449 was used \citep{Seth10b}, while for NGC~205 we use HIP 52877.   The final cubes were constructed via a combination of six (M32) and eight (NGC~205) dithered on-source cubes with good image quality after subtracting offset sky exposures.      The wavelength calibration used both an arc lamp image and the skylines in the science exposures with an absolute error of $\sim$2~\kms.  
   
%%%%%%%%%%% Table 2 of Gemini/NIFS and SINFONI kinematic DATA for 4 galaxies %%%%%%%%%%%%   
\begin{table*}[ht] 
\centering
\caption{Gemini/NIFS and VLT/SINFONI Spectroscopic Data}
\begin{tabular}{cccccccccc}
\hline \hline
Object& Instrument &   UT Date   & Mode/ &       Exptime   &            Sky      &Pixel Scale&$\overline{\rm FWHM}$&$\overline{\rm FWHM}$&PID\\ 
          &                     &                   &  Telescope               &         (s)          &                        &  (\se/pix)   &              (\AA)               &               (\kms)             &      \\
 (1)    &         (2)        &       (3)        &        (4)       &          (5)         &            (6)        &      (7)       &                (8)                 &                  (9)                & (10)\\[1mm]   
\hline 
M32         &     NIFS  &2005 Oct 23& AO + NGS &$ 6\times600$ & $ 5\times600$&    0.05      &             4.2                   &                57.0                &GN-2005B-SV-121\\
NGC~205  &     NIFS  &2008 Sep 19& AO + LGS &$ 8\times760$ & $ 6\times760$&    0.05      &             4.2                   &                55.7                &GN-2008B-Q-74\\
NGC~5102&SINFONI &2007 Mar 21&UT4-Yepun&$12\times600$& $12\times600$&   0.05      &             6.2                   &                82.2                &078.B-0103(A)\\
NGC~5206&SINFONI &2011 Apr 28&UT4-Yepun &$ 3\times600$ & $ 3\times600$&    0.05      &             6.2                   &                82.2                &086.B-0651(B)\\
NGC~5206&SINFONI &2013 Jun 18&UT4-Yepun&$ 3\times600$ & $ 3\times600$&    0.05      &             6.2                   &                82.2                &091.B-0685(A)\\
NGC~5206&SINFONI &2014 Mar 22&UT4-Yepun&$ 3\times600$ & $ 3\times600$&   0.05       &             6.2                   &                82.2                &091.B-0685(A)\\
\hline
\end{tabular}
\tablenotemark{}
\tablecomments{\normalsize   Column 1: galaxy name. Column 2: Instrument used. Column 3: the Universal Time date when the observations were processing. Column 4: additional information on AO/telescope used.  Column 5:  the exposure times of the observations. Column 6:  offset sky exposures used. Column 7: the pixel scale of each camera.  Columns 8 and 9: the median FWHM (in wavelength and velocity) of the LSF. Column 10: the program identification numbers.}
\label{spec_tab}
\end{table*}
%%%%%%%%%%%%%%%%%%%%%%%%%%%%%%%%%%%%%%%%%%%%%%%%%%%%%%%%%%%%%%%%%%%%%%
%%%%%%%%%%% Table 3 of Gemini/NIFS and SINFONI kinematic PSF for 4 galaxies %%%%%%%%%%%%  
\begin{table}[ht]
\caption{Gemini/NIFS and VLT/SINFONI PSF models}
\hspace{-2mm}
\begin{tabular}{cccccc}
\hline \hline 
Object&Observation&Gaussian 1&Gaussian 2&Moffat\\ 
          &        &  (\se/frac.) &  (\se/frac.) & (\se/frac.)\\
 (1)    &  (2)  &   (3)   &    (4)  &  (5)  \\[1mm]   
\hline 
M32$\tablenotemark{a}$&   NIFS  &0.250/45\%&        ...        &0.85/55\% \\
NGC~205                        &   NIFS   &0.093/68\%&        ...        &0.92/32\% \\
\hline
NGC~5102                      &SINFONI&0.079/35\%&0.824/65\%$\tablenotemark{b}$& ... \\   
NGC~5206                      &SINFONI&0.117/60\%&0.422/40\%$\tablenotemark{b}$& ... \\
                                      &              &0.110/53\%&0.487/47\%$\tablenotemark{c}$& ...  \\
\hline
\end{tabular}
\tablenotemark{}
\tablecomments{\normalsize   Column 1: galaxy name. Column 2: Instrument. Columns 3 and 4: the FWHMs/light fractions of components of the kinematic PSFs. Column 5:  the Moffat profile $r_d$ (half-light radius)/light fraction. \footnotetext[1]{\normalsize \citet{Seth10b}} \footnotetext[2]{\normalsize Scattered light subtraction (default).} \footnotetext[3]{\normalsize Half scattered light subtraction.}}
\label{kinepsf_tab}
\end{table}
%%%%%%%%%%%%%%%%%%%%%%%%%%%%%%%%%%%%%%%%%%%%%%%%%%%%%%%%%%%%%%%%%%%%%%
%%%%%%%%%%% 
\subsubsection{VLT/SINFONI Spectroscopy} \label{sssec:sinfoni}
NGC~5102 and NGC~5206 were observed with SINFONI \citep{Eisenhauer03, Bonnet04} on the UT4 (Yepun) of the European Southern Observatory's Very Large Telescope (ESO VLT) at Cerro Paranal, Chile. NGC~5102 was observed in two consecutive nights in March 2007 as part of the SINFONI GTO program (PI: Bender). A total of 12 on-source exposures of 600s integration time were observed in the two nights using the 100 mas pixel scale. NGC~5206 was observed in service mode in three different years (2011, 2013, and 2014; PI: Neumayer) with three on-source exposures of 600s in each of the runs. Both of the targets were observed using the laser guide star for the adaptive optics correction and used the NSC itself for the tip-tilt correction. The spectra were taken in $K$--band (1.93--2.47$\mu$m) at a spectral resolution of R~$\sim$~4000, covering a field of view of $3\arcsec\times3\arcsec$. The data were reduced using the ESO SINFONI data reduction pipeline, following the steps (1) sky-subtraction, (2) flat-fielding, (3) bad pixel correction, (4) distortion correction, (5) wavelength calibration, (6) cube reconstruction, and finally (8) telluric correction.  Details on these steps are given in \citet{Neumayer07}. The final data cubes were constructed via a combination of the individual dithered cubes. This leads to a total exposure time of 7200s for NGC~5102 and 5400s for NGC~5206.  

%%%%%%%%%%% 
\subsection{ Point-Spread Function Determinations}\label{ssec:psf} 
Our analysis requires careful characterization of the point-spread functions (PSFs) in both our kinematic and imaging data. The \hst~PSFs are used in fitting the two-dimensional (2D) surface brightness profiles of the galaxies \citep[GALFIT;][]{Peng10}, while the Gemini/NIFS and VLT/SINFONI PSFs are used for the dynamical modeling. 

For the~\hst~PSFs, we create the model PSF for each \hst~exposure using the {\tt Tiny Tim} routine for each involved individual filter for each object.   The PSFs are created using the $\tt{tiny1}$ and $\tt{tiny2}$ tasks \citep{Krist95, Krist11}.     We next insert these into corresponding {\tt c0m} (M32, NGC~5102, and NGC~5206) or {\tt flt} (NGC~205) images at the positions of the nucleus in each individual exposure to simulate our observations and use {\tt Astrodrizzle} to create our final PSF as described in \citet{denBrok15} and \citet{Nguyen17}.    

For the PSFs of our integral field spectra, we convolve the \hst~images with an additional broadening to determine the two component PSF expected from adaptive optics observations.   The Gemini/NIFS PSFs  for M32 and NGC~205 are estimated as described in \citet{Seth10a}.  First, the outer shape of the PSF was constrained using images of the telluric calibrator Moffat profile ($\Sigma(r)=\Sigma_0/[(1+(r/r_d)^2)]^{4.765}$).  We then convolved the \hst~images with a inner Gaussian $+$ outer Moffat function (with fixed shape but free amplitude) to best match the shape of the NIFS continuum image; we refer to these as G + M PSFs.  The full widths at half maximum (FWHM) and light fractions of components are presented in Table~\ref{kinepsf_tab}.

The VLT/SINFONI data cubes of NGC~5102 and NGC~5206 contain an apparent scattered light component that affects our data in two ways (1) it creates a structured non-stellar background in the spectra which we subtract off before measuring kinematics, and (2) it is uniform across the field and therefore  creates a flat outer profile that cannot be explained with a reasonable PSF.  Because this scattered light does not contribute to our kinematic measurements, we do our best to remove this before fitting the PSF.  Unfortunately, this component is degenerate with the true PSF.  To measure the level of this scattered light, we compare the surface brightness of the SINFONI data to a scaled version of the \hst.  Because the level of the background in the SINFONI continuum image is much larger than the expected background, we simply subtract of the difference to the \hst~background before fitting our PSFs.     However, as this likely results in somewhat of an oversubtraction, we also create an additional PSF where just half the original value was subtracted off the SINFONI data.  This had a negligible effect on the PSF of NGC~5102, however it was more significant for NGC~5206 ($\Delta{\rm FWHM}\sim18$\%).   We will discuss the impact of this PSF uncertainty on our dynamical modeling in Section~\ref{ssec:sinfoni_psf}.  For the functional form of these PSF, we test several functions and find that the best-fit PSFs are double-Gaussians (2G); their FWHMs and light fractions are presented in Table~\ref{kinepsf_tab}.  We note that this is different from the NIFS PSF, which we found was better characterized by a Gauss+Moffat function for M32 and NGC~205.

%%%%%%%%%%%%%%%%%%%%%%%%%%%%%%%%%%%%%%%%%%%%%%%%%%%%%%%%%%%%%%%%%%%%%%
% THE SECTION OF NUCLEAR STAR CLUSTERS: SERSICS PROFILES & MASSES 
%%%%%%%%%%%%%%%%%%%%%%%%%%%%%%%%%%%%%%%%%%%%%%%%%%%%%%%%%%%%%%%%%%%%%%
\section{Galaxies Sample Properties}\label{sec:hosts}

Our sample of galaxies was chosen from known nucleated galaxies within 3.5~Mpc.    We look at the completeness of this sample by examining the number of ETGs (numerical Hubble T~$\leq0$) with total stellar masses between $5\times10^8$\Msun~$\leq~M_{\star}\leq1\times10^{10}$\Msun~using the updated nearby galaxy catalog \citep{Karachentsev13}.     To calculate the total stellar masses of galaxies in the catalog, we use their $K_s$ band luminosities and assume an uniform \ml$_{K}\sim1$ (\Msun/\Lsun) for our sample. Only five ETGs are in this stellar mass range and within 3.5 Mpc, including the four in this work (M32, NGC~205, NGC~5102, and NGC~5206) -- the one other galaxy in the sample is NGC~404, which was the subject of our previous investigation  \citep{Nguyen17}.     We note that within the same luminosity/mass and distance range, there are 17 total galaxies; the additional galaxies are all LTGs (including the LMC, M33, NGC~0055, and NGC~2403).   Based on this, the four galaxies in this sample plus NGC~404 form a complete, unbiased sample of ETGs within 3.5~Mpc.  However, we note that these ETGs live in a limited range of environments from isolated (NGC~404) to the Local Group and the Cen~A group, and thus our sample does not include the dense cluster environment in which many ETGs live.  

Before continuing with our analysis, we discuss the properties of the four galaxies in our sample in this section, summarizing the result in Table~\ref{host_tab}.  The listed total stellar masses are based on our morphological fits to a combination of ground-based and \hst~data, presented in Section~\ref{ssec:1d} and \ref{ssec:2d}.  Specifically, we take the total stellar luminosities of each best-fit component in our GALFIT models, and multiply by a best-fit \ml~based on the color using the relations of \citet{Roediger15}; we note that using color-\ml~relations based on a Salpeter IMF give higher masses by a factor of $\sim$2 \citep[e.g.][]{Bell01,Bell03}.   We consider these total masses to be accurate estimates for the bulge/spheroidal masses as well, as no significant outer disk components are seen in our fits.  We note how these compare to previous estimates for each galaxy below.
 
\subsection{M32}\label{ssec:m32}
M32 is a dense dwarf elliptical or compact elliptical, with the highest central density \citep[$\rho>10^7$\Msun~pc$^{-3}$ within $r\lesssim0.1$~pc;][]{Lauer98} in the Local Group.  It has a prominent NSC component visible in the surface brightness (SB) profile at radii $\lesssim$$5\arcsec$ (20~pc) that accounts for $\sim$10\% of the total mass of the stellar bulge \citep{Graham09}.      This NSC light fraction is much larger than typical ETG NSCs \citep[e.g.,][]{Cote06}.    Despite this clear SB feature, there is no evidence of a break in stellar kinematics or populations \citep{Seth10b}.     Specifically, the rotational velocity decreases slowly with radius \citep{Dressler88, Seth10b}, while the stellar population age increases gradually outwards, with average ages from 3~Gyr (nucleus) to 8~Gyr (large radii) with metallicity values at the center of [Fe/H]$~=~$0--0.16 and declining outwards \citep{Worthey04, Rose05, Coelho09,Villaume17b}.       

The central BH has previously been measured using SAURON stellar kinematics combined with STIS kinematics by \citet{Verolme02}.   Their axisymmetric Schwarzschild code gives a best fit mass of \Mbh~$=(2.5\pm0.5)\times10^6$\Msun.   This value was accurately and independently confirmed by \citet{vandenBosch10} using a triaxial Schwarzschild modeling code.  Accretion onto this BH is indicated by a faint X-ray and radio source \citep[][]{Ho03, Yang15}, which suggests the source is accreting at $\sim$$10^{-9}$ of its Eddington limit.  However, \citet{Seth10b} finds nuclear emission from hot dust in $K$--band with a luminosity more than 100$\times$ the nuclear X-ray luminosity.

Our total stellar luminosity in $I$--band is $5.5\times10^8$\Lsun, and we estimate an average photometric $M/L_I=1.7$ based on the color--\ml~relation \citep{Roediger15} of the full galaxy.  This gives a photometric stellar mass of the galaxy/bulge of $9.4\times10^8$\Msun~(Section~\ref{ssec:2d} and Table~\ref{seric_nsc_tab}).   This photometric estimate is in relatively good agreement with the previous dynamical estimates \citet{Richstone72,Haring04, Kormendy13}.  In \citet{Graham09}, the bulge luminosity is $\sim$38\% lower due possibly to separation of a disk component dominating their fits at radii $>$100$\arcsec$; combined with the much lower \ml~($M/L_I = 0.92$) they assume based on the result from \citet{Coelho09}, their bulge mass is just $2.6\times10^8$\Msun.

%%%%%%%%%%% Table 4 of literater properties of 4 galaxy + nuclei + nucleus kinematic of 4 galaxies %%%%%%%%%%%%     determined through profile fitting
\begin{table*}[ht]
\caption{Host Galaxies Properties}
\begin{tabular}{cccccc}
\hline \hline                                                                          
                                                                          &    M32         &     NGC~205         &       NGC~5102      &       NGC~5206   &  Unit\\                                          
\hline 
Distance                                                    &     0.79 [1]   &        0.82 [2]        &           3.2 [3]        &            3.5 [4]   & (Mpc)\\
$m-M$                                                      &     24.49      &          24.75         &           27.52          &            27.72    & (mag)\\
Physical Scale                                          &        4.0       &           4.3            &           16.0            &             17.0     & (pc/$\arcsec$)\\
$M_{B,0}$, $M_{V,0}$, $M_{I,0}$ &$-16.3$ [5], ..., $-16.7$ [6]&$-15.0$ [7], ..., $-16.5$ [7]&$-16.5$ [5], ..., ... & $-16.2$ [5], ..., ... & (mag)\\
%Total Stellar Mass (\Msun) &1.0$\times10^9$ [8]&1.0$\times10^9$ [11]&$7.0\times10^9$ [9]&2.8$\times10^9$ [CGS]\\
Photometric Total Stellar Mass  &1.00$\times10^9$ [$\star$]&9.74$\times10^8$ [$\star$]&6.0$\times10^9$ [$\star$]&2.4$\times10^9$ [$\star$] & (\Msun)\\
Dynamical Total Stellar Mass  &1.08$\times10^9$ [$\star$]&1.07$\times10^9$ [$\star$]&6.9$\times10^9$ [$\star$]&2.5$\times10^9$ [$\star$] & (\Msun)\\
Effective Radius (${\rm r}_{\rm eff.}^{\rm galaxy}$)& 30$\arcsec$/120 [8, $\star$]&121$\arcsec$/520 [9, $\star$]  & 75$\arcsec$/1200 [10, 11, $\star$]   &  58$\arcsec$/986 [$\star$]  & ($\arcsec$ or pc)\\
%$\sigma_{\star}$ (peak/average) (\kms)&120/80 [6] &21/18 [14]&58.4/42.3 [15, $\star$]& 45.3/35.4 [$\star$] & \\
$v_{\rm sys.}^{\rm NED}$ or $v_{\rm sys.}^{\rm Measured}$ &$-200$/$-201$&$-241$/$-241$&$+473$/$+472$& $+571$/$+573$ & (\kms)\\
%Position Angle$^{\rm NED}$ (\deg)&45.5&170.0&48.0&45.0 \\
Inclination       &  70.0 [12] & ...  & 86.0 [12]& ...  & (\deg) \\[1mm]
\hline
\end{tabular}
\tablenotemark{}
\tablecomments{\normalsize   The subscripts $_{B, V, I}$ indicate the measurements in $B-$, $V-$, and $I-$band.  NED: NASA/IPAC Extragalactic Database.  \tiny{[CGS]}\small: indicates the Carnegie-Irvine Galaxy Survey color profiles for total stellar masses assuming \citet{Roediger15} color--\ml~relation.    References -- [1]: \citet{Welch86}; [2]: \citet{McConnachie05}; [3]: \citet{vandenBergh76}; [4]: \citet{Tully15};  [5]: \href{HypeLEDA}{http://leda.univ-lyon1.fr/search.html}; [6]: \citet{Seth10b};  [7]: \citet{Geha06}; [8]: \citet{Graham02}; [9]: \citet{DeRijcke06}, [10]: \citet[][LGA]{Jarrett03}; [11]: \citet{Davidge08}; [12]:  \citet{Verolme02}; [$\star$]: estimates from this work.  To calculate properties from multiple references, we take their mean.  }     
\label{host_tab}  %[8]: \citet{Richstone72}; [9]: \citet{Davidge15}; [14]: \citet{Valluri05}; [15]: \citet{Mitzkus17}
\end{table*}   
%%%%%%%%%%%%%%%%%%%%%%%%%%%%%%%%%%%%%%%%%%%%%%%%%%%%%%%%%%%%%%%%%%%%%

\subsection{NGC~205}\label{ssec:n205}
NGC~205 is a nucleated dwarf elliptical galaxy hosting stars with a range of ages \citep{Davidge05, Sharina06}.    \hst~observations of the inner $29\arcsec\times26\arcsec$ around the nucleus of NGC~205 show a population of bright blue stars with ages ranging from 60--400~Myr and a metallicity $[M/H]\sim-0.5$ \citep{Monaco09}.  This confirms that multiple star formation episodes have occurred within the central $\sim$100~pc of the galaxy \citep{Cappellari99, Davidge03}.

\citet{Valluri05} and \citet{Geha06} use STIS spectra of the \ion{Ca}{2} absorption triplet region to measure the NGC~205 nuclear kinematics.    They find no rotation and a flat dispersion of $\sim$20~\kms~within the central $0\farcs3$ ($\sim$1~pc).     The combination of \hst~images, STIS spectra of the NGC~205 nucleus, and three-integral axisymmetric dynamical models give an upper limit mass of a putative BH at the center of the galaxy of $\sim$3.8$\times10^4$\Msun~\citep[within 3$\sigma$;][]{Valluri05}.    

Our total stellar luminosity in $I-$band is $5.6\times10^8$\Lsun, and we estimate an average photometric \ml$_I$ of the whole galactic body based on color--\ml~relation \citep{Roediger15} of 1.8 (\Msun/\Lsun) and photometric stellar mass of $9.8\times10^8$\Msun~(Section~\ref{ssec:2d} and Table~\ref{seric_nsc_tab}).   The total mass within $\sim$1~kpc (2$R_e$) was dynamically estimated to be $1.0\times10^9$~\Msun~by \citet{DeRijcke06}, with the stellar mass making up $\sim$60\% of this mass and thus matching quite closely our stellar mass estimate.  \citet{DeRijcke06} also track the kinematics of the galaxy out to 1.2~kpc, and find the dispersion slowly rises from $\sim$30 to 45~\kms, with rotation in the outer regions up to $\sim$20~\kms.

\subsection{NGC~5102}\label{ssec:n5102}
NGC~5102 is an S0 post-starburst galaxy in the nearby Cen A group \citep{DeHarveng97, Davidge08}.   It has an \ion{H}{1} disk extending to $>$5$\arcmin$ ($\sim$4.8~kpc) \citep{vanWoerden96} as well as extended ionized gas and dust emission \citep{McMillan94, Xilouris04a}.  Resolved stars and spectral synthesis studies show the presence of young stars and a significant intermediate age ($\sim$100~Myr--3~Gyr) population which appears to dominate the stellar mass near the center, with the mean age growing with radius \citep{Kraft05, Davidge08, Davidge15, Mitzkus17}.  \citet{Mitzkus17} finds that the nucleus has a mass-weighted age of 0.8~Gyr, with the best-fit suggesting a young $\sim$300~Myr Solar metallicity population overlying an old, metal-poor population ($>10$~Gyr, [Fe/H] $< -1$).   

\citet{Mitzkus17} also measure the kinematics of NGC~5102 ($r\lesssim30\arcsec$) with MUSE data and find a flat dispersion of $\sim$44~\kms~at radii larger than $0\farcs5$ and a dispersion peak of 60~\kms~at the center.  They also find a maximum rotation amplitude of 20~\kms~at a radius of $10\arcsec$.  A key characteristics of this galaxy is that it has two clear counter-rotating disks \citep{Mitzkus17}; this results in two dispersion peaks making it a ``2$\sigma$" galaxy.

A nuclear X-ray point source was detected by \citet{Kraft05}; this point source has a luminosity of $\sim$10$^{37}$ ergs s$^{-1}$ in the 0.5--2.0 keV band.   No previous estimates exist for its central BH mass.

Our total stellar luminosity in $V$--band is $2.2\times10^9$\Lsun, and we estimate an average photometric \ml$_V$ of the whole galactic body based on color--\ml~relation \citep{Roediger15} of 2.7 (\Msun/\Lsun)  and a photometric stellar mass of $6.0\times10^9$\Msun~(Section~\ref{ssec:2d} and Table~\ref{seric_nsc_tab}); this mass is 14\% lower than the previous estimates of the galaxy total stellar mass \citep{Davidge08}.

\subsection{NGC~5206}\label{ssec:n5206}
NGC~5206 is a poorly studied dE/S0 galaxy in the Cen~A group.  It was previously found by \citet{Caldwell87b} that the structure of the galaxy was not well fit just by a nucleus and a single component.  The Carnegie-Irvine color profiles show it has a bluer center suggesting younger stellar populations \citep{Li11}, while spectral synthesis fits to Xshooter data of its nucleus (N. Karcharov {\em et al., in prep}) suggest a wide range of ages in the NSC with the most recently formed stars being a population of Solar metallicity stars formed $\sim$1 Gyr ago. A previous measurement using has found a central velocity dispersion of $39\pm5$~\kms~\citep{Peterson93} and $41.5\pm6.5$~\kms~\citep{Wegner03}.

Our total stellar luminosity in $I$--band is $1.2\times10^9$\Lsun, and we  estimate an average photometric dynamical \ml$_I$ of the whole galactic body based on color--\ml~relation \citep{Roediger15} of 1.98 (\Msun/\Lsun) and photometric stellar mass of $2.4\times10^9$\Msun~(Section~\ref{ssec:2d} and Table~\ref{seric_nsc_tab}).

%%%%%%%%%%%%%%%%%%%%%%%%%%%%%%%%%%%%%%%%%%%%%%%%%%%%%%%%%%%%%%%%%%%%%%
\begin{figure*}[ht]
\minipage{0.5\textwidth}
  \includegraphics[width=\linewidth]{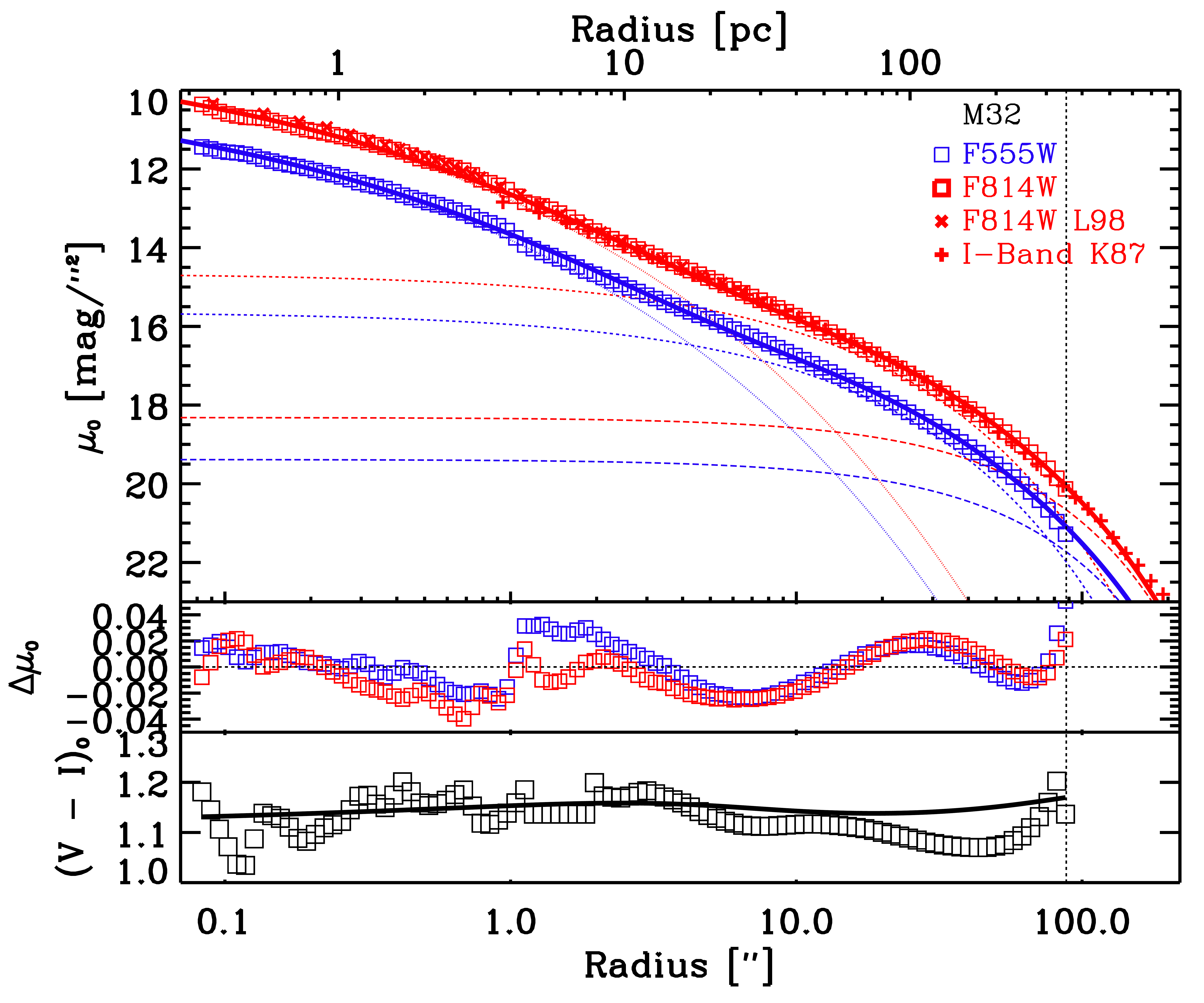}
\endminipage\hfill
\minipage{0.5\textwidth}
  \includegraphics[width=\linewidth]{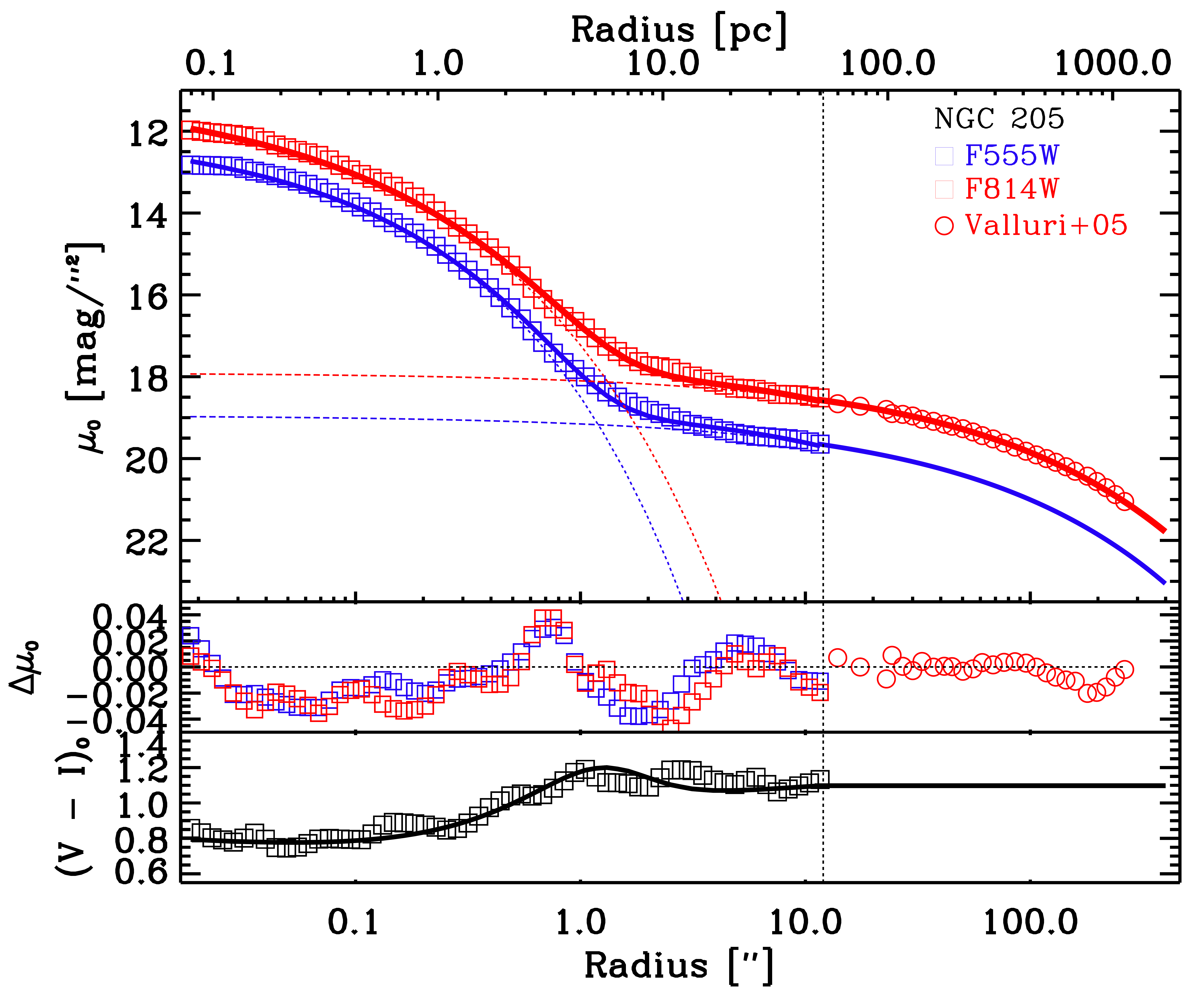}
\endminipage\hfill
\vspace{2mm}
\minipage{0.5\textwidth}
  \includegraphics[width=\linewidth]{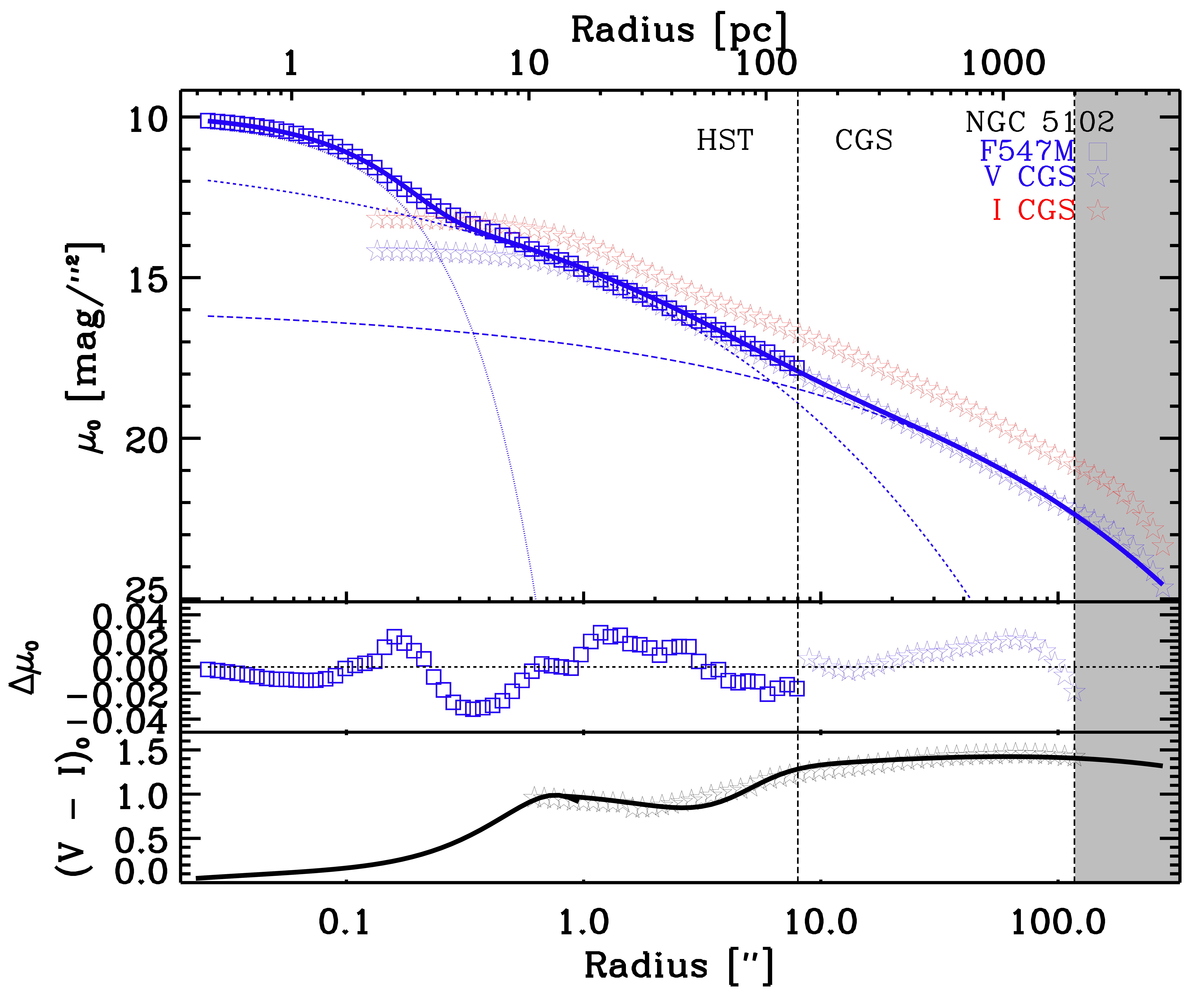}
\endminipage\hfill
\minipage{0.5\textwidth}
  \includegraphics[width=\linewidth]{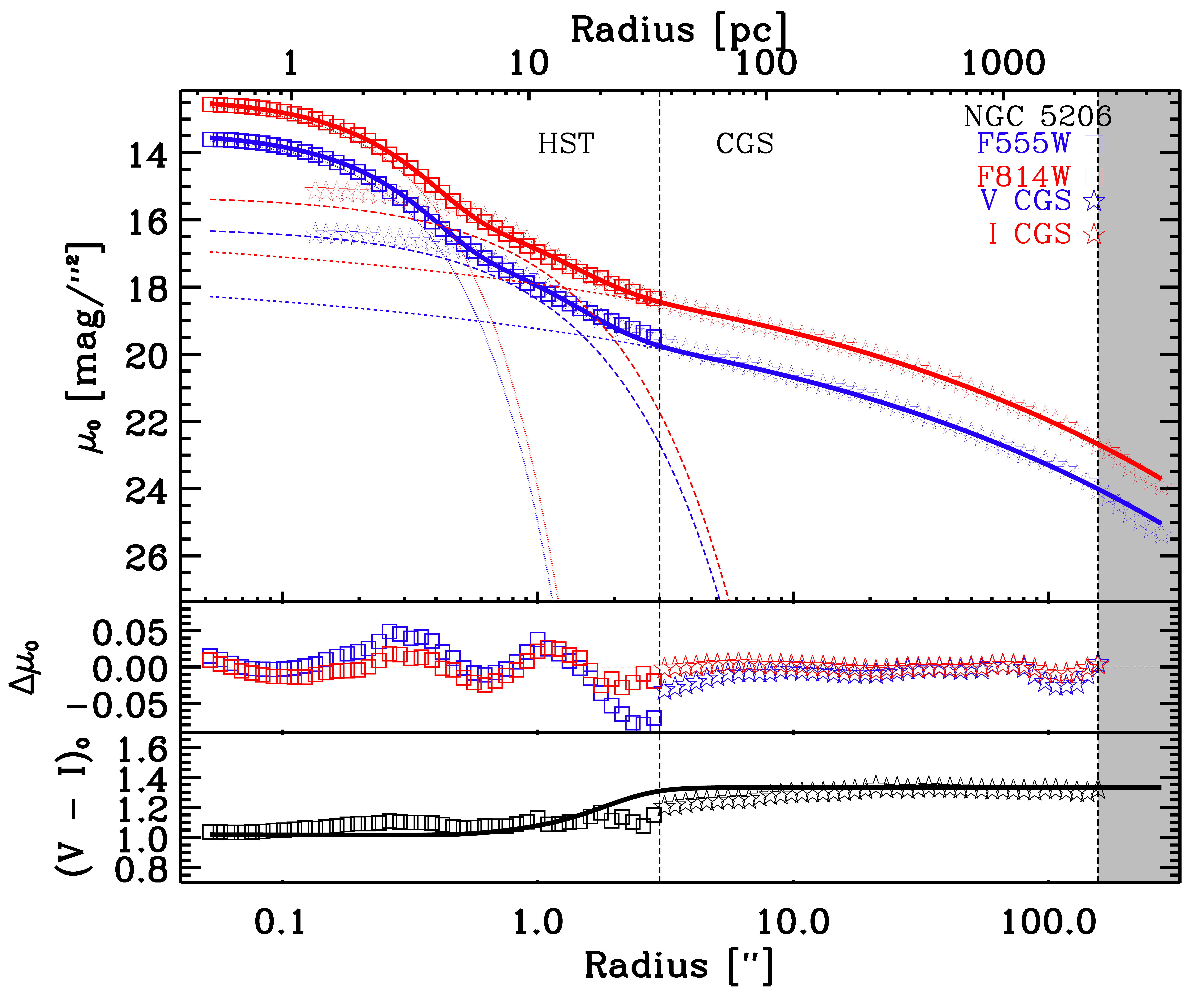}
\endminipage\hfill
\caption{\normalsize The surface brightness profiles of each galaxy studied here. The profiles are constructed from a combination of ground-based and \hst~imaging.  The blue lines show $V$/F555W/F547M profiles, while the red lines show $I$/F814W profiles.  All magnitudes/colors are corrected for foreground extinction.  Symbols show the data, while the best fitting 1D surface brightness profiles are shown as solid lines.   Each component of the best-fit models are also plotted for visualization (the innermost S\'ersic component is shown in dotted line, the second largest is shown as a dashed, and when present, the outermost component is shown in the long-dashed line).    The residuals of the fits are shown in the middle-panel plots and the $(V-I)_o$ color profiles are illustrated in the bottom-panel plots including the data (open-black symbols) and the best-fit model (black solid line).   Different symbols are plotted correspondingly to their data set, which are shown in the legend of each top panel.  The inner and outer vertical dashed lines show the ends of the \hst~and ground-based data, respectively.   The grey regions in the plots of NGC~5102 and NGC~5206 indicate the areas beyond our fitting radius.}
\label{sbp_lmp}
\end{figure*}
%%%%%%%%%%%%%%%%%%%%%%%%%%%%%%%%%%%%%%%%%%%%%%%%%%%%%%%%%%%%%%%%%%%%%%%

%%%%%%%%%%%%%%%%%%%%%%%%%%%%%%%%%%%%%%%%%%%%%%%%%%%%%%%%%%%%%%%%%%%%%%%
% THE SECTION OF 1D AND 2D SURFACE BRIGHTNESS PROFILES
%%%%%%%%%%%%%%%%%%%%%%%%%%%%%%%%%%%%%%%%%%%%%%%%%%%%%%%%%%%%%%%%%%%%%%%
\section{Surface Brightness Profiles}\label{sec:sbfs}
%%%%%%%%%%%%
\subsection{Large Scale Structure from the one-dimensional (1D) surface brightness (SB) Profiles}\label{ssec:1d}
To characterize the NSCs and create mass models for our targets, we first investigate the central SB profiles using \hst~imaging combined with larger scale SB profiles from the ground-based data.    After fitting for the large scale properties of the galaxy, we fit the smaller scale structure near the center in Section~\ref{ssec:2d} using  GALFIT.

For our 1D profiles, we use the IRAF \texttt{ellipse} \citep{Jedrzejewski87} routine to extract fluxes in the annuli as a function of the major semi-axis.     While extracting the fluxes, we allow the position angle (PA) and ellipticity ($\epsilon$) to vary.  

Due to the small field of view (FOV) of the \hst~data, the sky background is not easy to estimate.   To solve this problem, we use the existing ground-based data to estimate the sky backgrounds; the existing ground-based data include \citet{Kent87} and \citet{Lauer98} for M32, \citet{Kent87} and \citet{Valluri05} for NGC~205;  and Carnegie-Irvine Galaxy Survey \citep[CGS;][]{Ho11, Li11, Huang13} for NGC~5102 and NGC~5206.      We use data at intermediate radii to match the \hst~SB and ground-based SB with shifts of $<$0.15~mags.  Next we determine the sky background in the \hst~images by matching them out to larger radii.  In the end, our 1D SB profiles are calibrated in Vega magnitudes \citep{Sirianni05} and are corrected for Galactic extinction (Table~\ref{hst_data_tab}).

We fit the combined 1D SB profiles of each galaxy with multiple S\'ersic profiles using the nonlinear least-squares IDL \texttt{MPFIT} function \citep{Markwardt09}\footnote{available from {\tt http://purl.com/net/mpfit}}.   Because we do not do PSF convolution, we use these fits to constrain just the outer components of the fit, the best fits are shown in Figure~\ref{sbp_lmp} and Table~\ref{seric_nsc_tab}.  The NGC~205 SB profiles are well fit by a double S\'ersic function (NSC $+$ galaxy), however, the 1D SB of M32 requires two outer S\'ersics profiles (the outermost S\'ersic is an exponential) + NSC, while NGC~5102 and NGC~5206 require two NSC S\'ersic components + a galaxy component based on their 2D fits.   We note that our exponential disk component of M32 and the outer S\'ersis component of NGC~205 are fully consistent with \citet{Graham02} and \citet{Graham09}, and these components will be fixed in 2D GALFIT (Section~\ref{ssec:2d}).  The robustness of our 1D fits are tested by changing the outer boundaries in the ranges of 70$\arcsec$--90$\arcsec$, 100$\arcsec$--200$\arcsec$, 90$\arcsec$--110$\arcsec$, and 90$\arcsec$--120$\arcsec$ for M32, NGC~205, NGC~5102, and NGC~5206, respectively.    The standard deviation of the S\'ersic parameters from these fits was used to determine the errors; these errors are $<$10\% in all cases. The fits are performed in multiple bands, enabling us to model the color variation.

We show the 1D SB profile fits, residuals, and $(V-I)_{\rm o}$ color of each galaxy in the top- middle-, and bottom-panel of each plot in Figure~\ref{sbp_lmp}.   These models agree well within the data with \texttt{MEAN(ABS((data-model)/data)} \textless5\% for all four galaxies.  M32 shows no radial color gradient, consistent with previous observations \citep{Lauer98}.  The other three galaxies show bluer colors toward their centers.  We also note that we use these models only to constrain the outer S\'ersic components of the galaxies, the best-fit inner components are derived from 2D modeling of the \hst~images.

%%%%%%%%%%%%%%%%%%%%%%%%%%%%%%%%%%%%%%%%%%%%%%%%%%%%%%%%%%%%%%%%%%%%%%
\begin{figure*}[ht]
    \begin{minipage}{\linewidth}\hspace{-10mm}
       \includegraphics[width=0.37\linewidth,height=0.25\textheight,keepaspectratio=true]{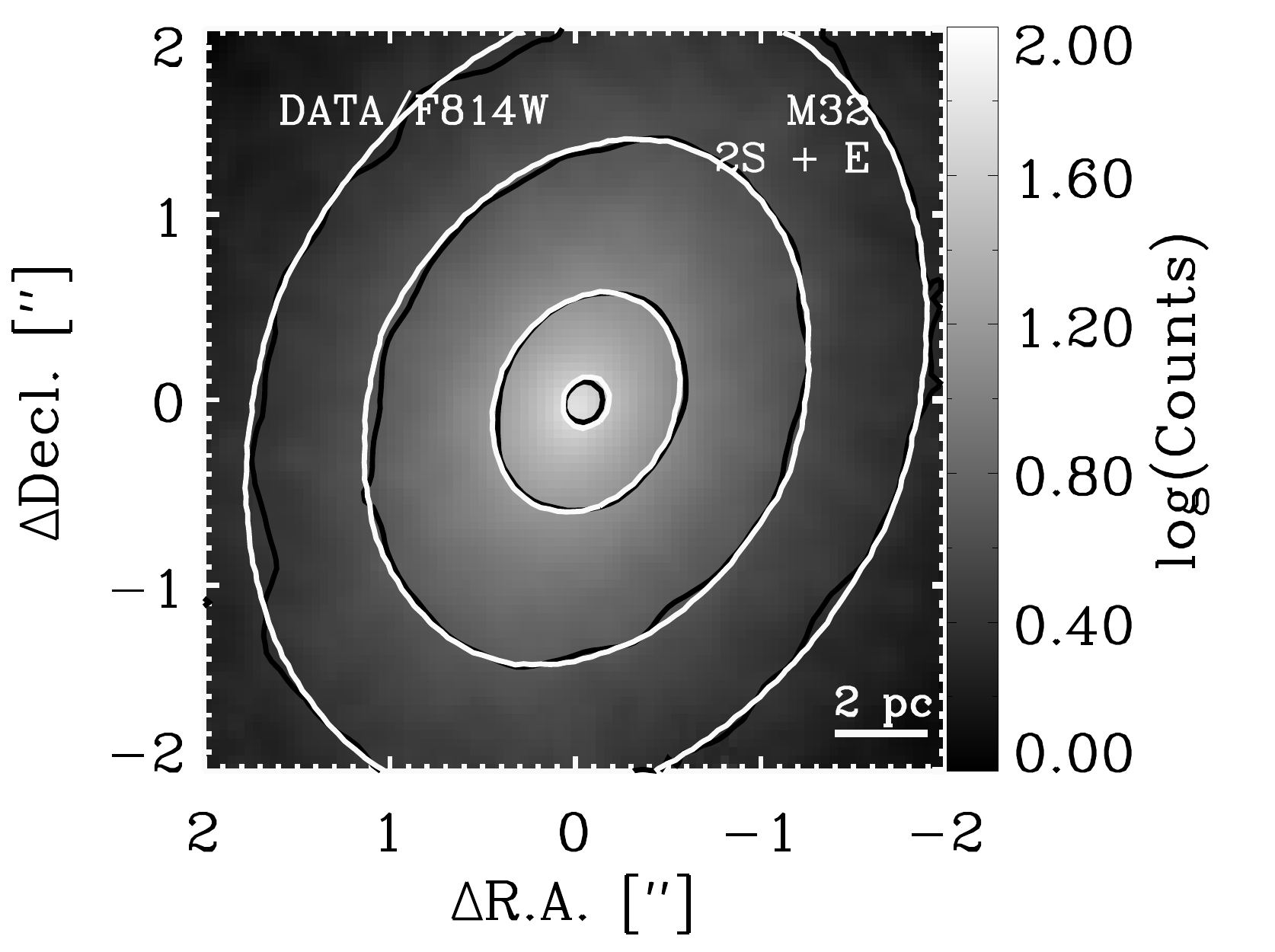}
       \hspace{-0.3cm}
       \includegraphics[width=0.37\linewidth,height=0.25\textheight,keepaspectratio=true]{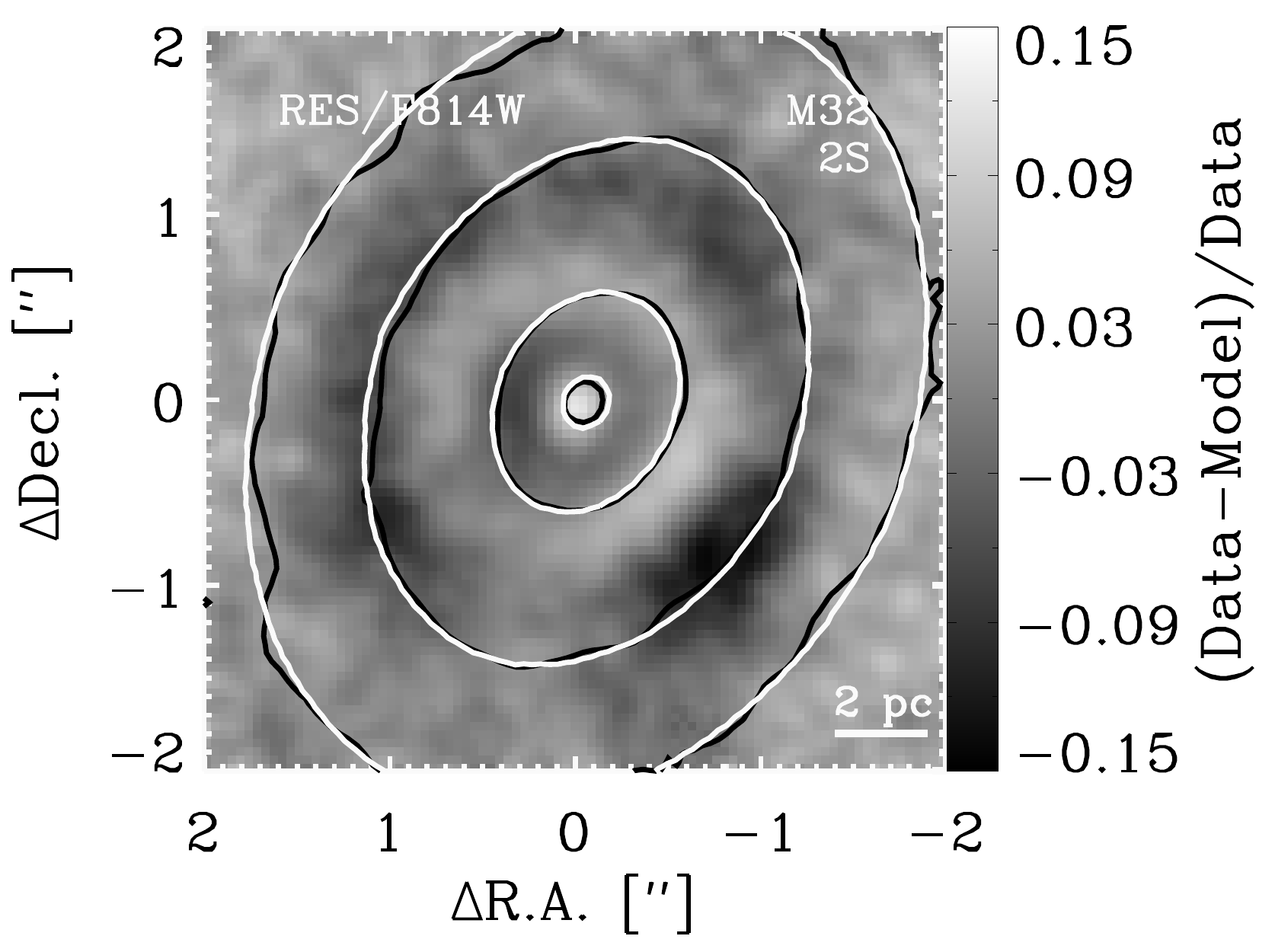}      
       \includegraphics[width=0.37\linewidth,height=0.25\textheight,keepaspectratio=true]{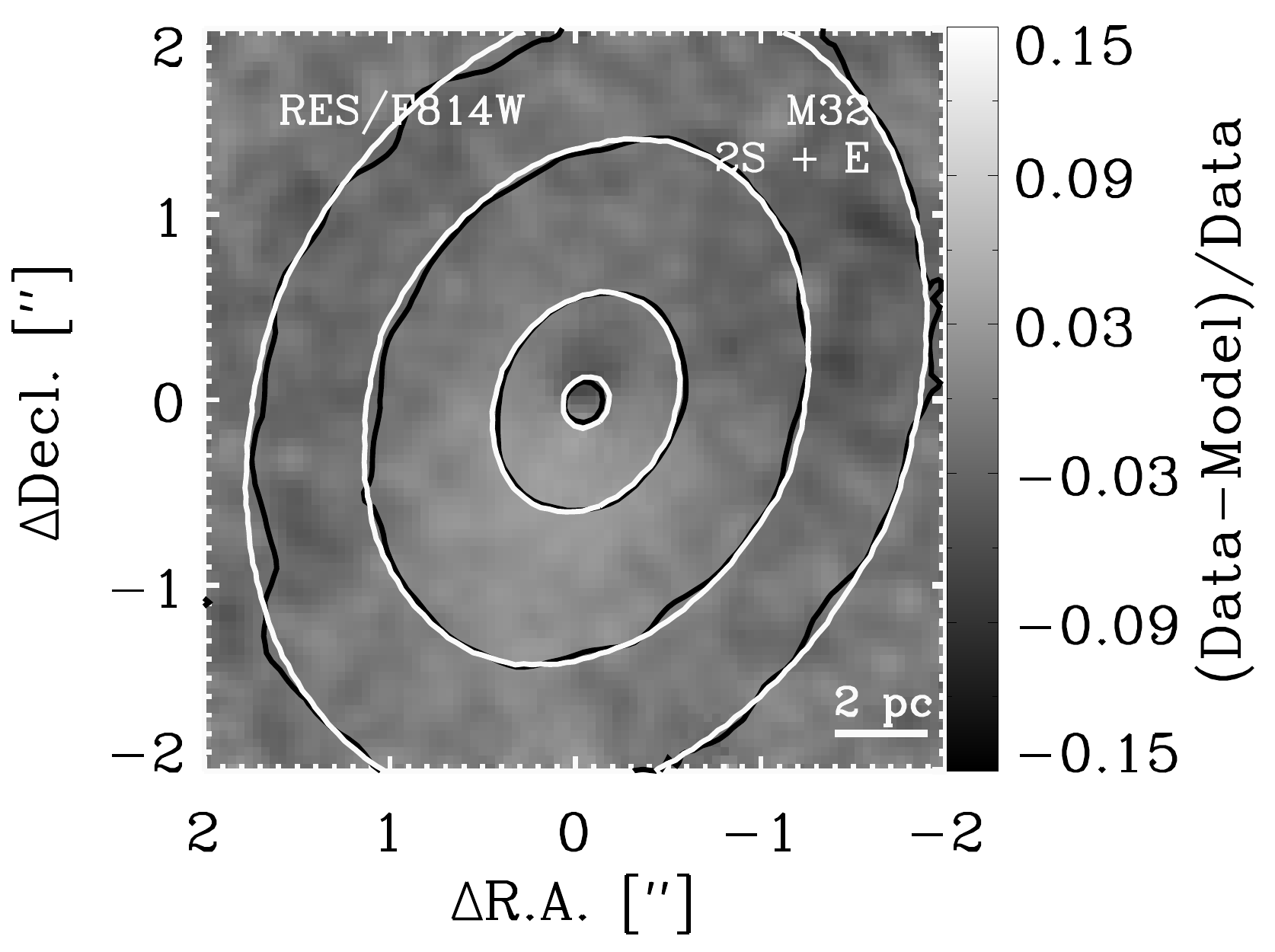}
    \end{minipage} 
    \begin{minipage}{\linewidth}\hspace{-10mm}\vspace{0.3cm}
         \includegraphics[width=0.37\linewidth,height=0.25\textheight,keepaspectratio=true]{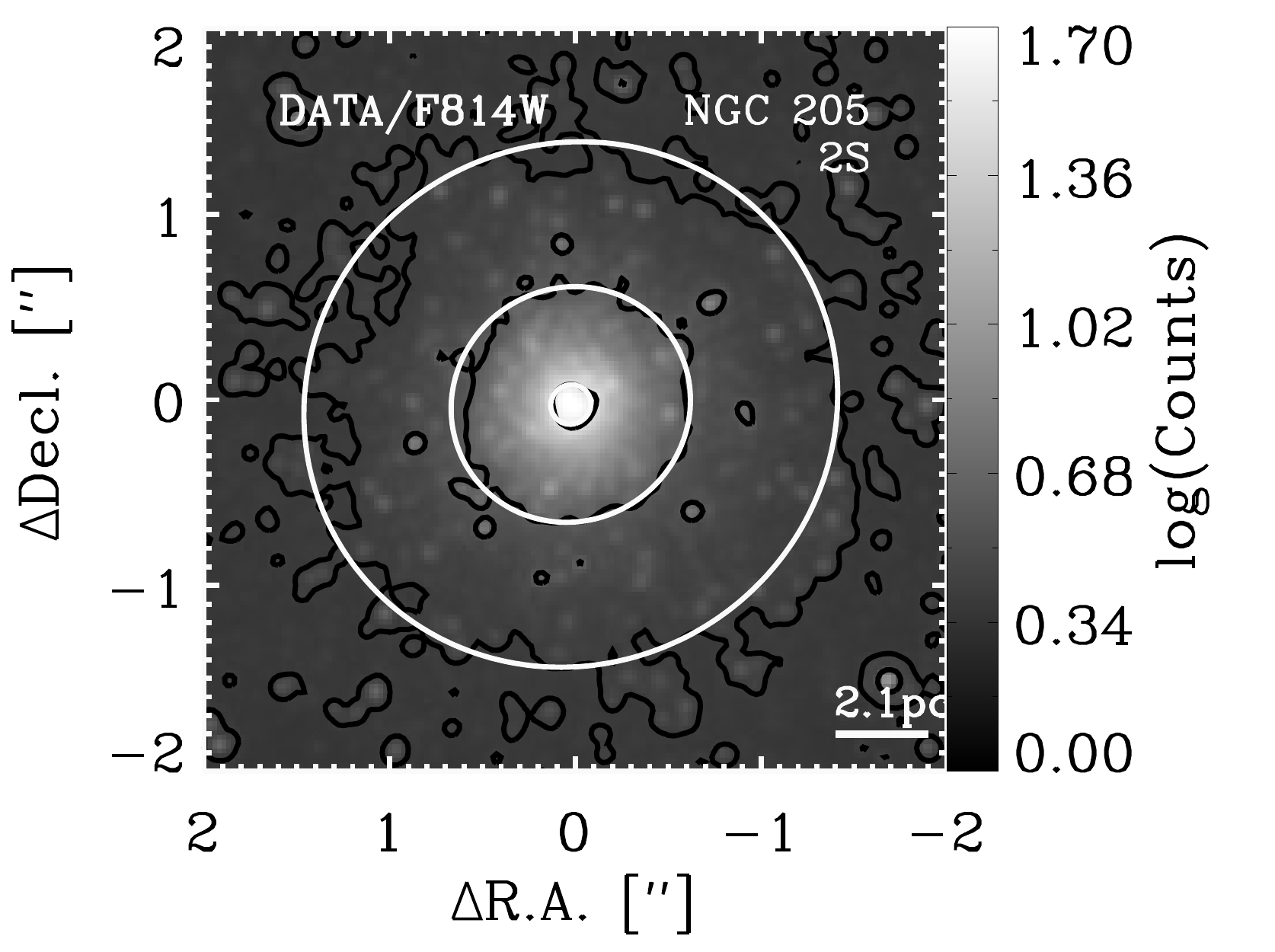}
       \hspace{-0.3cm}
       \includegraphics[width=0.37\linewidth,height=0.25\textheight,keepaspectratio=true]{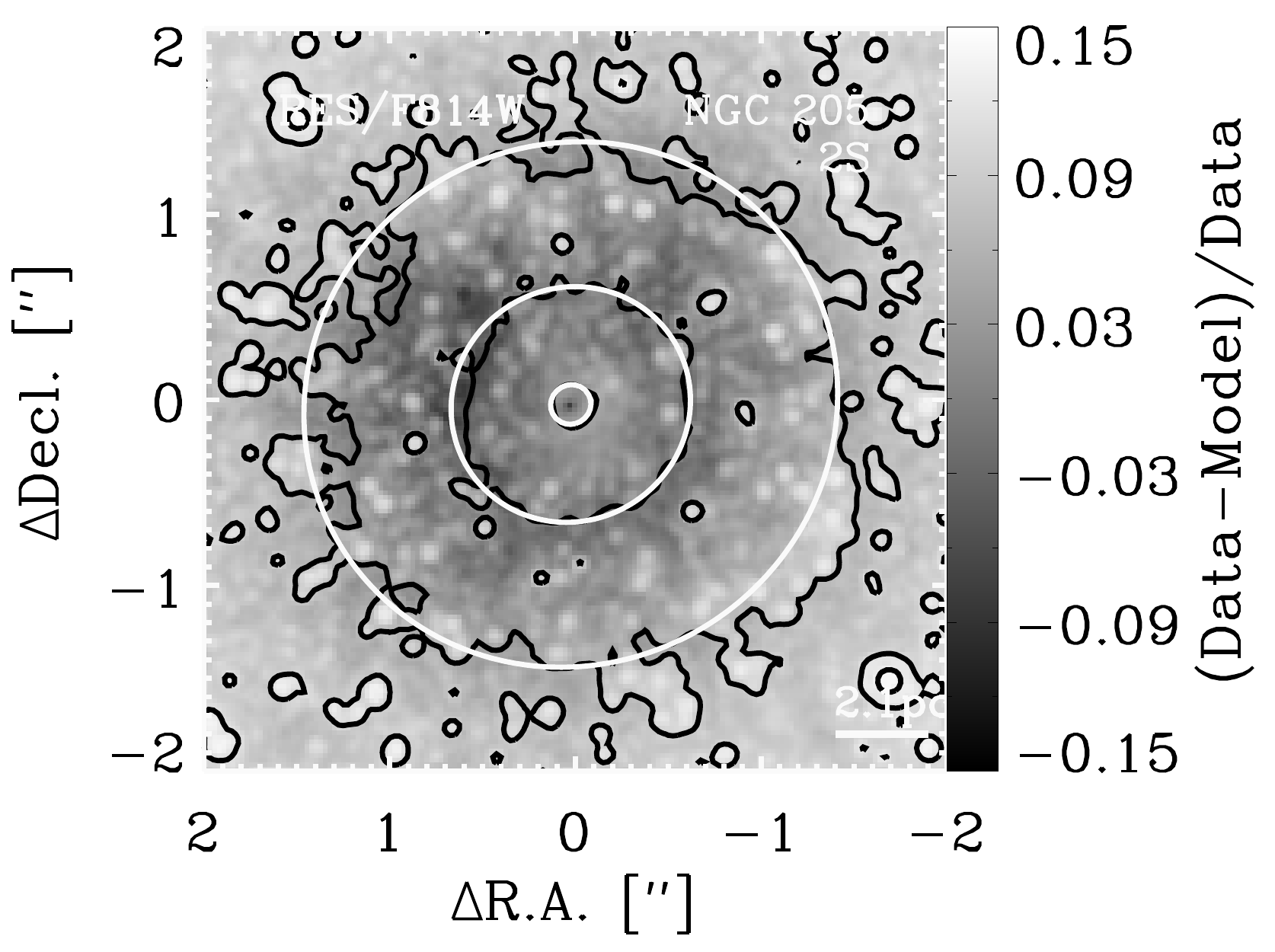}
    \end{minipage}
    \begin{minipage}{\linewidth}\hspace{-10mm}\vspace{0.3cm}
       \includegraphics[width=0.37\linewidth,height=0.25\textheight,keepaspectratio=true]{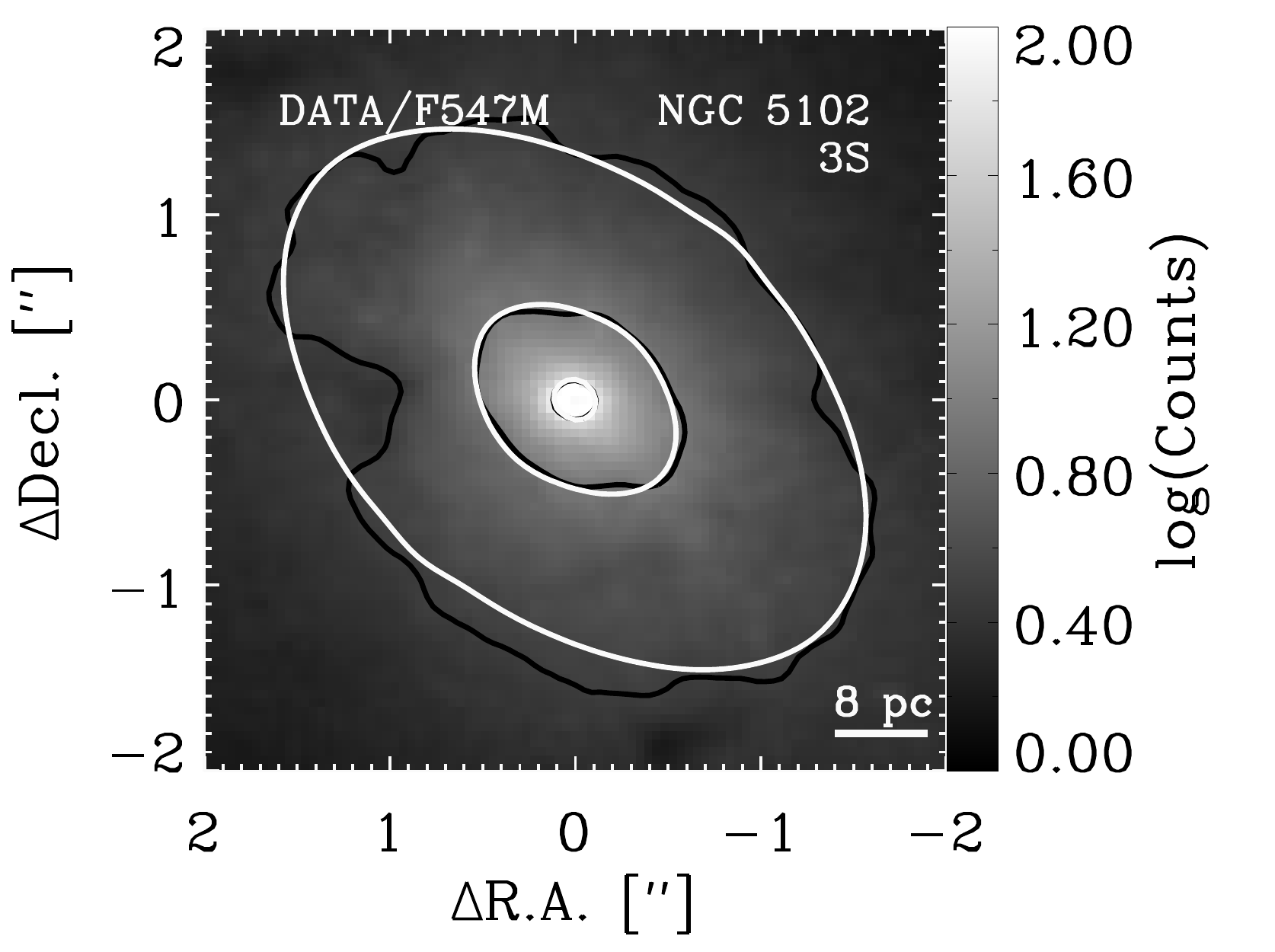}
       \hspace{-0.3cm}
       \includegraphics[width=0.37\linewidth,height=0.25\textheight,keepaspectratio=true]{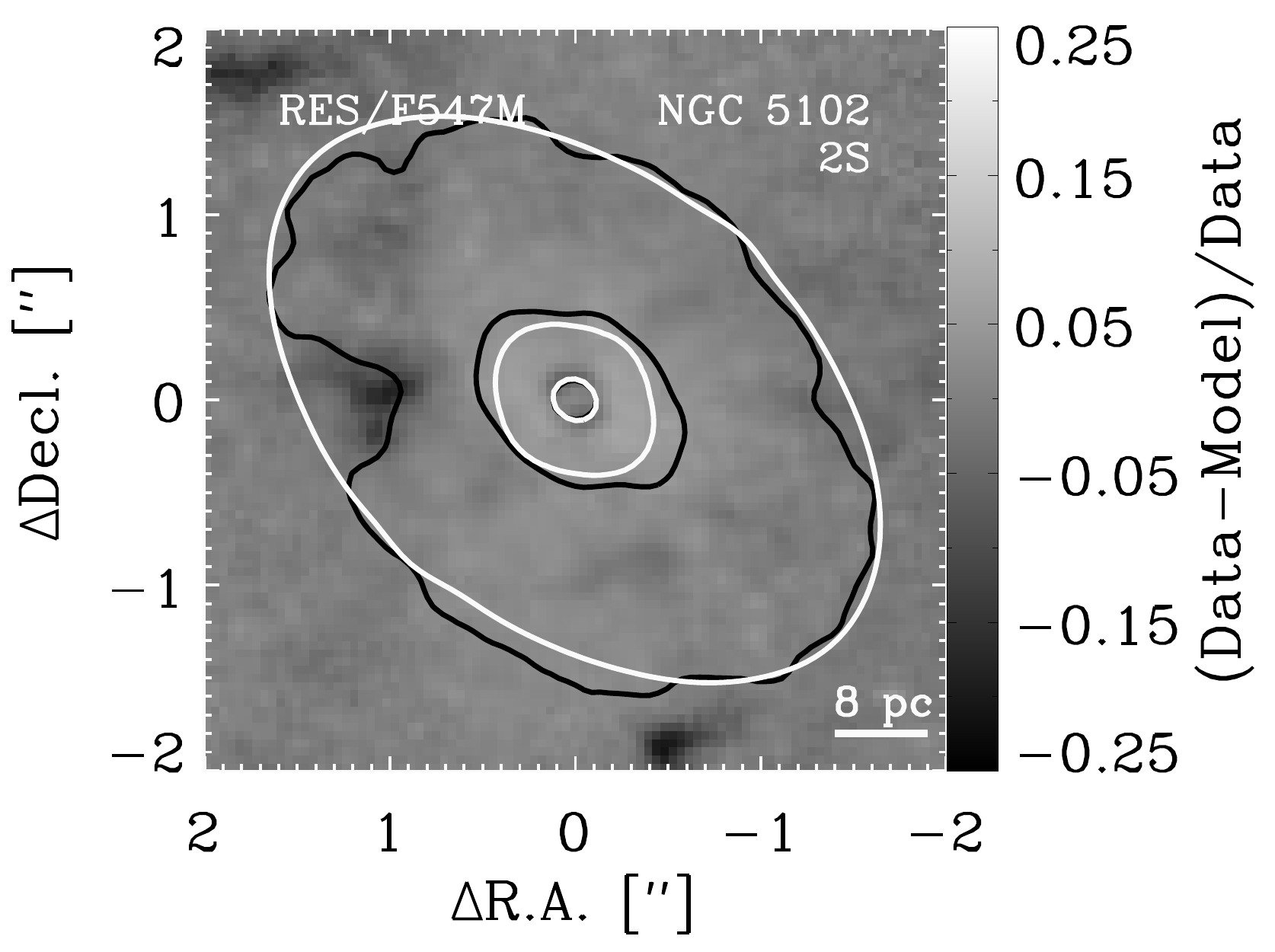}
        \includegraphics[width=0.37\linewidth,height=0.25\textheight,keepaspectratio=true]{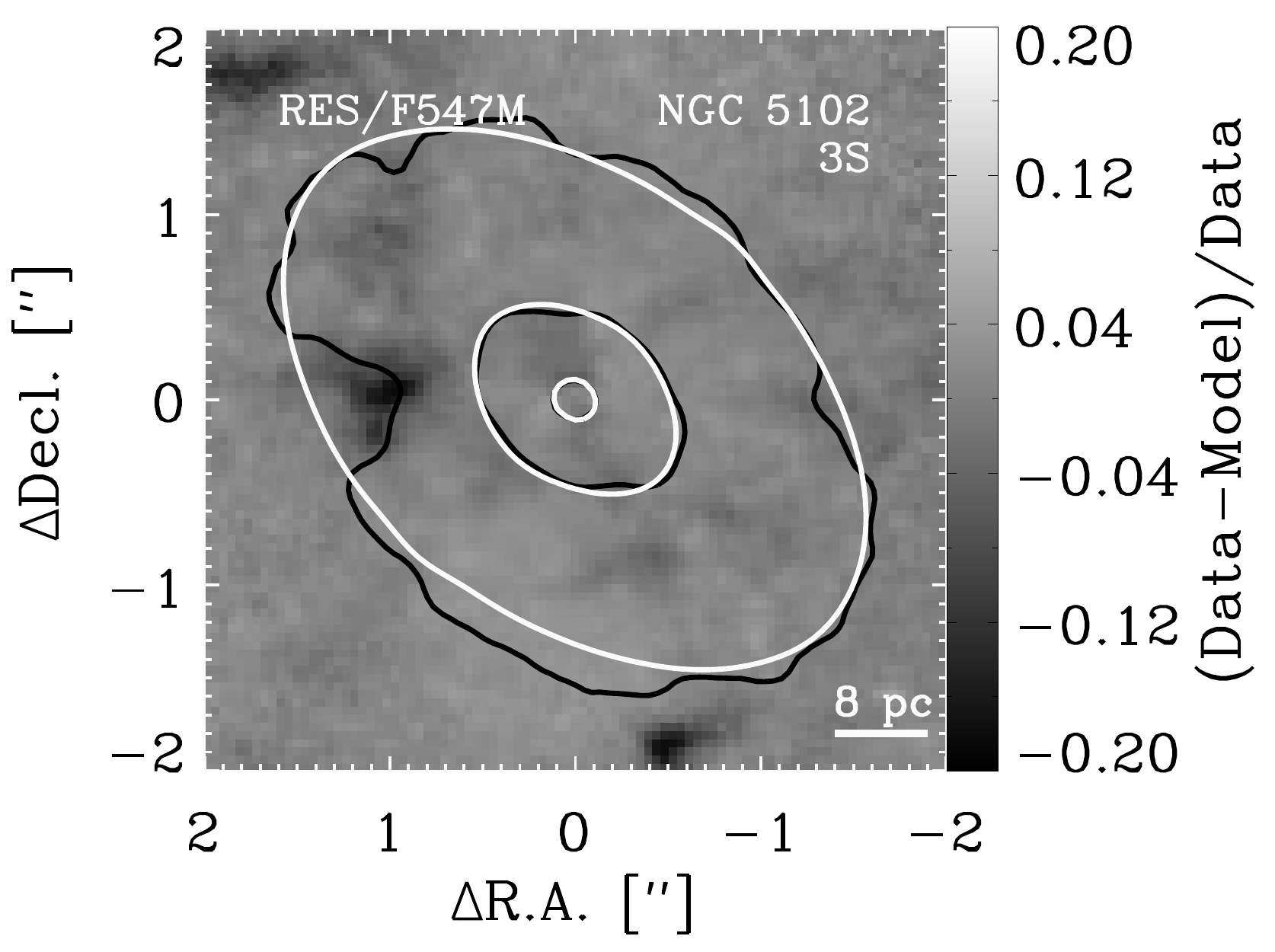}
    \end{minipage}
    \begin{minipage}{\linewidth}\hspace{-10mm}\vspace{0.3cm}
       \includegraphics[width=0.37\linewidth,height=0.25\textheight,keepaspectratio=true]{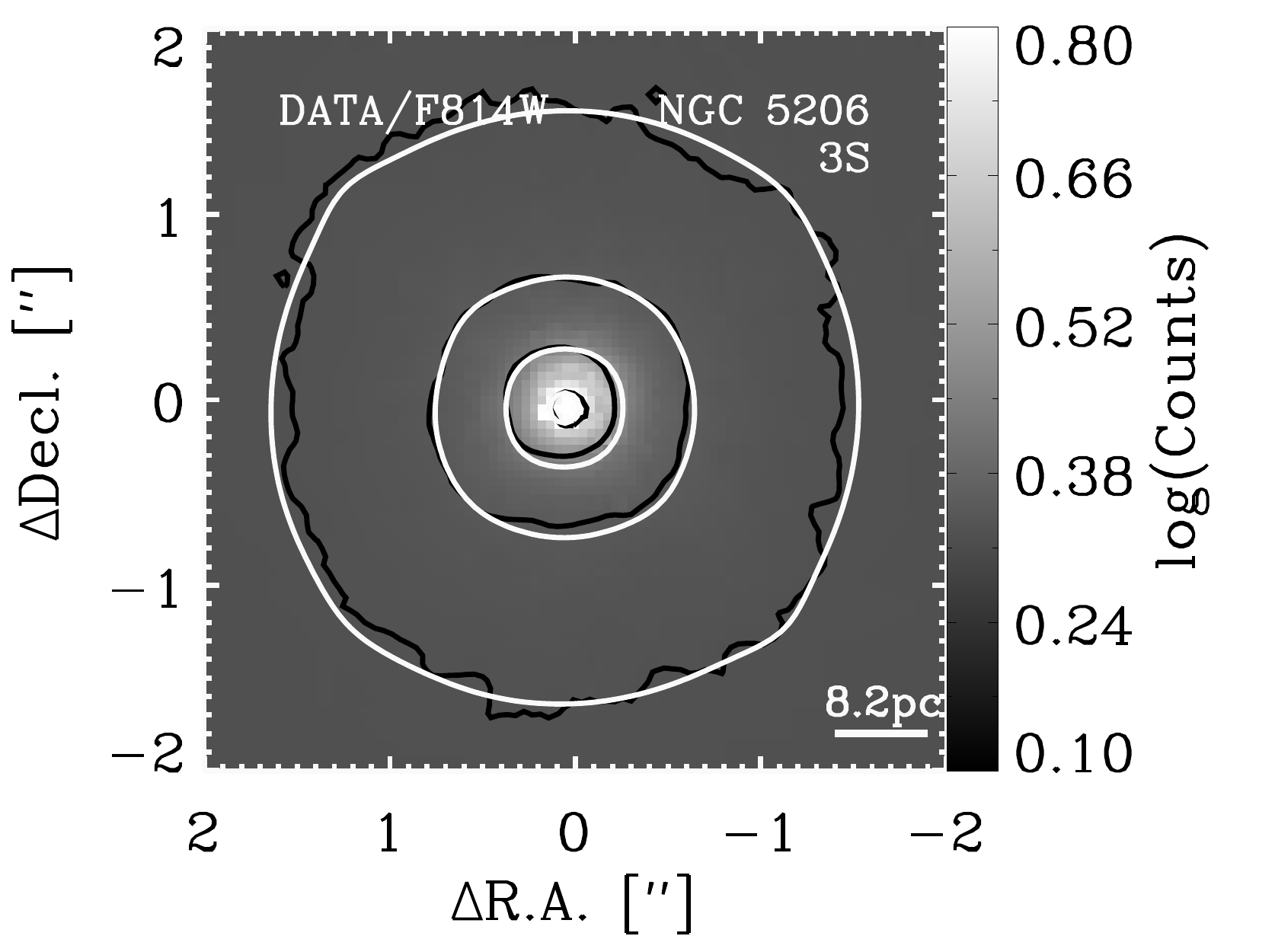}
       \hspace{-0.3cm}
       \includegraphics[width=0.37\linewidth,height=0.25\textheight,keepaspectratio=true]{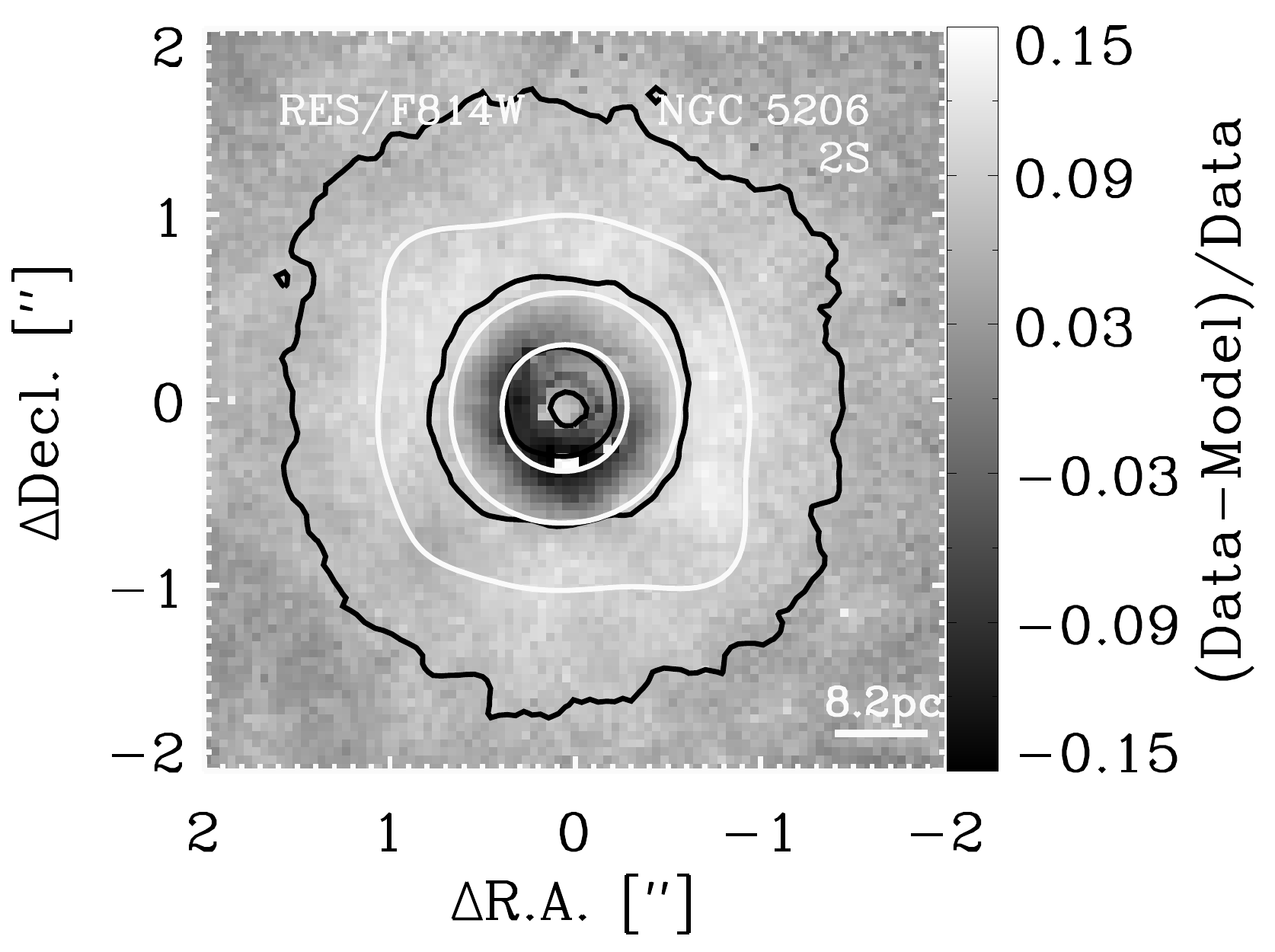}
       \includegraphics[width=0.37\linewidth,height=0.25\textheight,keepaspectratio=true]{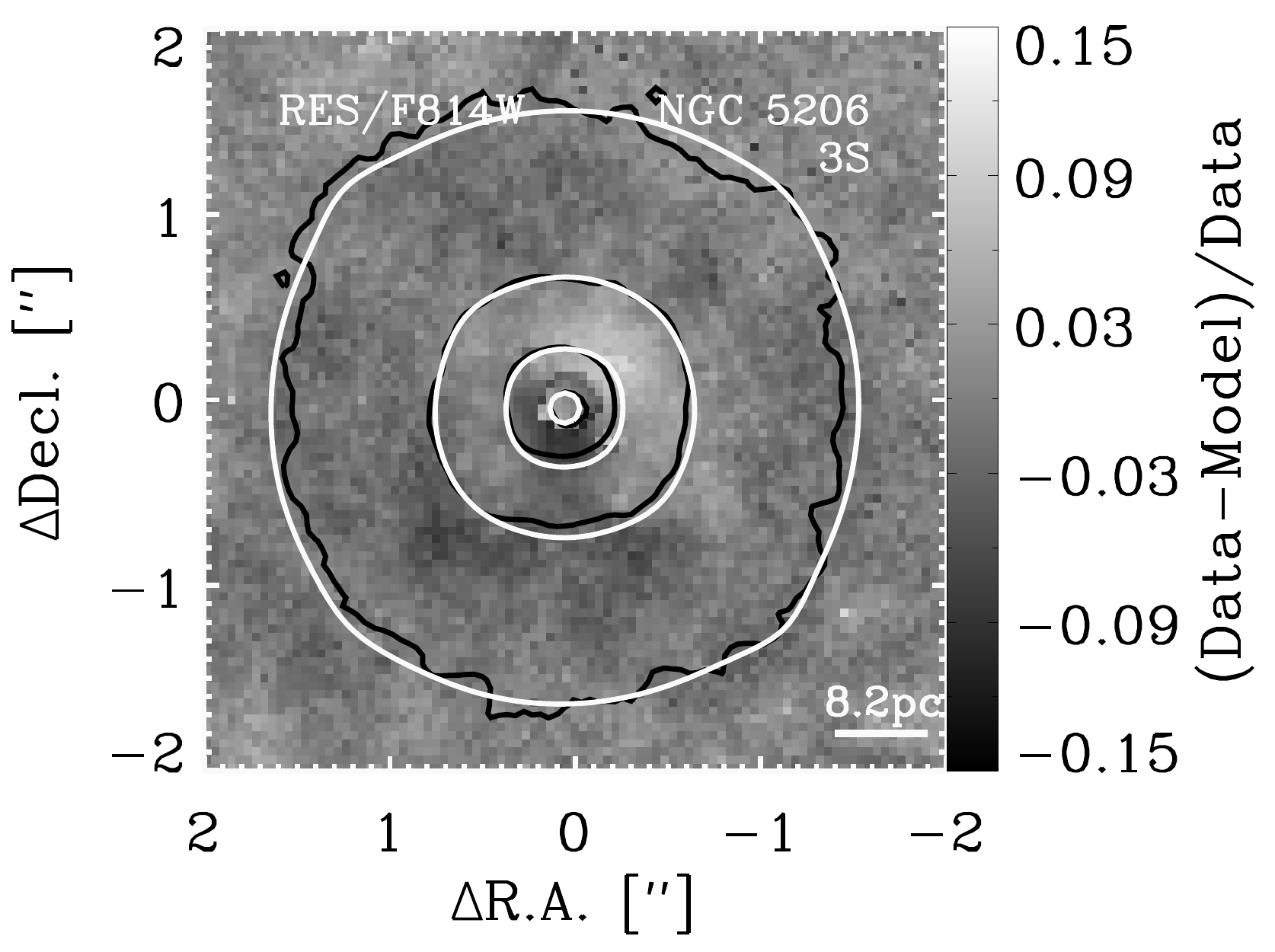}
    \end{minipage}
    \caption{\normalsize Comparison of \hst~images to their GALFIT models for each galaxy.      Left panels: the 2D \hst/WFPC2 PC (M32, NGC~5102, NGC~5206) or ACS/HRC (NGC~205) images.   Middle panels: the fractional residual \texttt{(Data-Model)/Data} between the \hst~imaging and their corresponding two-component GALFIT models.    Right panels: fractional residuals from the three-components GALFIT models.    The data and model contours are shown at the same levels of flux and radius in black and white repeating from the left to the right panels, respectively.  All figures show East to the left, and North up.}  
 \label{hst_gal_res}  
\end{figure*}
%%%%%%%%%%%%%%%%%%%%%%%%%%%%%%%%%%%%%%%%%%%%%%%%%%%%%%%%%%%%%%%%%%%%%%
%%%%%%%%%%% Table 5 of 1D and 2D SB parameters fits  of 4 galaxies %%%%%%%%%%%%  
\begin{table*}[ht]
\caption{GALFIT Models, Stellar Population Masses, and Luminosities of Galaxies in the Sample}
\hspace{-5mm}
\begin{tabular}{ccccccc|c|cccccc}
\hline\hline 
Object         &   Filter  &       SB       &        $n_i$    &$r_{\rm eff.,}$$_{ i}$&$r_{\rm eff.,}$$_{ i}$&      $m_{i}$   &        PA$_i$       &        b/a$_i$    &$\chi_{\rm r}^2$&    \Lstar$_{, i}$       &\Mstar$_{, i, \;\rm pop.}$&Comp.\\ 
                   &             &                   &                     &      ($\arcsec$)        &             (pc)            &        (mag)    &      (\deg)           &                        &                        &($\times10^7$\Lsun)&     ($\times10^7$\Msun)&          \\
      (1)         &    (2)    &       (3)        &         (4)       &            (5)                &             (6)              &          (7)       &         (8)             &        (9)            &          (10)       &              (11)            &                (12)                & (13)  \\
%% M32
\hline  \hline  
                  &             &                   &                     &                               &                                &                      &         Free PA     &                       &                        &                                 &                                      &       \\
\hline  
M32           &F814W&       2D         &2.7$\pm$0.3&    1.1$\pm$0.1       &      4.4$\pm$0.4      &11.0$\pm$0.1&$-$23.4$\pm$0.5&0.75$\pm$0.03&                       &     1.10$\pm$0.10       &   1.45$\pm$0.24          &NSC \\
                  &            &       2D        &1.6$\pm$0.1&    27.0$\pm$1.0     &       108$\pm$4        &7.0$\pm$0.1 &$-$24.7$\pm$0.4&0.79$\pm$0.07&        1.7          &        43.5$\pm$4.0    &     79.4$\pm$10.3        &Bulge\\
                  &            &1D$^{\star}$&         1         &         129                &               516            &       8.6         &$-$25.0$\pm$0.7&0.79$\pm$0.05&                       &              9.96             &     19.3$\pm$2.5          &Disk \\
\hline  
                  &            &                    &                    &                              &                                 &                      &        Fixed PA     &                         &                      &                                 &                                     &         \\
\hline   
M32          &F814W &       2D        &2.7$\pm$0.3&    1.1$\pm$0.1       &   4.4$\pm$0.4         &11.1$\pm$0.1  &         $-$25.0       &0.75$\pm$0.09&                      &    1.10$\pm$0.10       &      1.45$\pm$0.25      &NSC \\
                 &            &       2D         &1.6$\pm$0.1&    27.0$\pm$1.0     &    108$\pm$4          & 7.1$\pm$0.1  &         $-$25.0       &0.79$\pm$0.11&     2.0            &      43.8$\pm$4.1     &        78.0$\pm$10.4    &Bulge\\
                 &            &1D$^{\star}$&          1        &             129             &              516            &           8.6       &         $-$25.0       &0.79$\pm$0.08&                      &              10.00          &        19.3$\pm$2.5      &Disk \\[1mm]
\hline \hline 
                 &            &                    &                    &                               &                                 &                       &          Free PA     &                         &                      &                                 &                                    &      \\
\hline    
%% NGC 205
NGC~205 &F814W&         2D      &1.6$\pm$0.2&      0.3$\pm$0.1      &    1.3$\pm$0.4        & 13.6$\pm$0.4 &$-$37.1$\pm$1.4 &0.95$\pm$0.03&                    &   0.10$\pm$0.04    &      0.18$\pm$0.08    &NSC\\
                &            &1D$^{\star}$&           1.4    &             120             &              516            &            6.8       &$-$40.4$\pm$1.0 &0.90$\pm$0.07&        5.5       &             55.7           &      97.2$\pm$15.2      &Bulge\\
\hline
                &           &                    &                    &                                &                                 &                       &         Fixed PA     &                         &                      &                             &                                     &         \\
\hline 
NGC~205  &F814W&        2D       &1.6$\pm$0.2&      0.3$\pm$0.1     &       1.3$\pm$0.4     &  13.6$\pm$0.4  &       $-$40.4        &0.95$\pm$0.06&                     & 0.10$\pm$0.04  &     0.18$\pm$0.08      &NSC\\
                  &            &1D$^{\star}$&         1.4      &             120            &              516            &           6.8          &       $-$40.4        &0.91$\pm$0.05&            5.9    &           55.7         &    97.2$\pm$15.2         &Bulge\\[1mm]          
\hline \hline  
%% NGC 5102
                &            &                    &                   &                                 &                                 &                          &             Free PA  &                        &                     &                               &                                   &   \\
\hline
NGC~5102 &F547M&         2D      &0.8$\pm$0.2&   0.1$\pm$0.1       &    1.6$\pm$1.6        & 14.20$\pm$0.30 &    55.1$\pm$1.5  &  0.68$\pm$0.06&                    &  1.81$\pm$0.51      &0.71$\pm$0.22        &NSC$_{\rm 1}$\\
                 &           &         2D      &3.1$\pm$0.1&   2.0$\pm$0.3         &     32.0$\pm$4.8      & 12.27$\pm$0.21&    51.1$\pm$1.7  &  0.59$\pm$0.04&      3.3        &  10.7$\pm$1.98        &5.8$\pm$0.6             &NSC$_{\rm 2}$\\
                 &           &1D$^{\star}$&        3          &             75               &            1200             &            9.05         &               50.0      &  0.60$\pm$0.07&                   &             210             &592$\pm$83             &Bulge\\
\hline
                 &            &                   &                   &                                &                                  &                          &        Fixed PA     &                          &                   &                                 &                                  &     \\
\hline
NGC~5102&F547M &         2D      &0.8$\pm$0.2&   0.1$\pm$0.1      &    1.6$\pm$1.6         & 14.20$\pm$0.30&            50.5         &  0.68$\pm$0.05&                   &  1.81$\pm$0.51      &0.71$\pm$0.22          &NSC$_{\rm 1}$\\
                 &           &        2D       &3.1$\pm$0.1&   2.0$\pm$0.3        &    32.0$\pm$4.8        & 12.27$\pm$0.22&            50.5         &  0.60$\pm$0.08&        3.6     &   10.7$\pm$1.98      &5.8$\pm$0.6              &NSC$_{\rm 2}$\\
                 &           &1D$^{\star}$&        3          &             75              &               1200           &          9.05          &            50.5         &  0.63$\pm$0.10&                  &               210            &592$\pm$83              &Bulge\\[1mm]
\hline \hline 
%% NGC 5206
                 &            &                   &                   &                               &                                   &                         &      Free PA          &                           &                  &                                &                                 &        \\
\hline   
NGC~5206&F814W&         2D      &0.8$\pm$0.1 &   0.2$\pm$0.1    &       3.4$\pm$1.7       & 16.9$\pm$0.5   &    36.0$\pm$0.1  &  0.96$\pm$0.03   &                  &  0.094$\pm$0.045  &  0.17$\pm$0.10      &NSC$_{\rm 1}$\\   
                   &           &         2D      &2.3$\pm$0.3 &   0.6$\pm$0.1    &      10.5$\pm$1.7      & 14.8$\pm$0.2  &    38.5$\pm$1.4   &  0.96$\pm$0.02  &      2.4       &  0.65$\pm$0.12       & 1.28$\pm$0.6      &NSC$_{\rm 2}$\\    
                   &           &1D$^{\star}$&         2.57     &           58            &                986            &        10.1          &              38.6        &  0.98$\pm$0.01  &                  &           123.3            &  241$\pm$47          &Bulge\\
\hline
                   &           &                   &                     &                           &                                  &                          &       Fixed PA        &                            &                   &                                &                               &        \\
\hline
NGC~5206&F814W&        2D      &0.8$\pm$0.1  &   0.2$\pm$0.1   &        3.4$\pm$1.7      &  16.9$\pm$0.5   &              38.3         &0.96$\pm$0.03&                  &   0.094$\pm$0.045  &   0.17$\pm$0.10   &NSC$_{\rm 1}$\\   
                  &           &         2D      &2.3$\pm$0.3 &    0.6$\pm$0.1  &       10.2$\pm$1.7      &  14.8$\pm$0.2   &              38.3         &0.97$\pm$0.03&      2.7       &   0.65$\pm$0.12     &   1.28$\pm$0.27    &NSC$_{\rm 2}$\\    
                  &           &1D$^{\star}$&          2.57    &            58          &                986             &           9.1           &              38.3         &0.98$\pm$0.01&                  &            122.3           &   241$\pm$47        &Bulge\\[1mm]
\hline \hline  
\end{tabular}
\tablecomments{\normalsize   Column 1: galaxy name.    Columns 2: Filter.    Column 3: source of the S\'ersic parameters from either 1D SB fits or 2D GALFIT; GALFIT models were run with the 1D parameters fixed.    Column 4:  the S\'ersic index of each component (Section 4.1 and 4.2 for 1D and 2D respectively).     Columns 5--12: parameters of each component including the effective radius (half-light radius) in arc-second (Columns 5) and in pc (Columns 6) scale, total apparent magnitude, which is Galactic Extinction corrected (Columns 7), position angle, flattening ($b/a$), overall reduced chi-square of best-fit GALFIT model, luminosity, and photometric mass estimated from stellar populations of each of each S\'ersic component integrated to $\infty$, respectively.    Photometric masses in Column 12 are calculated assuming the R15 color--\ml~relation within the bulge.  Column 13: component identification.  To convert the total apparent magnitude of each S\'ersic component into its corresponding total luminosity, we used the absolute magnitude of the Sun in the Vega system of \hst/WFPC2 F814W (M32 and NGC~5206, 4.107 mag), \hst/WFPC2 F555W (approximate F547M for NGC~5102, 4.820 mag), and \hst/HRC ACS F814W (NGC~205, 4.096 mag)\footnote{available at {http://www.baryons.org/ezgal/filters.php}}. } 
\label{seric_nsc_tab}
\end{table*} 
%%%%%%%%%%%%%%%%%%%%%%%%%%%%%%%%%%%%%%%%%%%%%%%%%%%%%%%%%%%%%%%%%%%%%%%
%%%%%%%%%%%%%%%%%%%%%%%%%%%%%%%%%%%%%%%%%%%%%%%%%%%%%%%%%%%%%%%%%%%%%%%
\begin{figure*}[ht] 
\minipage{0.25\textwidth}
  \includegraphics[width=\linewidth]{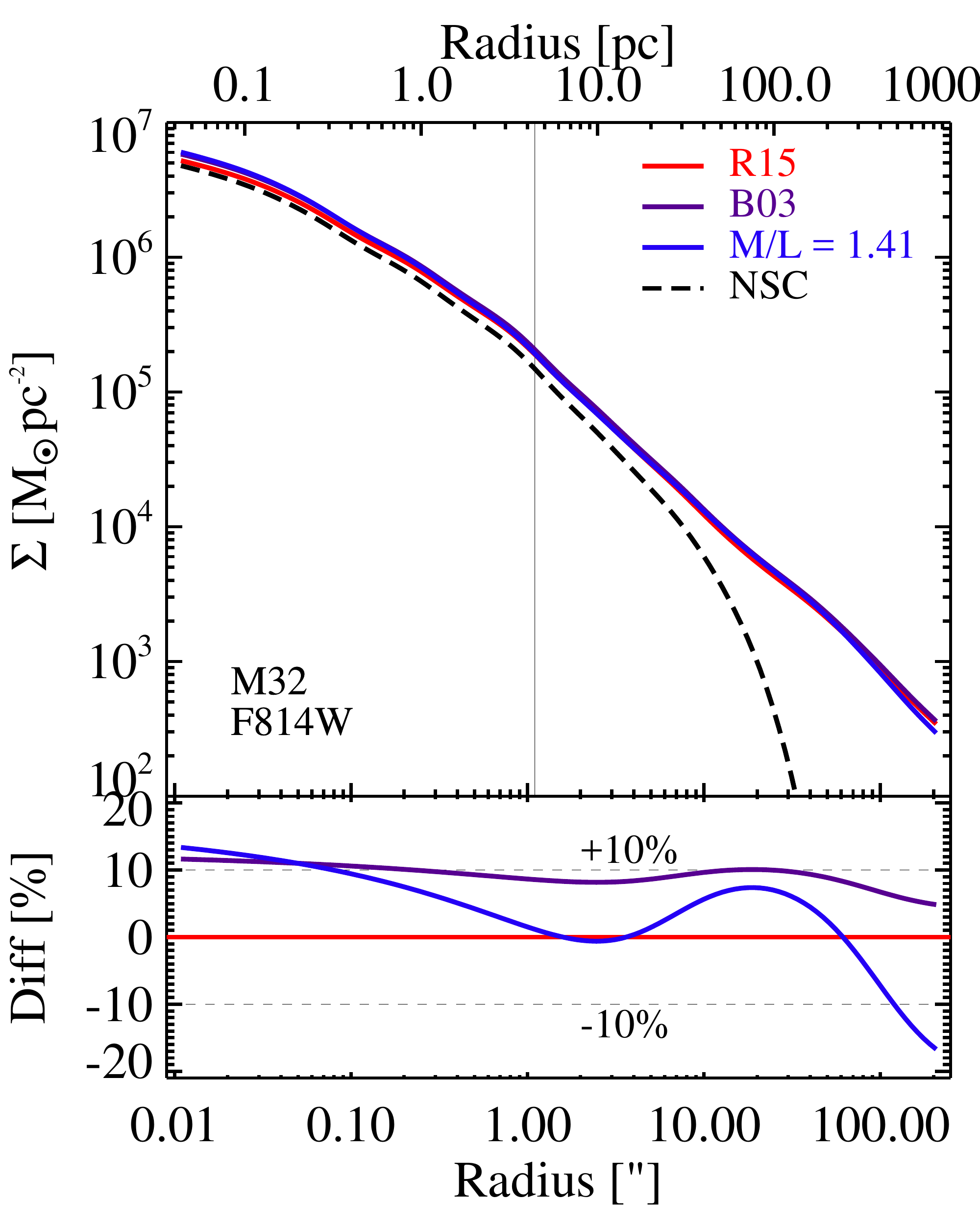}
 \endminipage\hfill
\minipage{0.25\textwidth}
  \includegraphics[width=\linewidth]{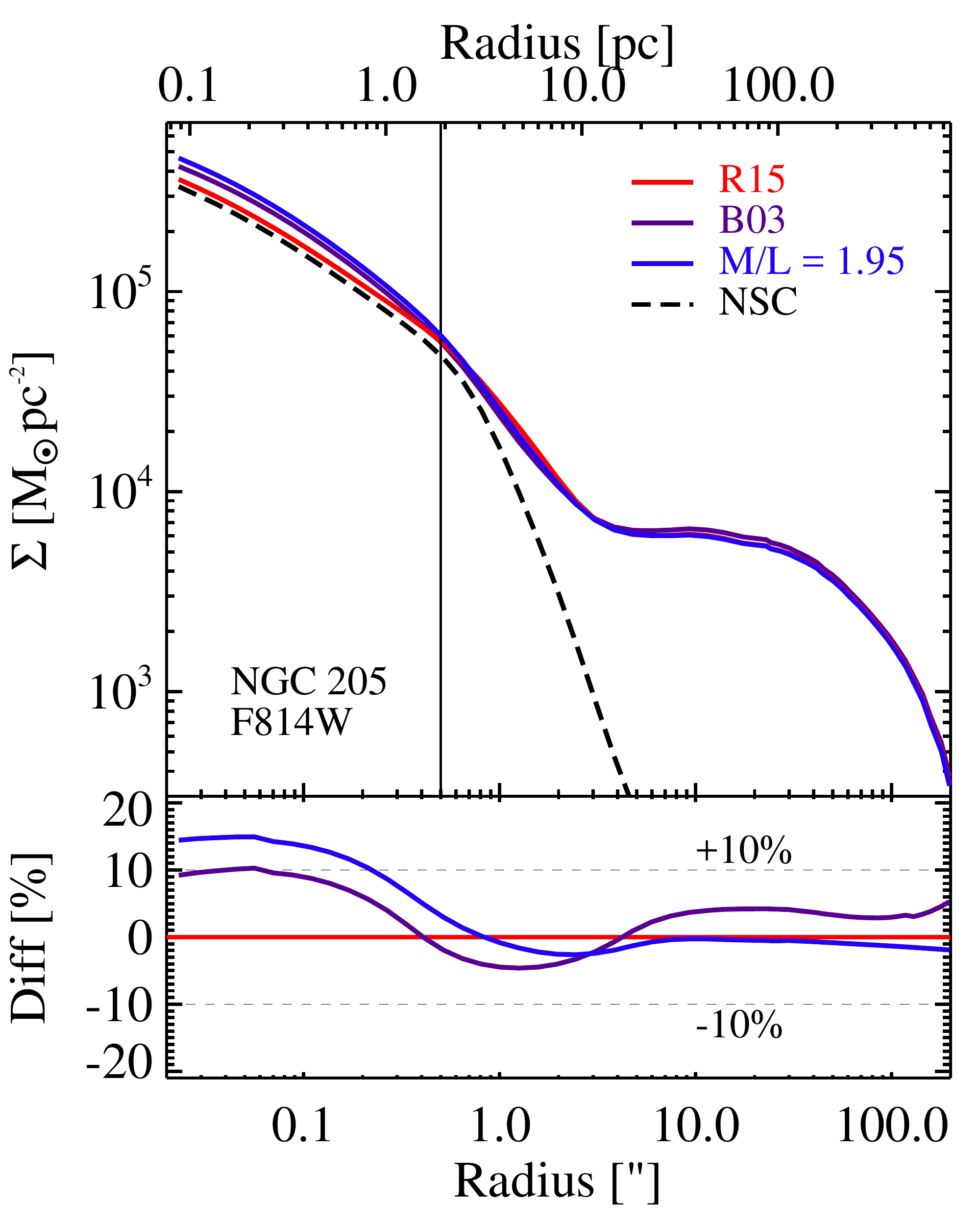}
\endminipage\hfill
\minipage{0.25\textwidth}
 \includegraphics[width=\linewidth]{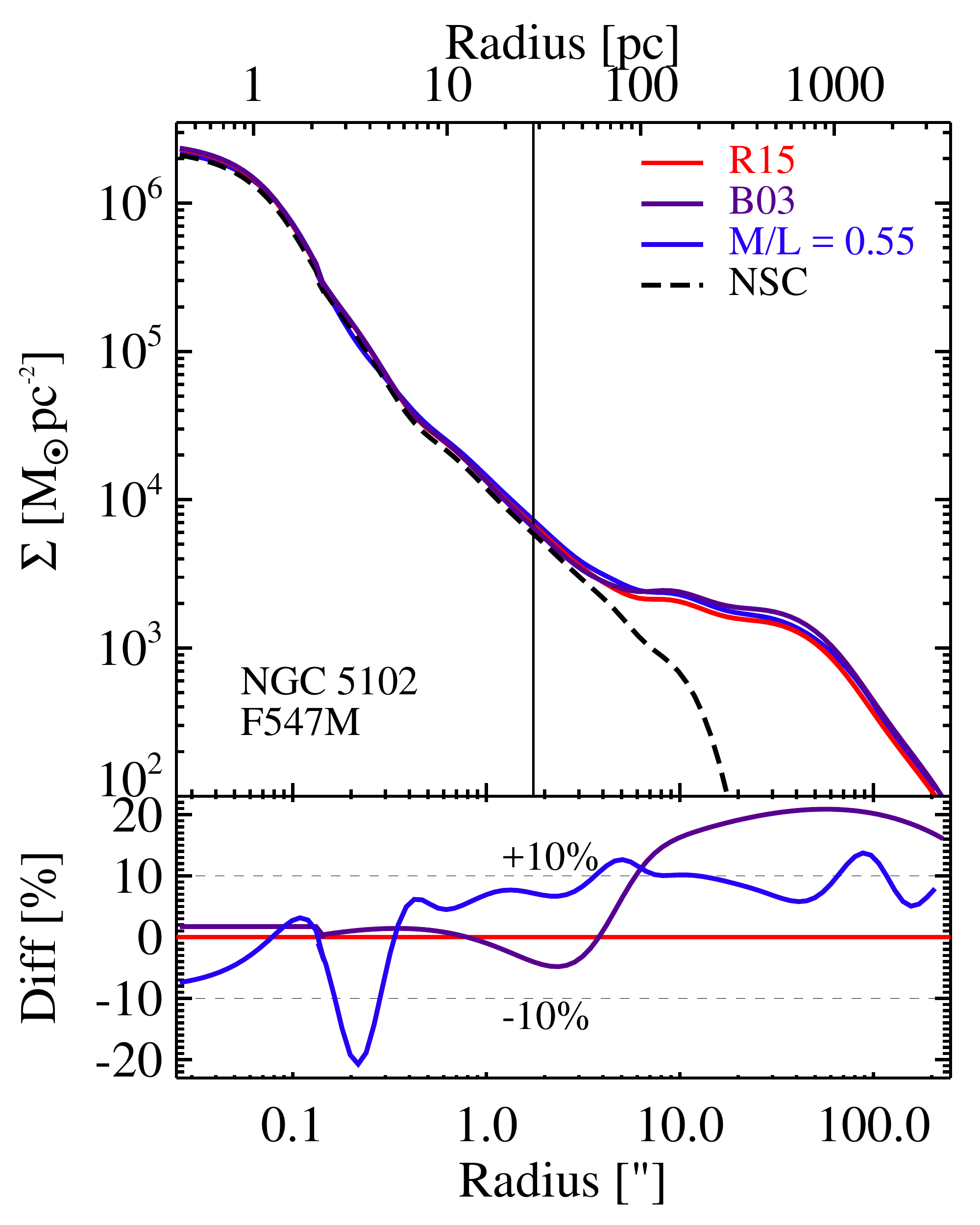}
\endminipage\hfill
\minipage{0.25\textwidth}
 \includegraphics[width=\linewidth]{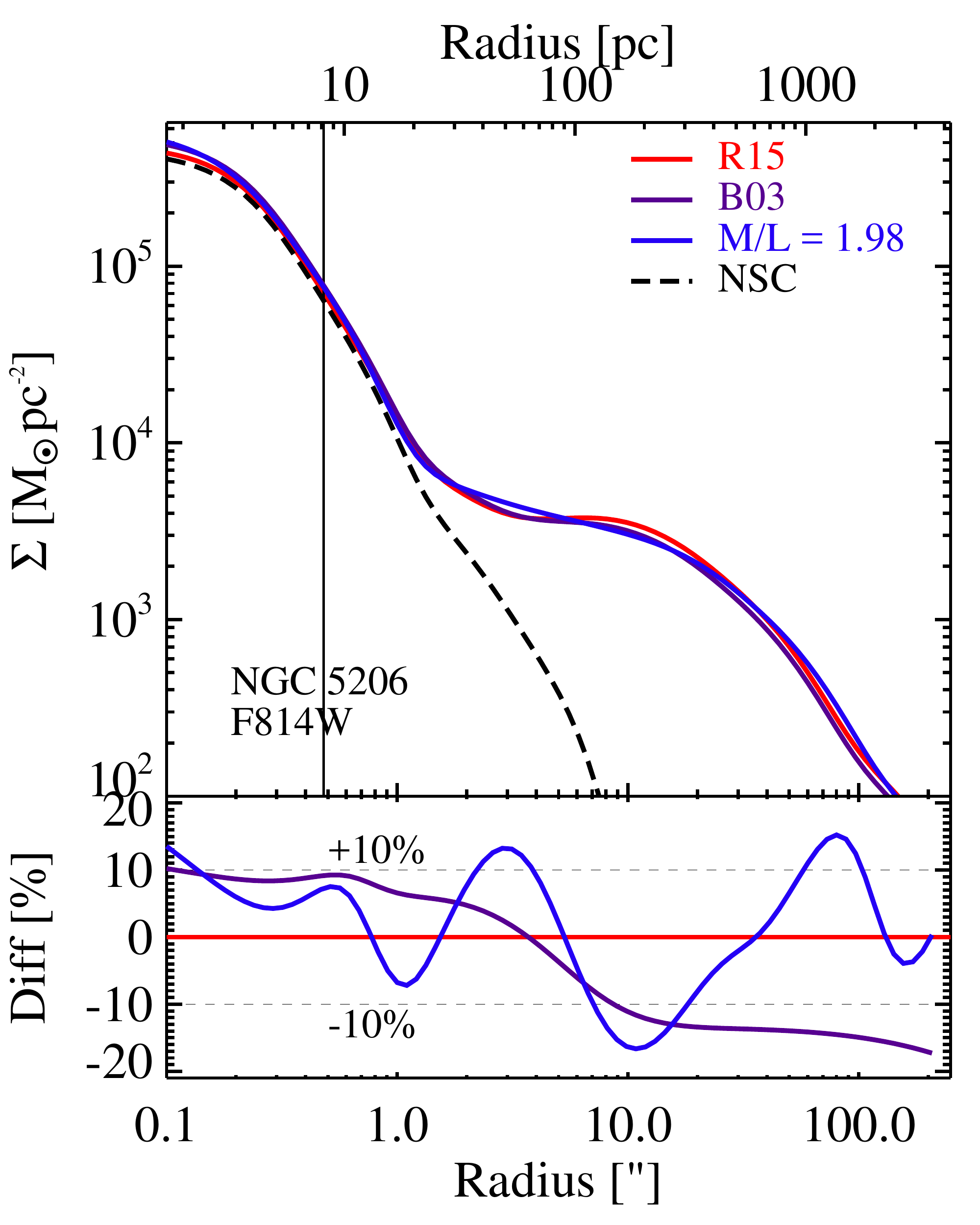}
\endminipage\hfill
\caption{\normalsize Top panels:  Comparison of the mass surface density models in each galaxy.  These models are created using MGEs constructed from our GALFIT models. These MGEs are then multiplied by a \ml~to get a mass surface density. For the constant \ml~model (dark blue) the chosen \ml~is discussed in Section~\ref{ssec:colorm2ls}.  For the other mass models, an \ml~is assigned to each MGE based on the color of the galaxy at the radius of the MGE using the R15 or B03 color--\ml~relation (red/purple lines); we use the R15 model as our default.  Dashed black lines show the NSCs in the R15 model; their effective radii are shown by vertical lines.   Bottom panels:  To compare the relative mass distribution of the three different mass profiles for each galaxy, we plot their fractional difference relative to the R15 color--\ml~relation.}
   \label{massden}     
\end{figure*}
%%%%%%%%%%%%%%%%%%%%%%%%%%%%%%%%%%%%%%%%%%%%%%%%%%%%%%%%%%%%%%%%%%%%%%%%
%\vspace{-5mm}
%%%%%%%%%%%%%%%%%%%%%%%%%%%%%%%%%%%%%%%%%%%%%%%%%%%%%%%%%%%%%%%%%%%%%%%
\begin{figure*}[ht] 
\hspace{-7mm}
\minipage{0.33\textwidth}
  \includegraphics[width=\linewidth]{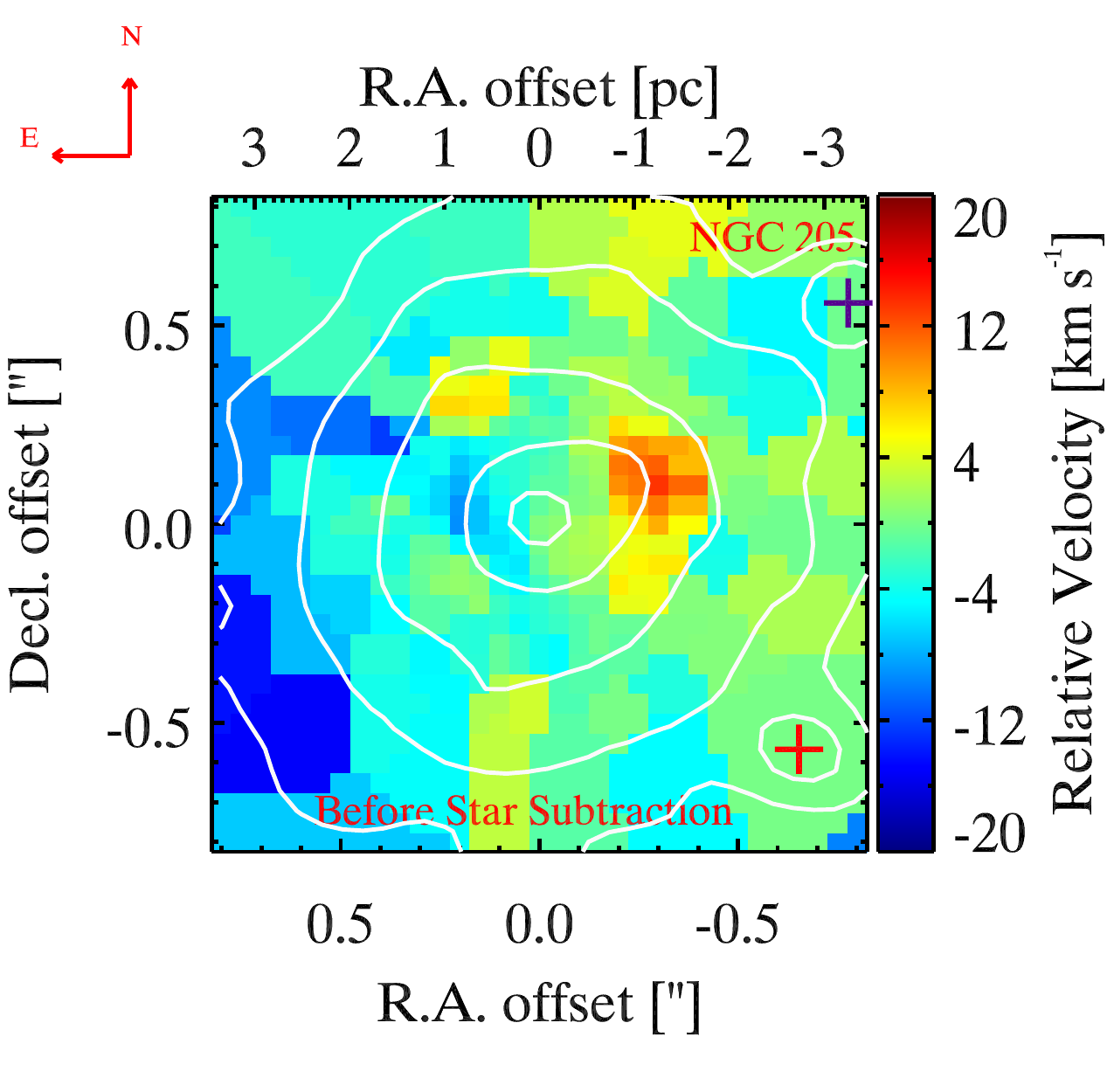}
 \endminipage\hfill
 \minipage{0.36\textwidth}
  \includegraphics[width=\linewidth]{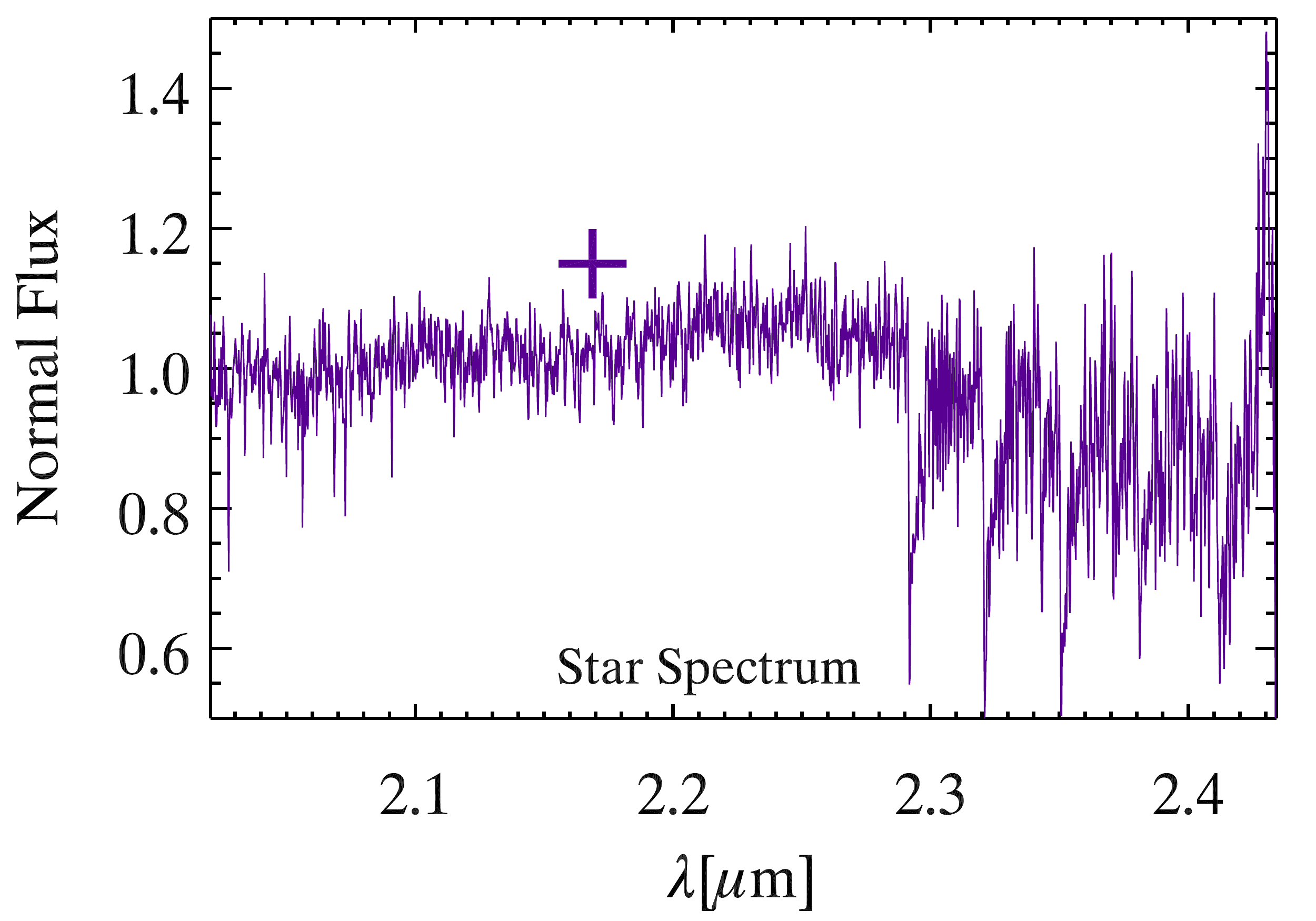}
\endminipage\hfill
\minipage{0.33\textwidth}
  \includegraphics[width=\linewidth]{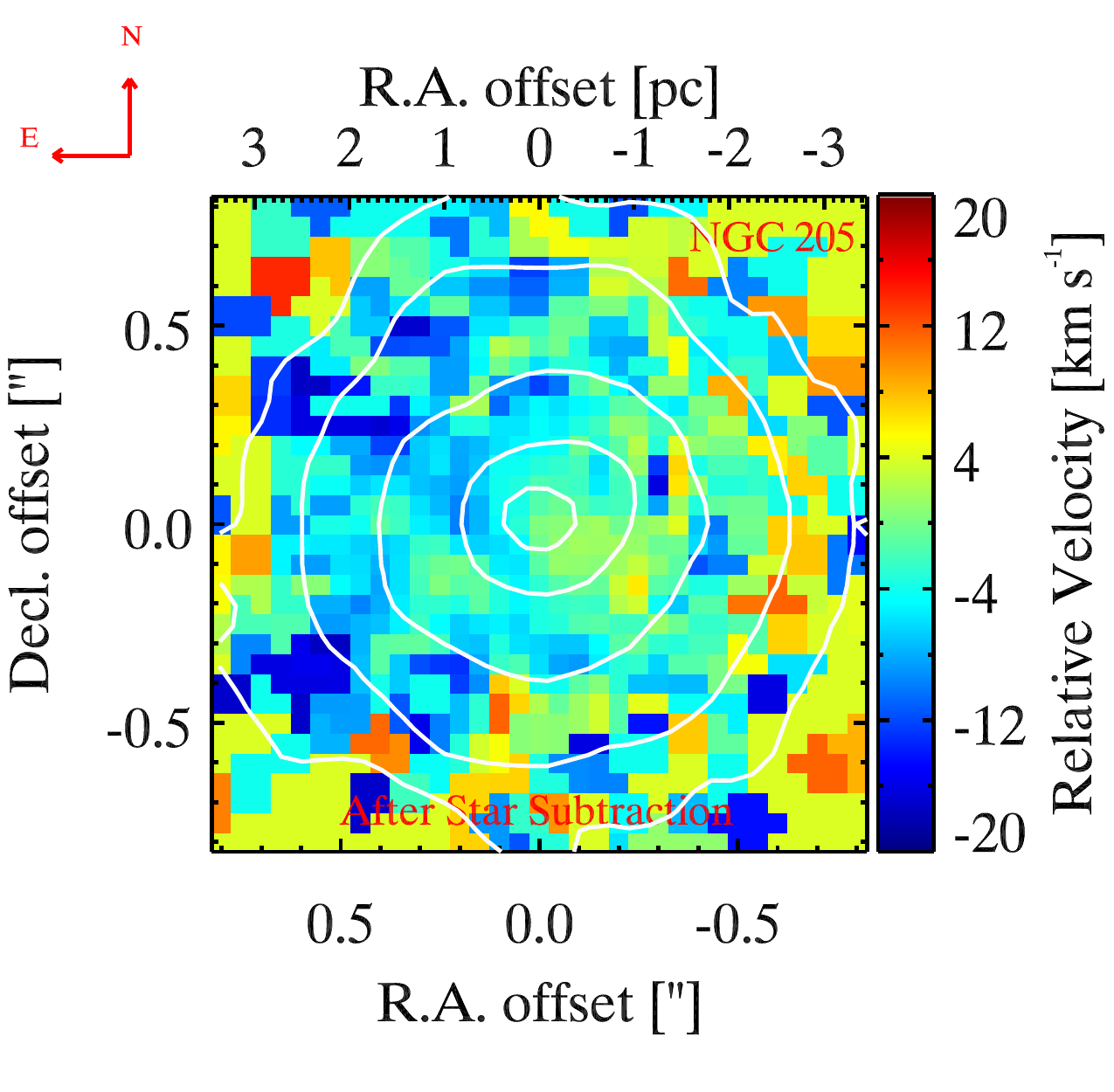}
\endminipage\hfill
\caption{\normalsize Left panel:  The radial velocity map derived from CO bandhead spectroscopy from Gemini/NIFS of NGC~205 before central resolved stars subtraction. The white intensity contours are plotted overlay on top of the map show those resolved star in the FOV with the two brights on north-west (purple plus) and south-west (red plus) of the map. Middle panel:  The spectrum of the brightest purple plus star is shown in the same color respectively.  Right panel: The radial velocity map derived from CO bandhead spectroscopy from Gemini/NIFS of NGC~205 after central resolved star subtraction (Section~\ref{ssec:Kinn205}).  The white intensity contours after the subtraction of resolved stars are now smoother.}
   \label{n205_restars}     
\end{figure*}
%%%%%%%%%%%%%%%%%%%%%%%%%%%%%%%%%%%%%%%%%%%%%%%%%%%%%%%%%%%%%%%%%%%%%%%%

%%%%%%%%%%%%%%%%%%%%%%%%%%%%%%%%%%%%%%%%%%%%%%%%%%%%%%%%%%%%%%%%%%%%%%%%
\begin{figure*}[ht]
\centering
%%%%% NGC 205   
   \begin{minipage}{\linewidth}
       \includegraphics[width=0.33\linewidth,height=0.42\textheight,keepaspectratio=true]{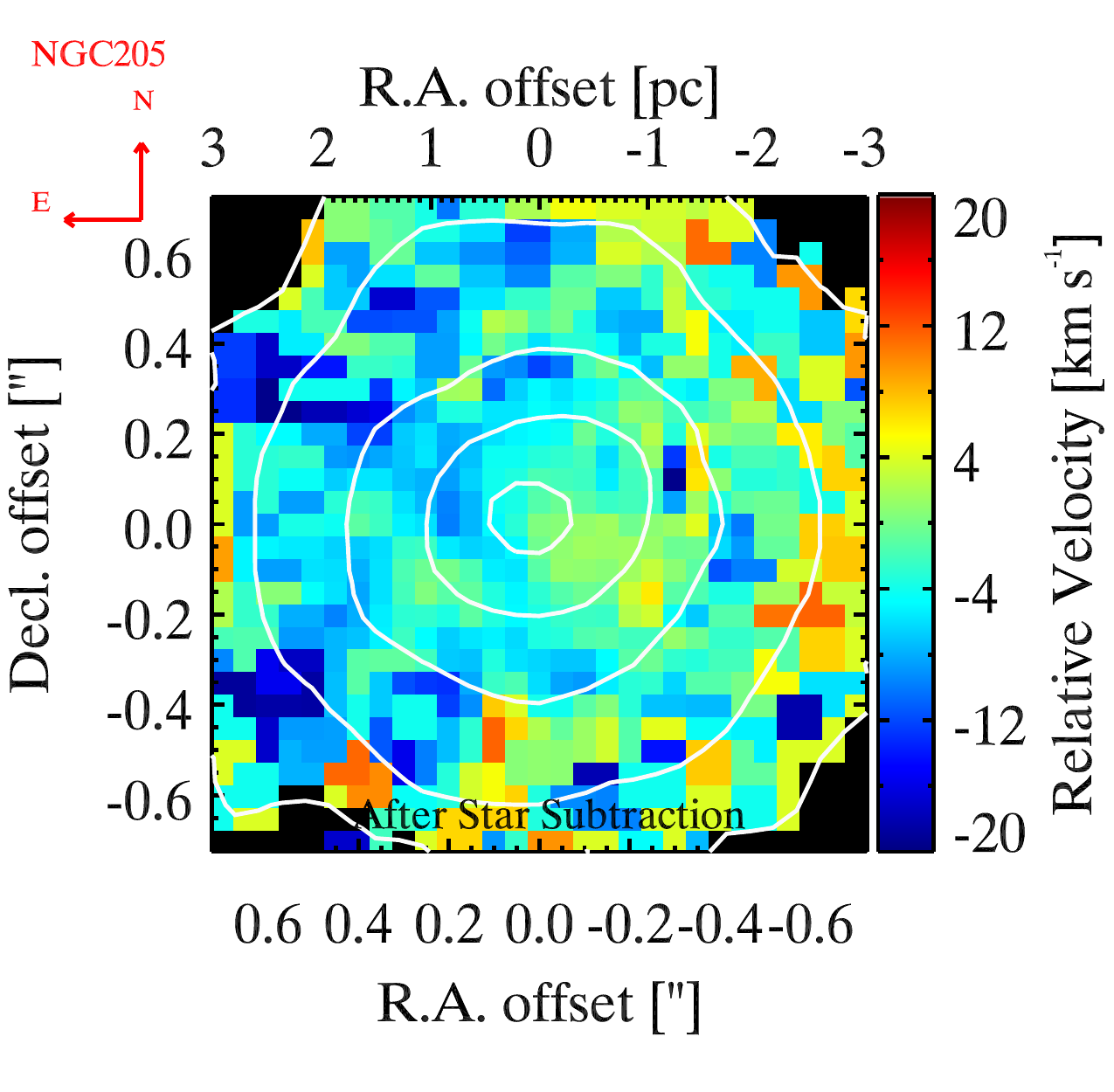}
       \hspace{0.1cm}  
       \includegraphics[width=0.33\linewidth,height=0.42\textheight,keepaspectratio=true]{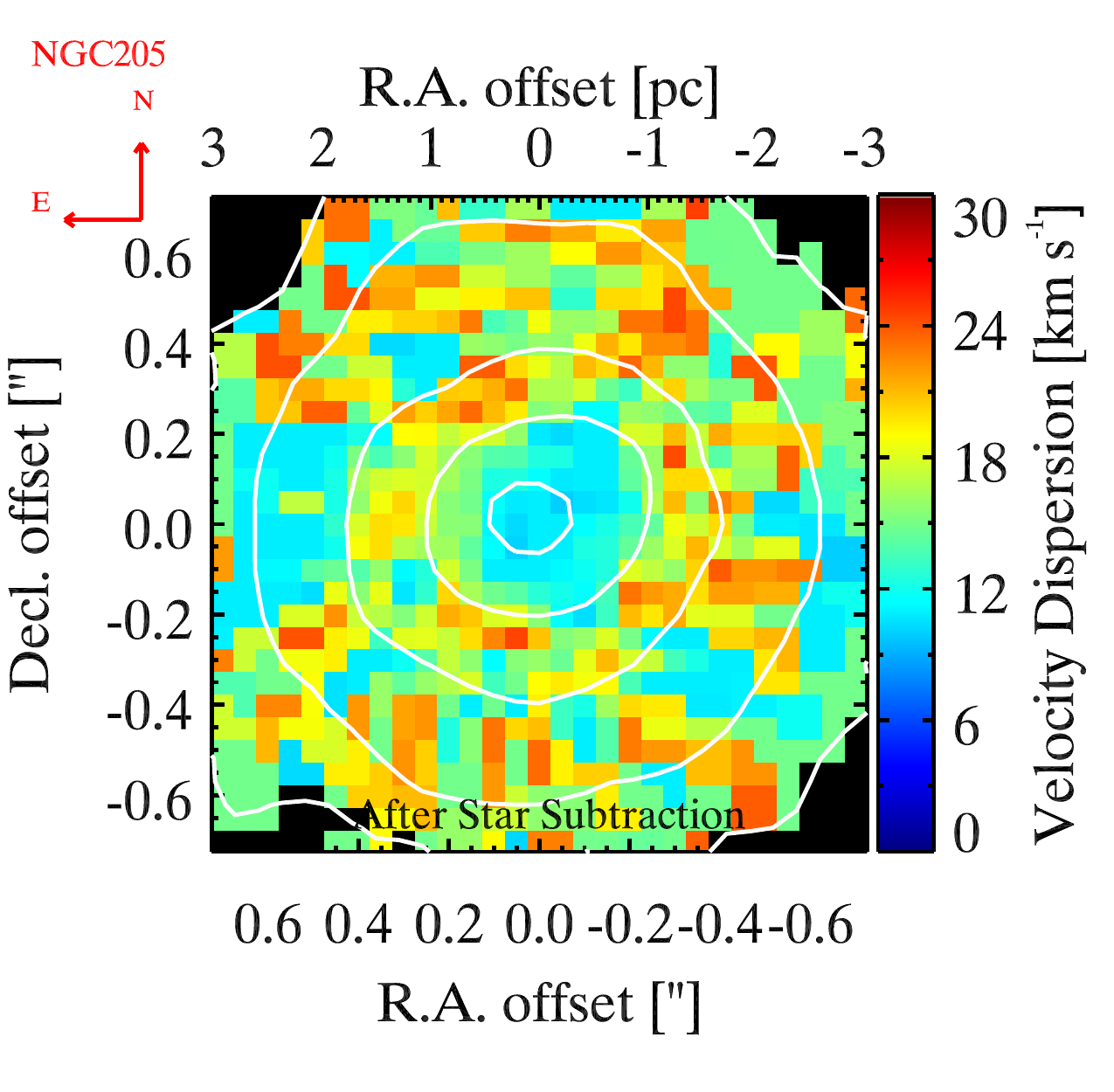}
       \hspace{0.1cm}
       \includegraphics[width=0.3\linewidth,height=0.4\textheight,keepaspectratio=true]{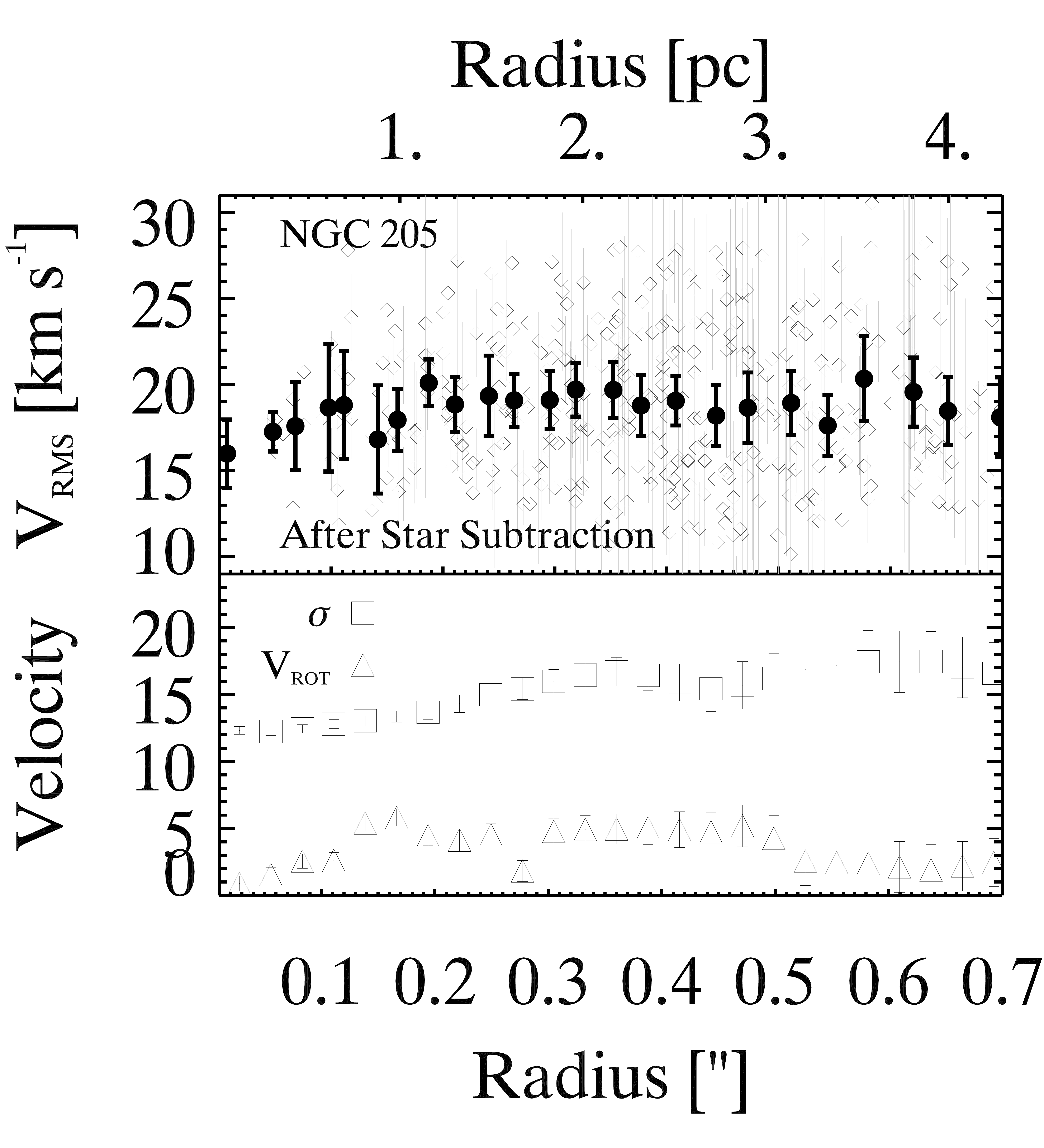}
    \end{minipage}
  %%%%% NGC 5102     
    \begin{minipage}{\linewidth}
       \includegraphics[width=0.33\linewidth,height=0.42\textheight,keepaspectratio=true]{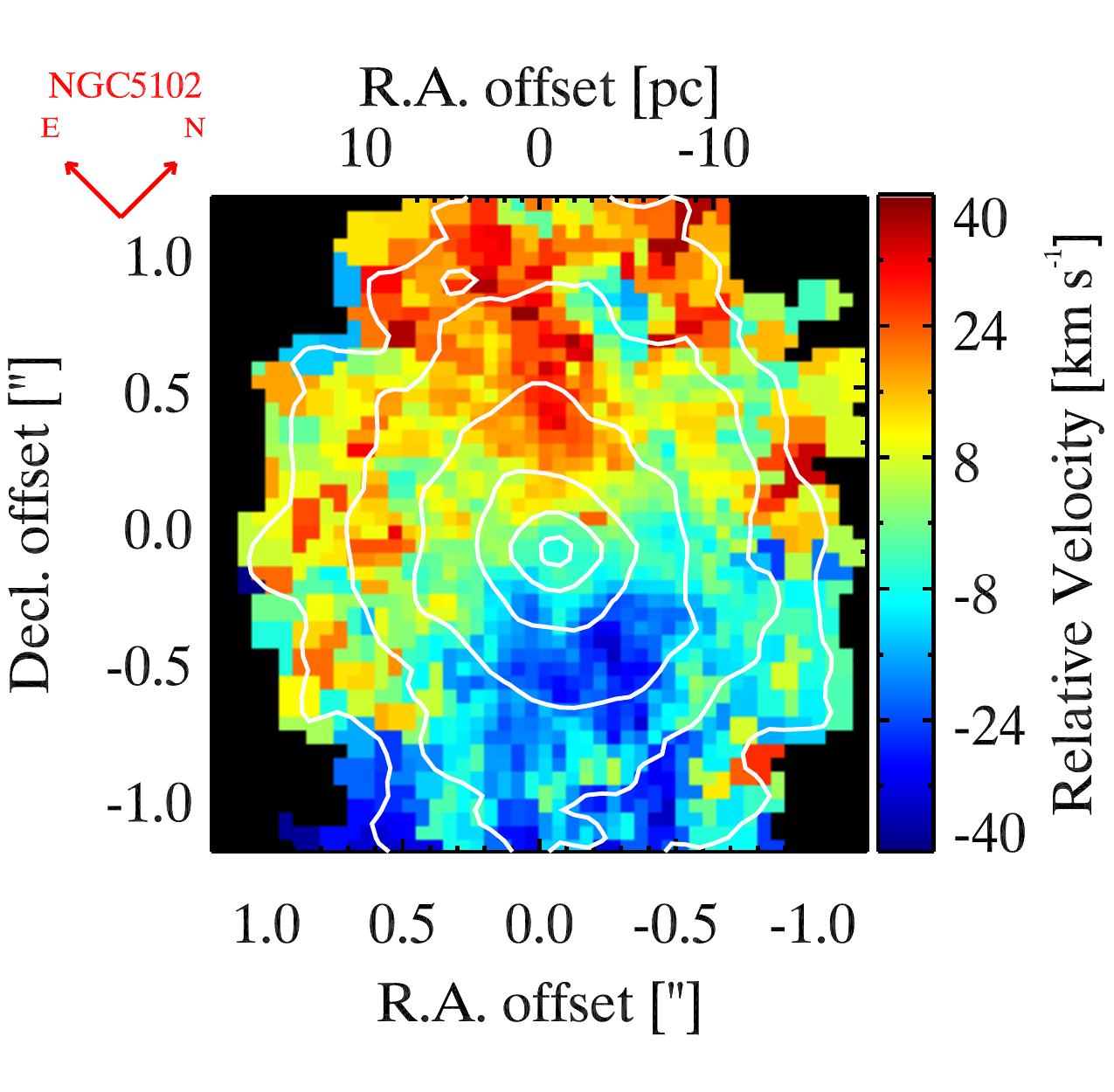}
       \hspace{0.1cm}
       \includegraphics[width=0.33\linewidth,height=0.42\textheight,keepaspectratio=true]{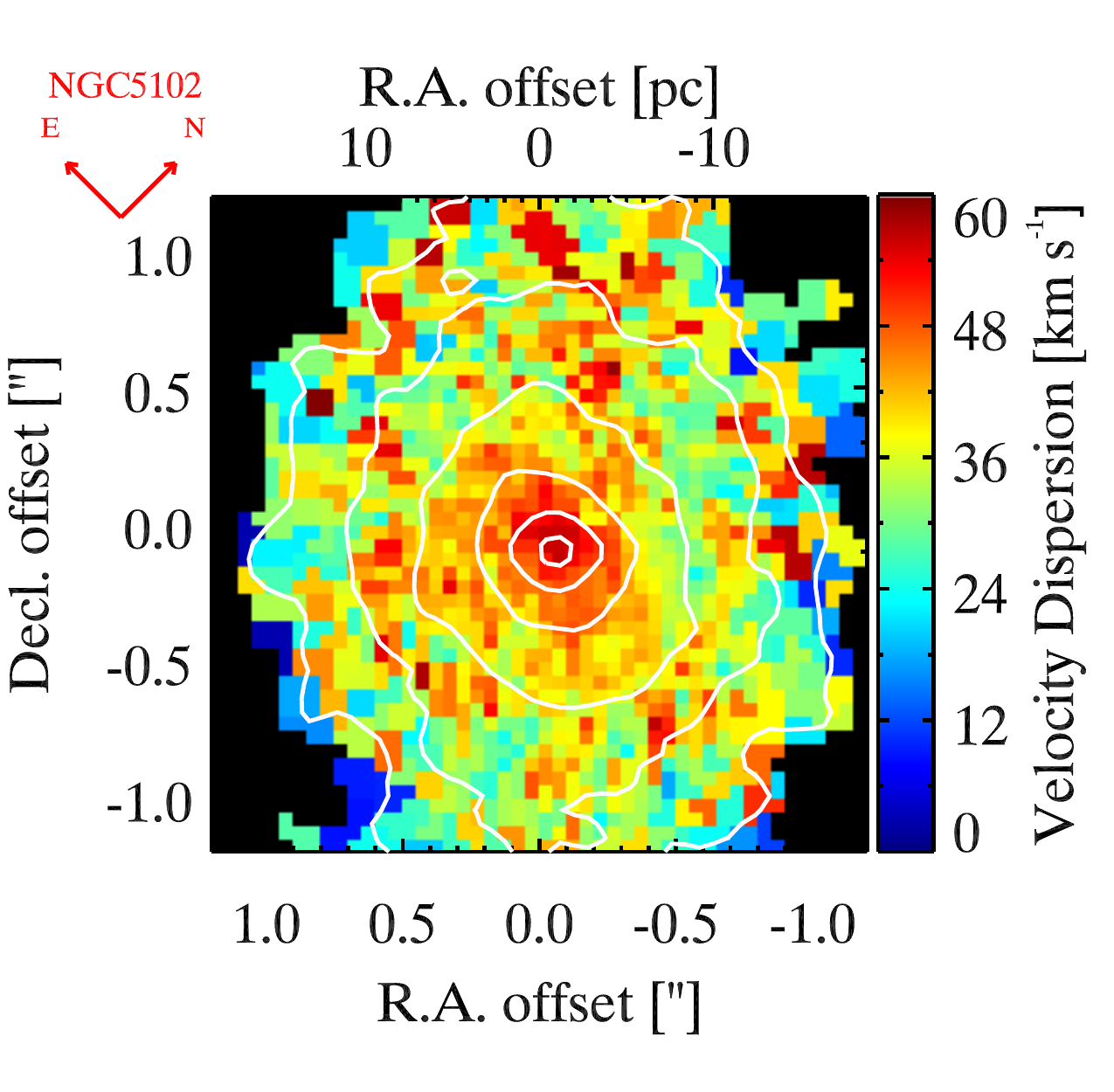}
      \hspace{0.2cm}
       \includegraphics[width=0.29\linewidth,height=0.38\textheight,keepaspectratio=true]{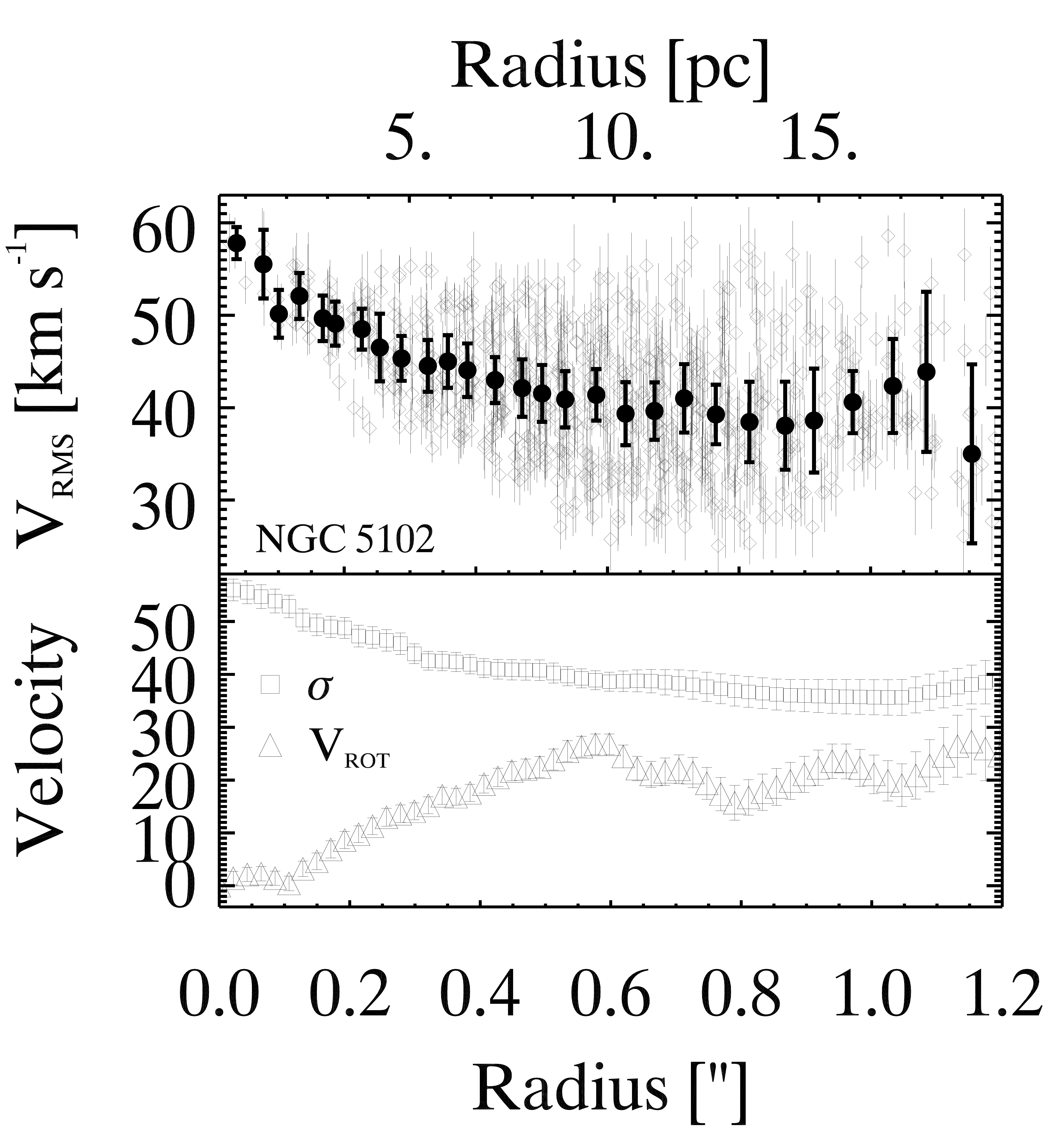}
    \end{minipage}
 %%%%% NGC 5206   
    \begin{minipage}{\linewidth}
       \includegraphics[width=0.33\linewidth,height=0.42\textheight,keepaspectratio=true]{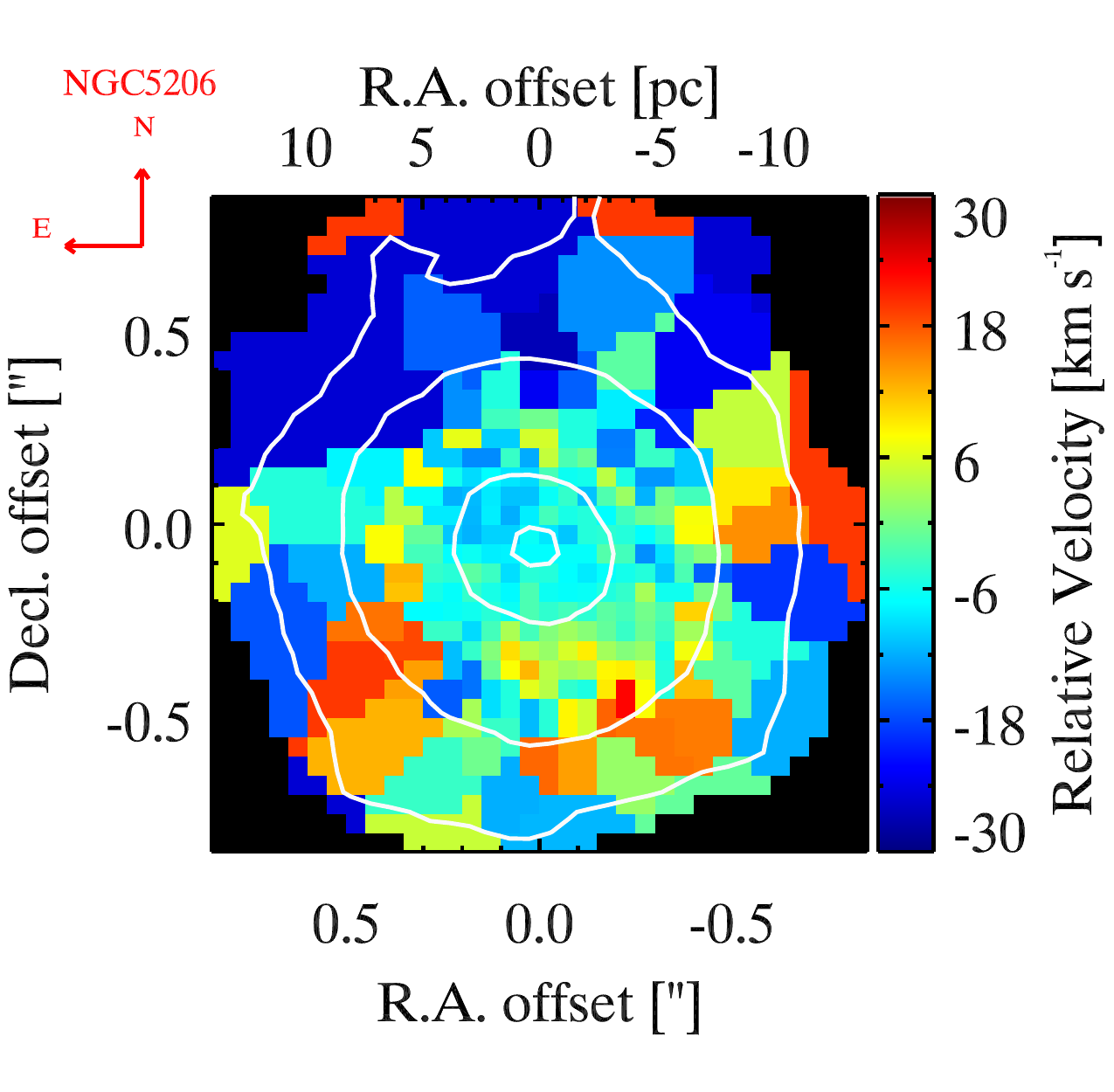}
       \hspace{0.1cm}
       \includegraphics[width=0.33\linewidth,height=0.42\textheight,keepaspectratio=true]{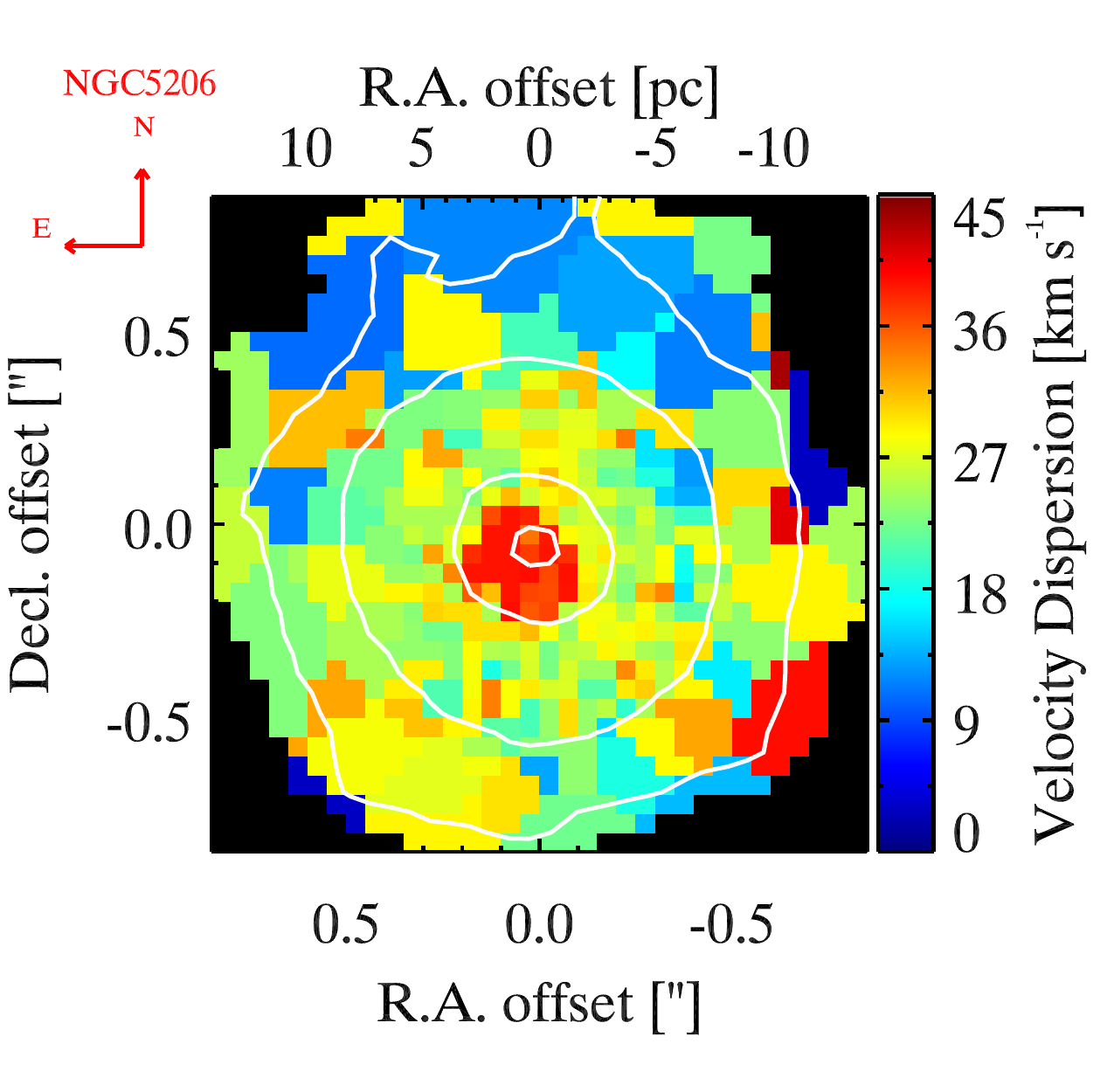}
       \hspace{0.2cm}
       \includegraphics[width=0.29\linewidth,height=0.38\textheight,keepaspectratio=true]{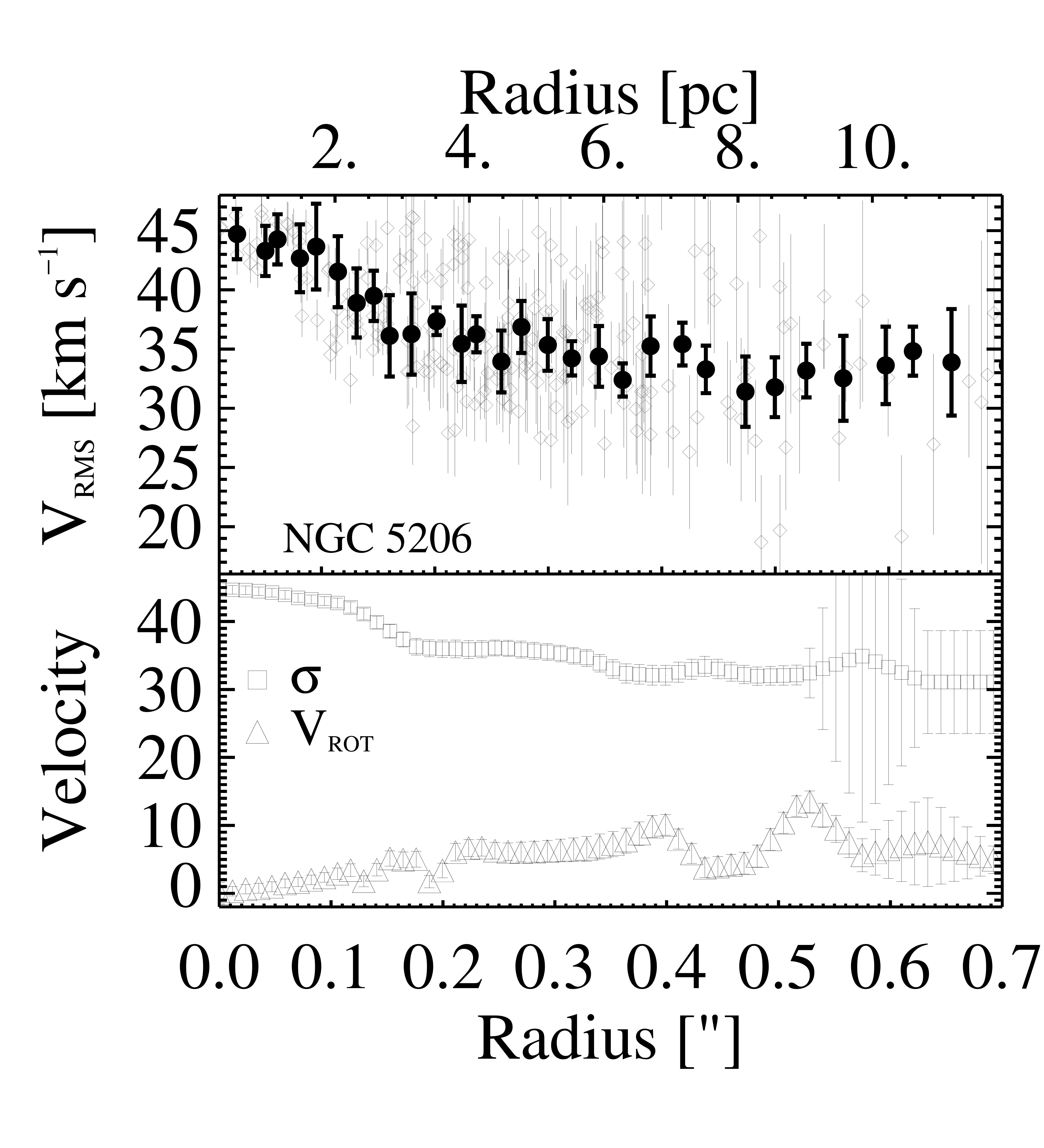}
    \end{minipage}
 \caption{\normalsize Stellar kinematic maps derived from CO bandhead spectroscopy from Gemini/NIFS (NGC~205) and VLT/SINFONI (NGC~5102 and NGC~5206) spectroscopic data. Radial velocity maps are shown in the left column, and dispersion maps in the middle column.  The radial velocity is shown relative to the systemic velocity of $-241\pm2$~\kms (NGC~205), $472\pm2$~\kms (NGC~5102), $573\pm5$~\kms (NGC~5206).   The top plots show the NGC~205 stellar kinematics after removing individual bright stars (Section~\ref{ssec:Kinn205}).     White contours show the stellar continuum; and the  red arrows indicate the orientation.  Kinematics are only plotted out to the radii where they are reliable; black pixels indicate data not used in the JAM modeling.  In the right plots, the top-panel shows the $V_{\rm RMS}=\sqrt{V^2+\sigma^2}$ (open diamonds), with the filled black dots showing the bi-weighted $V_{\rm RMS}$ in the circular annuli.  The error bars are $1\sigma$ deviations of the kinematic measurements of the Voronoi bins in the same annuli.   The KINEMETRY decomposition \citep{Krajnovic06} of dispersion (open squares) and rotation (open triangles) curves are shown in the bottom-panel.  } 
\label{kine_map} 
\end{figure*}
%%%%%%%%%%%%%%%%%%%%%%%%%%%%%%%%%%%%%%%%%%%%%%%%%%%%%%%%%%%%%%%%%%%%%%%
%%%%%%%%%%%%%%%%%%%%%%%%%%%%%%%%%%%%%%%%%%%%%%%%%%%%%%%%%%%%%%%%%%%%%%%
\begin{figure*}[ht]
\begin{minipage}{\linewidth}\hspace{-7mm}
  \includegraphics[width=2.5\linewidth,height=0.225\textheight,keepaspectratio=true]{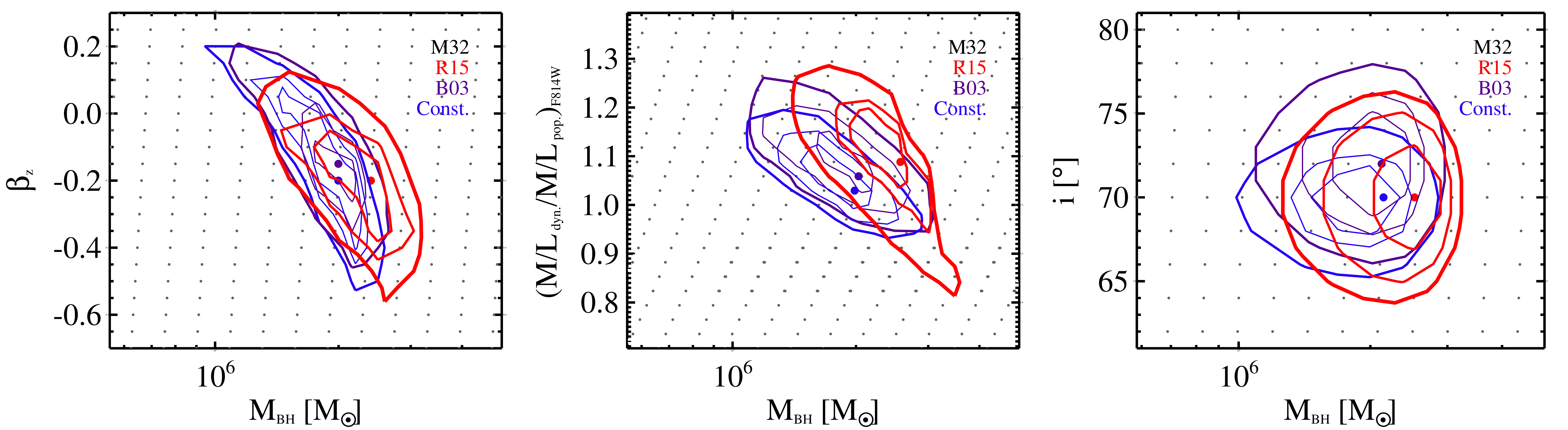}
 \end{minipage} 
 \begin{minipage}{\linewidth}\hspace{-7mm}
   \includegraphics[width=2.5\linewidth,height=0.225\textheight,keepaspectratio=true]{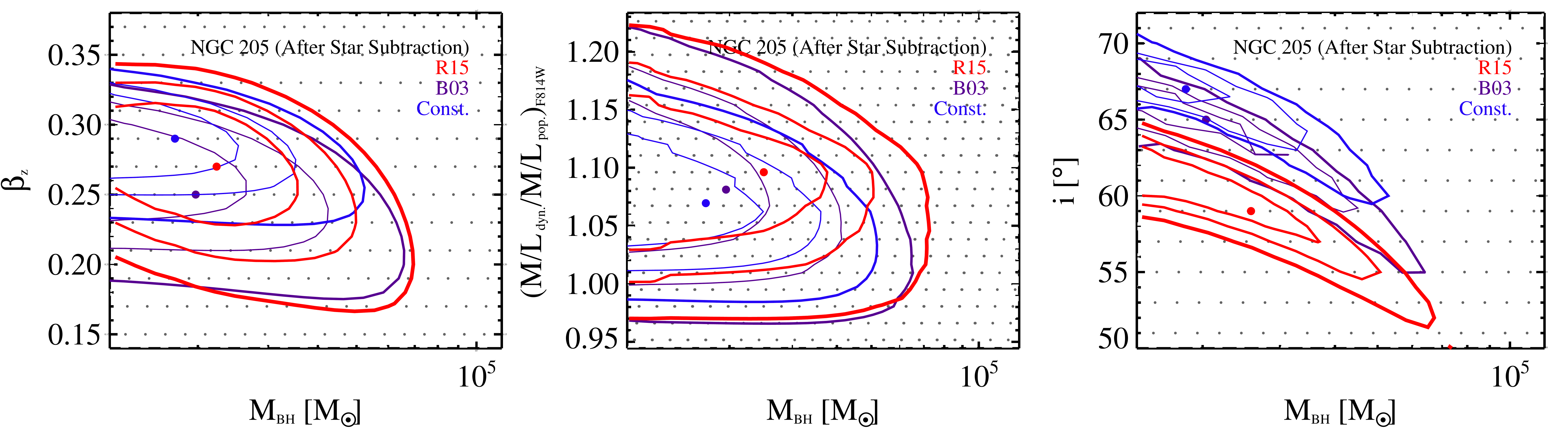}
 \end{minipage}
 \begin{minipage}{\linewidth}\hspace{-7mm}
    \includegraphics[width=2.5\linewidth,height=0.225\textheight,keepaspectratio=true]{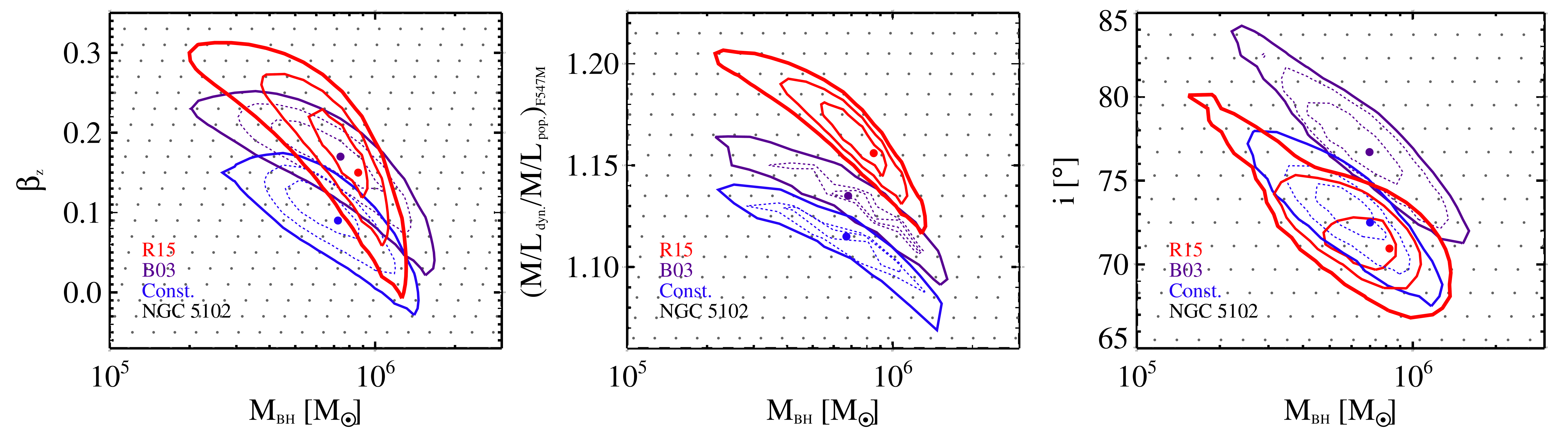}
 \end{minipage}
   \begin{minipage}{\linewidth}\hspace{-7mm}
    \includegraphics[width=2.5\linewidth,height=0.225\textheight,keepaspectratio=true]{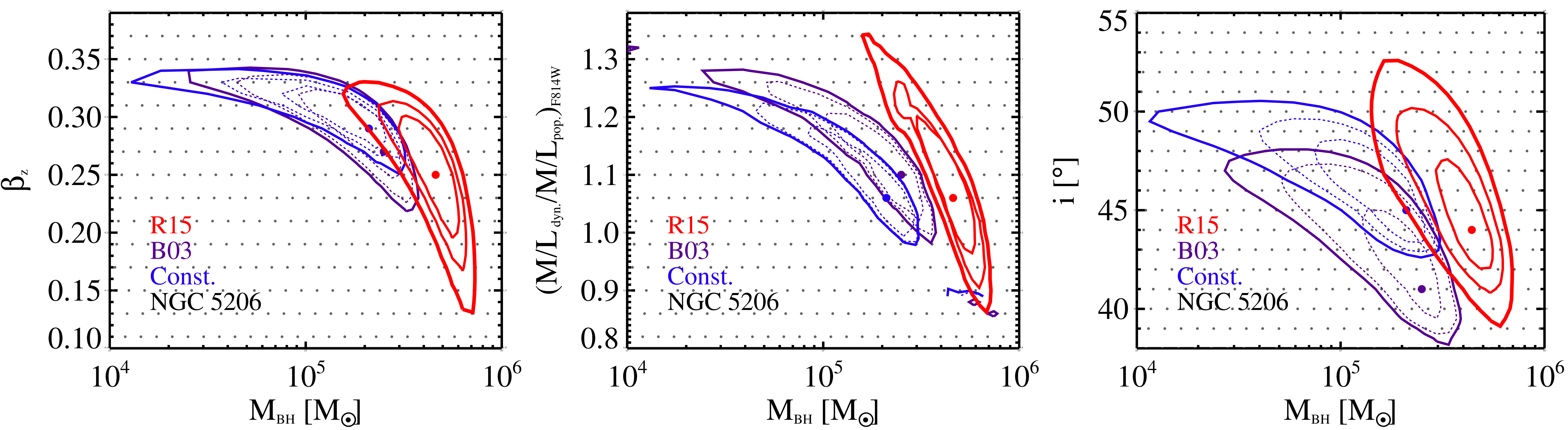}
 \end{minipage}
   \caption{\normalsize Best-fit JAM models.  The red contours and dot show our reference model in each galaxy based on the R15 color--\ml~relation.  The blue and purple contours and dots give the B03 and constant~\ml~mass models.  Contours show $\Delta\chi^2$ = 2.30, 6.18 (dashed lines), and 11.83 (solid line) corresponding to 1$\sigma$, 2$\sigma$, and 3$\sigma$ confidence levels for two parameters after marginalizing over the other two parameters.  The gray dots show the grid of models.  The columns show the BH mass vs.~the anisotropy, $\beta_z$ (left), the mass scaling factor, $\gamma=M/L_{\rm dyn.}/M/L_{\rm pop.}$ (center), and inclination, $i$ (right). } 
\label{jam_plots} 
\end{figure*}
%%%%%%%%%%%%%%%%%%%%%%%%%%%%%%%%%%%%%%%%%%%%%%%%%%%%%%%%%%%%%%%%%%%%%%%%
%%%%%%%%%%%%%%%%%%%%%%%%%%%%%%%%%%%%%%%%%%%%%%%%%%%%%%%%%%%%%%%%%%%%%%%%
\begin{figure*}[ht] 
\minipage{0.49\textwidth}\centering
  \includegraphics[width=\linewidth]{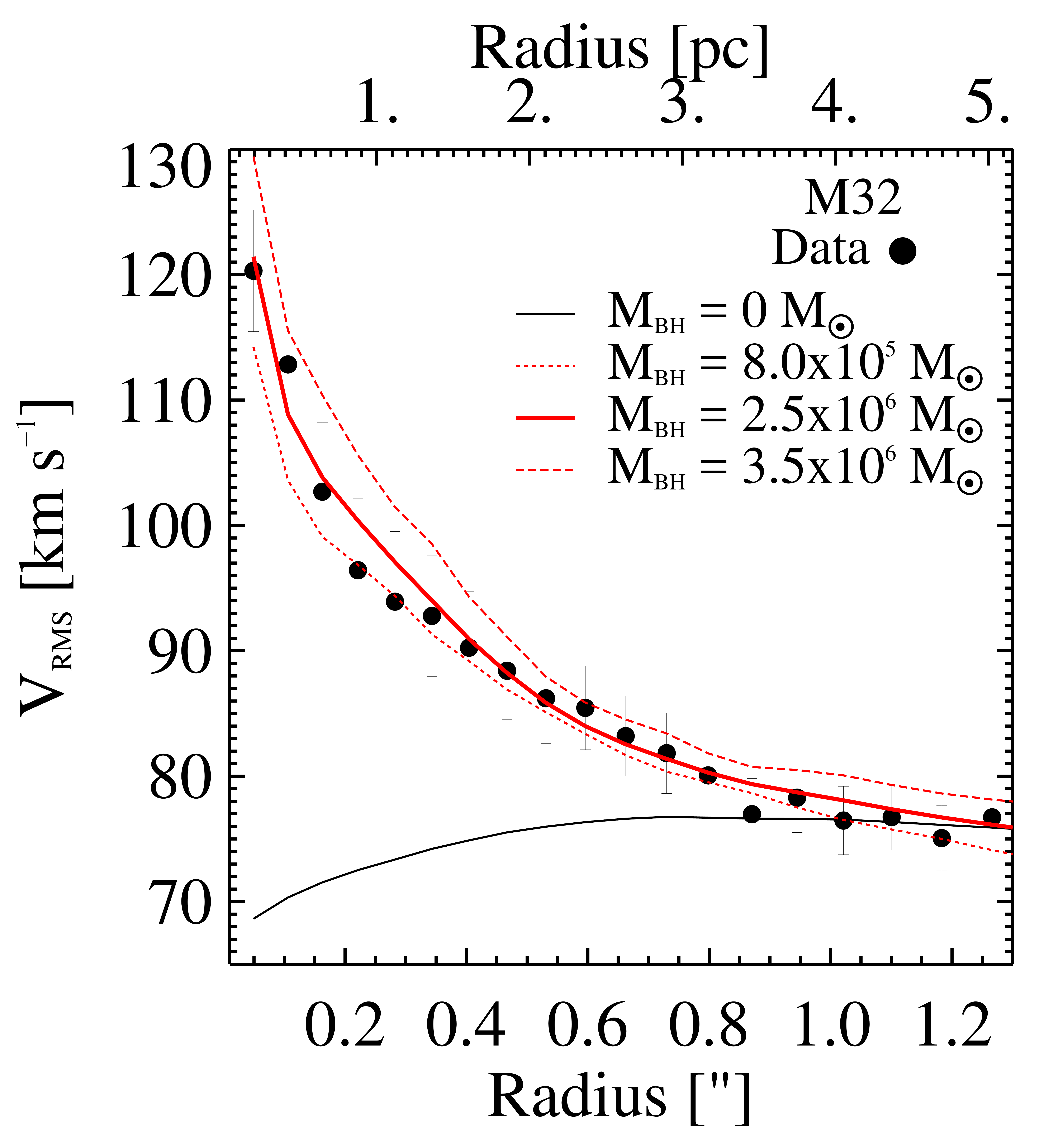}
 \endminipage\hfill
\minipage{0.49\textwidth}\centering
  \includegraphics[width=\linewidth]{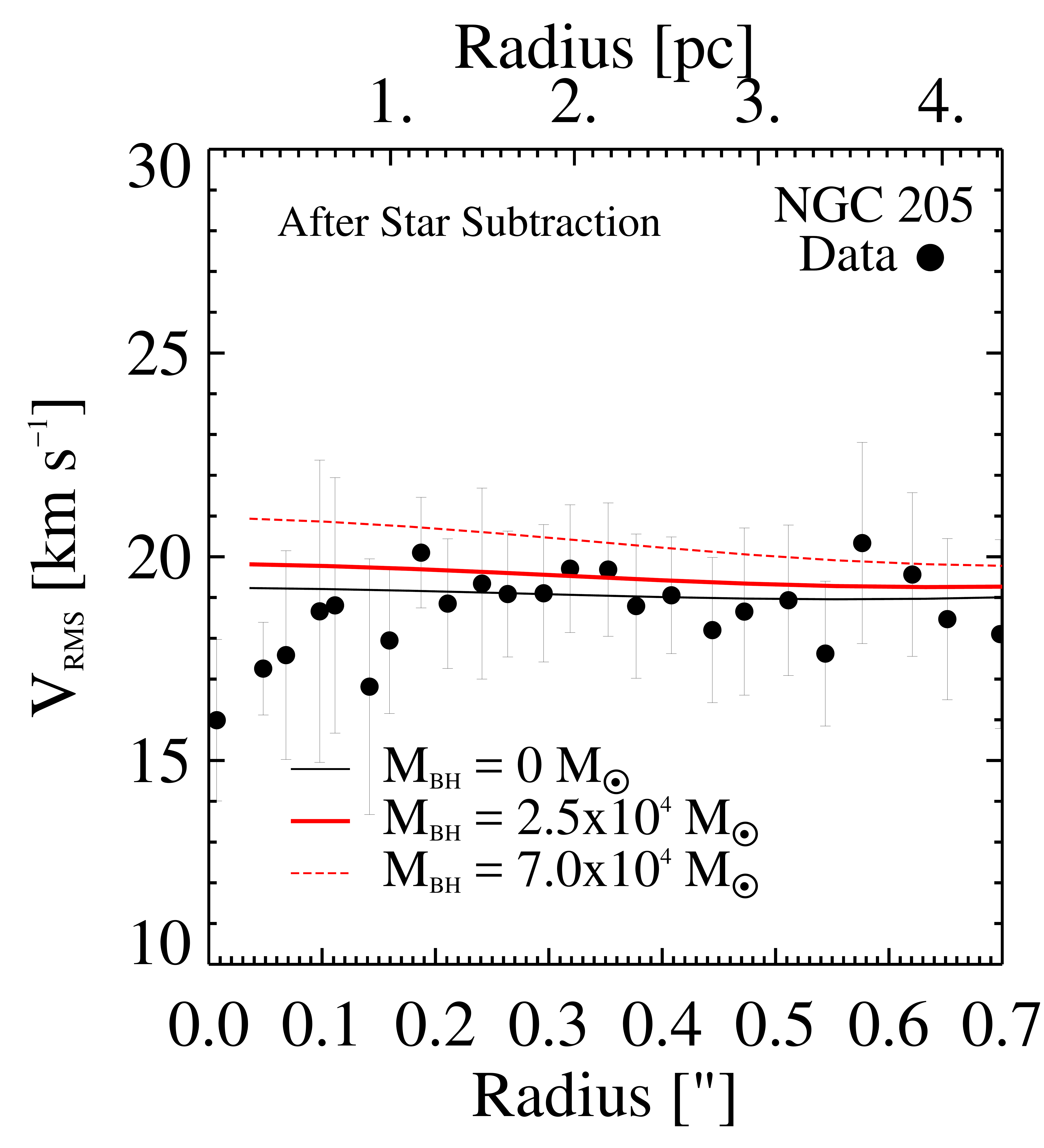}
\endminipage\hfill \vspace{5mm}
\minipage{0.49\textwidth}\centering
 \includegraphics[width=\linewidth]{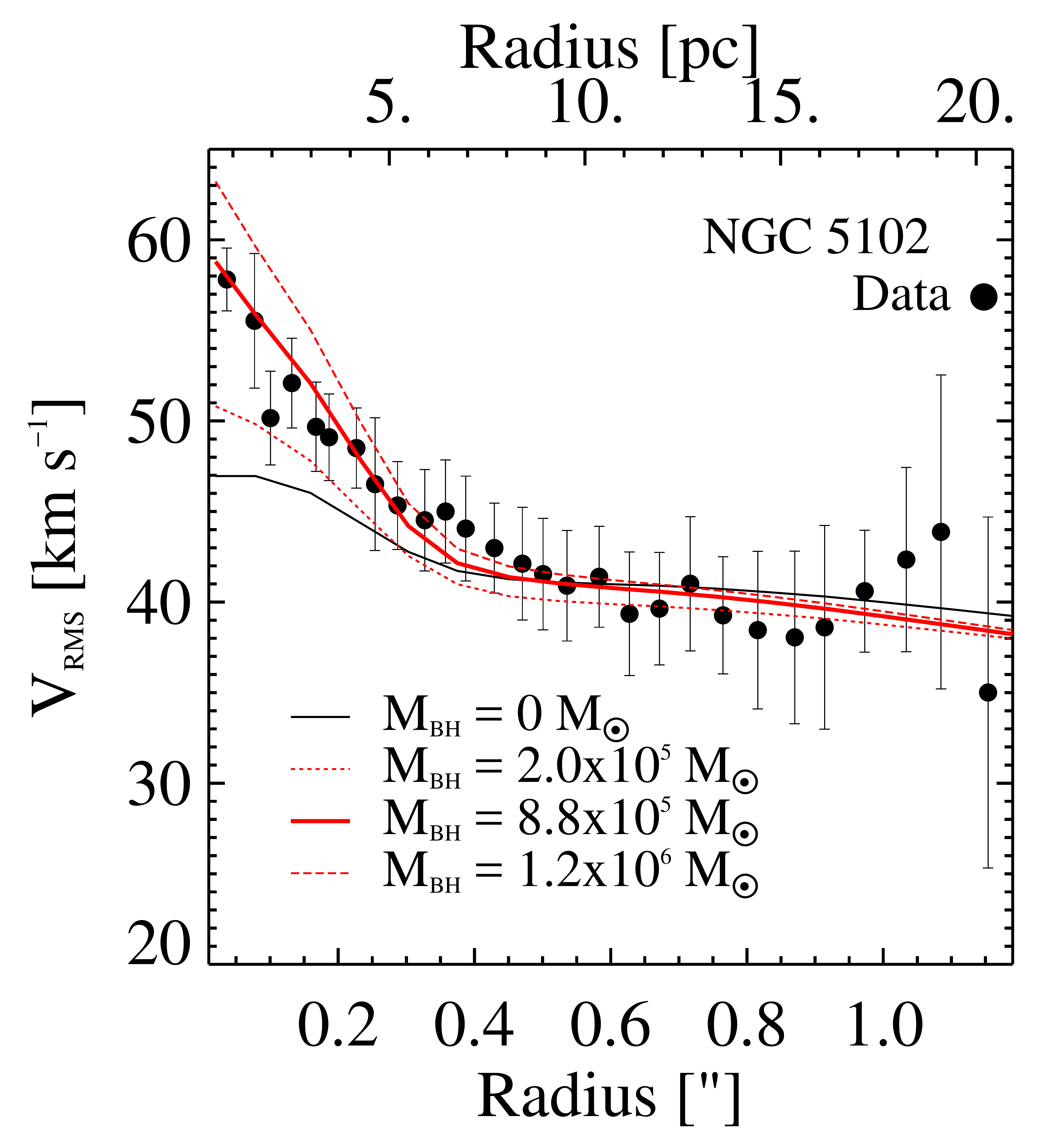}
\endminipage\hfill
\minipage{0.49\textwidth}\centering
 \includegraphics[width=\linewidth]{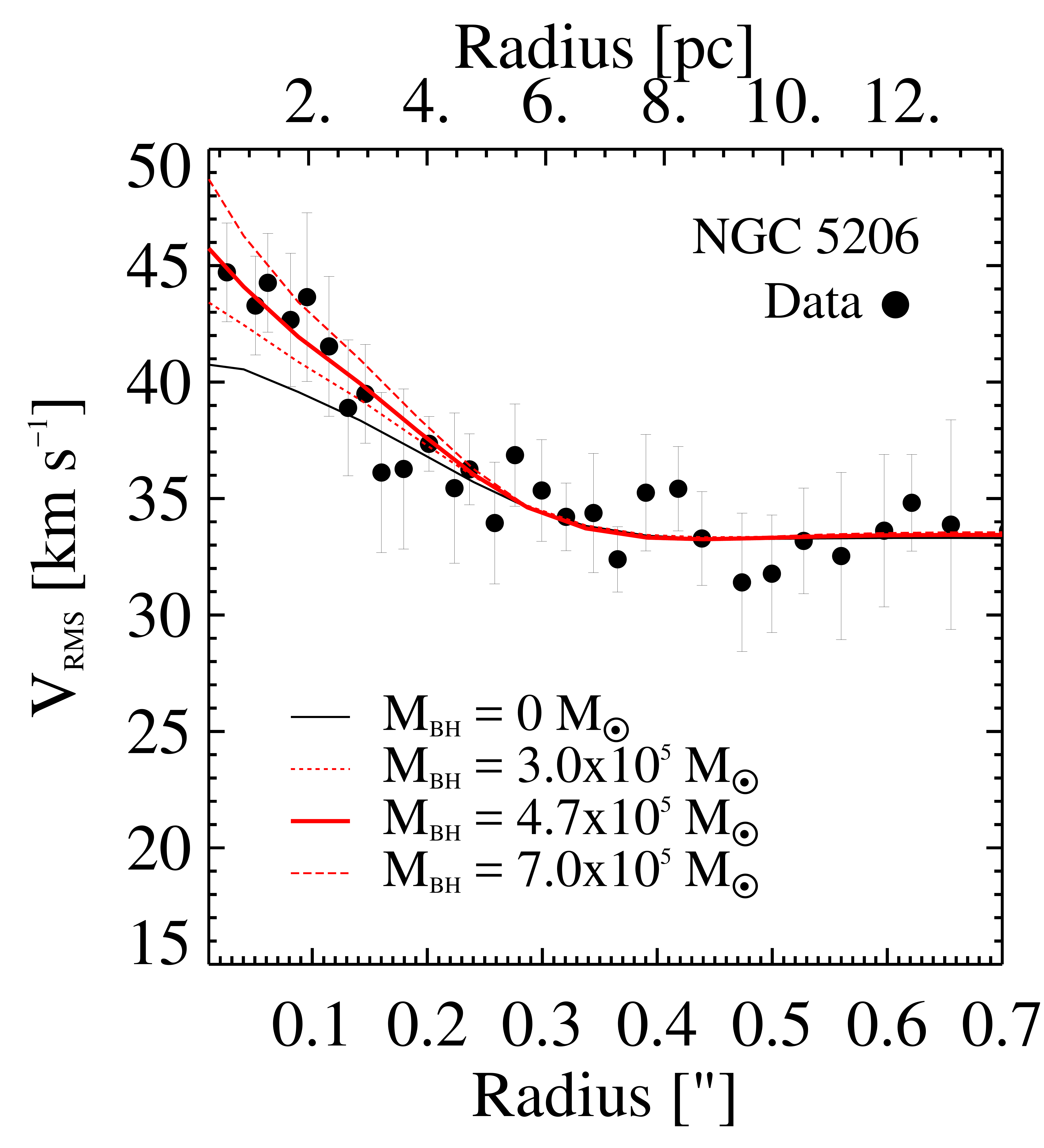}
\endminipage\hfill
  \caption{\normalsize 1D $V_{\rm RMS}$ vs. JAM predictions of mass models with varying BH masses.  The red solid lines show our best-fit JAM models.   Long-dashed and short-dashed lines indicate the upper and lower range of BH masses within 3$\sigma$ (see Figure~\ref{jam_plots} and Table~\ref{kine_jeanmodel_tab}), while the black line shows the best-fit JAM models without a BH.    All models are fixed to the corresponding best-fit inclination angles of galaxies (column 9, Table~\ref{kine_jeanmodel_tab}) but varying for anisotropy, mass scaling ratio, and BH mass, which will be written in the form of ($\beta_z$, $\gamma$, \Mbh, $i$).  The specific zero-mass and best-fit BH models plotted have (0.10,  1.20, 0\Msun, 70\deg) and ($-$0.2, 1.08, $2.5\times10^{6}$\Msun, 70\deg) for M32,  (0.27, 1.12, 0\Msun, 63\deg) and (0.27, 1.10, $7.0\times10^{4}$\Msun, 55\deg) for NGC~205, (0.26, 1.19, 0\Msun, 71.5\deg) and (0.15, 1.15, $8.8\times10^{5}$\Msun, 71.5\deg) for NGC~5102, and (0.30, 1.16, 0\Msun, 44.5\deg) and (0.25, 1.06, $4.7\times10^{5}$\Msun, 45.5\deg) for NGC~5206.  Both $V_{\rm RMS}$ profiles of spectroscopic data and their corresponding JAM models predictions are binned radially in the same manner for all galaxies as presented in the right panels of Figure~\ref{kine_map}.  The error bars are $1\sigma$ deviations of the kinematic measurements of the Voronoi bins in the same annuli.}
 \label{jam1d}     
\end{figure*}
%%%%%%%%%%%%%%%%%%%%%%%%%%%%%%%%%%%%%%%%%%%%%%%%%%%%%%%%%%%%%%%%%%%%%%%%%%%%%
%%%%%%%%%%%%%%%%%%%%%%%%%%%%%%%%%%%%%%%%%%%%%%%%%%%%%%%%%%%%%%%%%%%%%%%%%%%
\begin{figure*}[ht] 
\minipage{1.1\textwidth}\hspace{-10mm}
  \includegraphics[width=\linewidth]{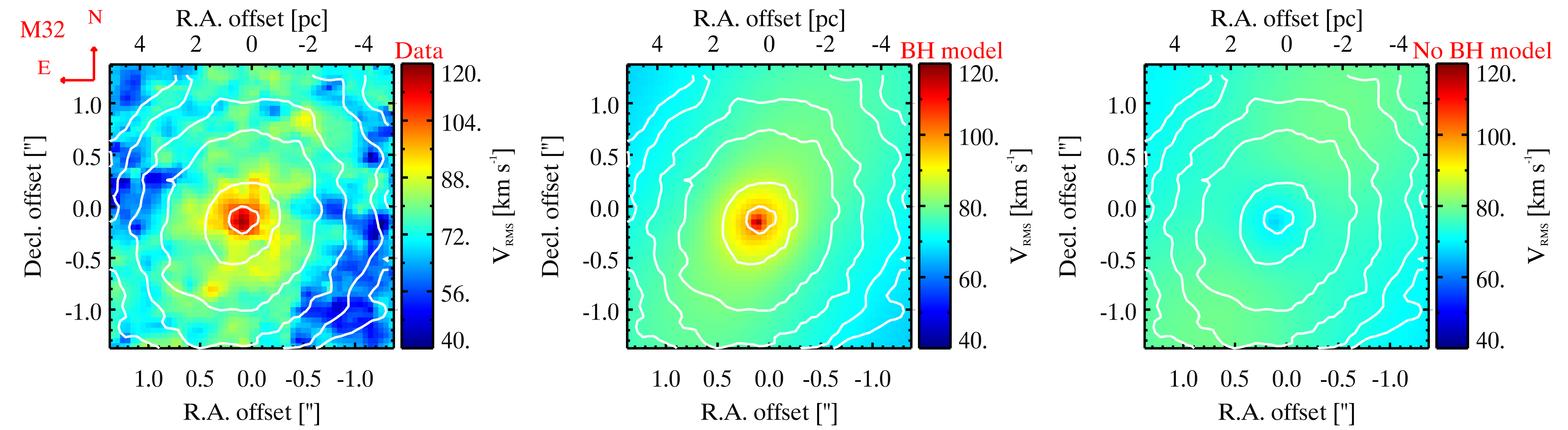}
 \endminipage\hfill
 \minipage{1.1\textwidth}\hspace{-10mm}
  \includegraphics[width=\linewidth]{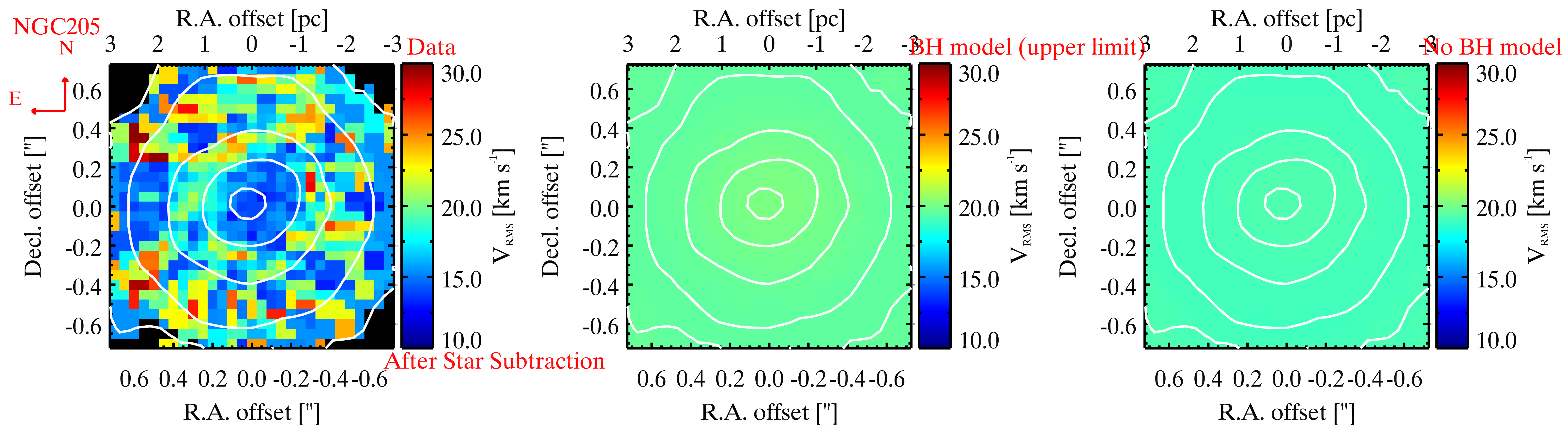}
\endminipage\hfill 
\minipage{1.1\textwidth}\hspace{-10mm}
 \includegraphics[width=\linewidth]{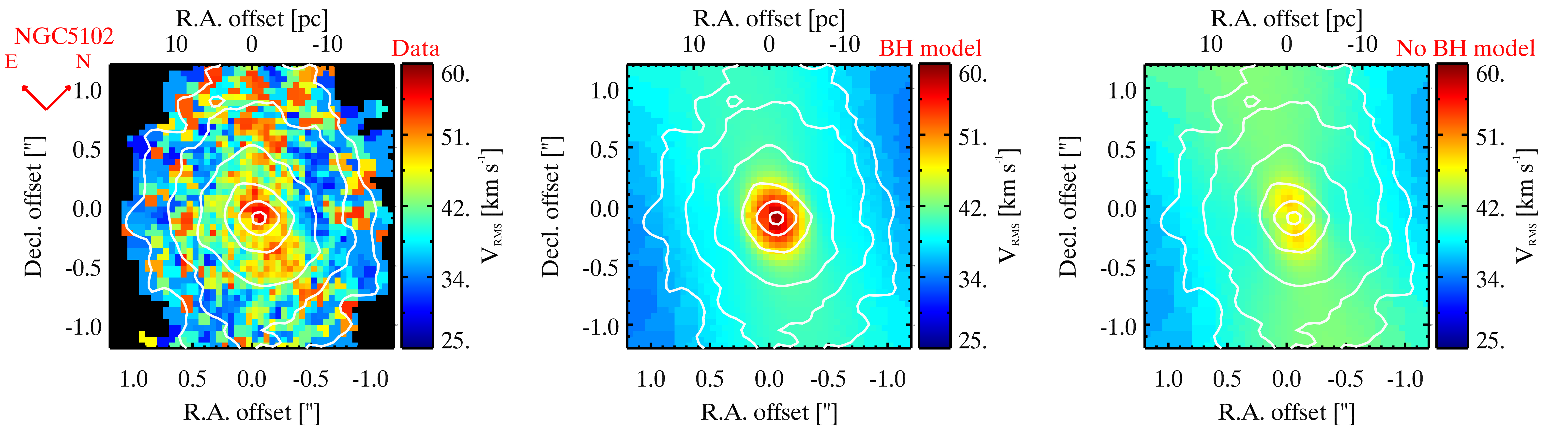}
\endminipage\hfill 
\minipage{1.1\textwidth}\hspace{-10mm}
 \includegraphics[width=\linewidth]{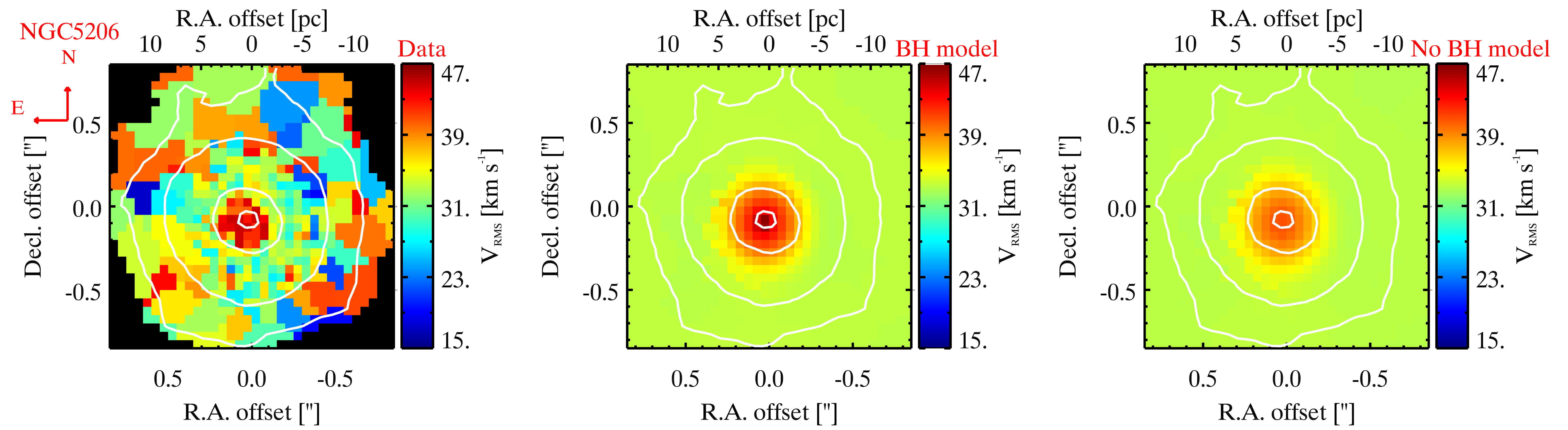}
\endminipage\hfill
  \caption{\normalsize A 2D data-model comparison of JAM models with and without BHs.  The left panels show maps of $V_{\rm RMS}$ data in each galaxy, the middle panels show the best-fit JAM model 2D maps, and the right panels best-fit models without a BH.   These models are the same as shown in 1D in Figure~\ref{jam1d}.   The white contours show the continuum and red arrows the orientation.}
 \label{jam2d}     
\end{figure*}
%%%%%%%%%%%%%%%%%%%%%%%%%%%%%%%%%%%%%%%%%%%%%%%%%%%%%%%%%%%%%%%%%%%%%%%%%%%%%
%%%%%%%%%%%%
\subsection{NSC morphology from 2D GALFIT Models}\label{ssec:2d}
Our dynamical models rely on the accurate measurements of the 2D stellar mass distribution near the centers of each galaxy.   Moreover, the 2D SB profile is also important for quantifying the morphology of the NSCs.   We model the \hst~images around the nucleus using GALFIT.   GALFIT enables fitting with convolved models (using a PSF from {\tt Tiny Tim}, Section~\ref{ssec:psf}), and excluding pixels using a bad pixel mask.    The bad pixel mask is obtained using an initial GALFIT run without a mask; we mask all pixels with absolute pixel values \textgreater$3\sigma$ than the median value in the residual images \texttt{(Data-Model)}.    Based on our 1D SB profile, we chose to fit NGC~205 with a double-S\'ersic function (2S),  M32 is fitted with a 2S + Exponential disk (2S + E).  For NGC~5102 and NGC~5206 we find the nuclear regions require two components (Figure~\ref{hst_gal_res}), and thus these are fitted with triple-S\'ersic functions (3S).  The initial guesses were input with the best-fit parameters from the 1D SB fits with fixed parameters for the outermost S\'ersic or E component except for the PA and axis ratio ($b/a$).  For the purpose of creating mass models for dynamical modeling of BH masses measurements, we repeat these 2D fit with fixed PA in all S\'ersic components because the Jeans Anisotropic Models \citep[JAM;][]{Cappellari08} assume axisymmetry, which implies a constant PA.   

The left column of Figure~\ref{hst_gal_res} show the F814W images of M32, NGC~205, and NGC~5206 and F547M image of NGC~5102, while the middle column show their relative errors between the \hst~images and their 2S models, \texttt{(Data-Model)/Data}.   The right column also show these relative errors between the \hst~images and their 2S + E (M32) or 3S (NGC~5102 and NGC~5206) models.     The contours on each panel show both data (black) and model (white) at the same radius and flux level to highlight the regions of agreement and disagreement between data and models.  Excluding masked regions (and point sources in NGC~205), the maximum errors on the individual fits are \textless9\%, 15\%, 13\%, and 10\% for M32, NGC~205, NGC~5102, and NGC~5206, respectively.  The parameters of the best-fit GALFIT models are shown in Columns 4--9 of Table~\ref{seric_nsc_tab}; errors are scaled based on the 1D errors in the components estimated in Section~\ref{ssec:1d} due to their dominant over the GALFIT errors.  We note that the 2D GALFIT models give S\'ersic parameters consistent to that of 1D SB profiles fit, especially for the middle S\'ersic components where the PSF effects are minimal. We note that our NSC component in M32 is significantly less luminous than the 1D SB profile given in \citet{Graham09}; we believe this is at least in part due to their normalization in the SB profile, which is nearly a magnitude higher than that derived here.  

Next, we use the final fixed-PA 2D GALFIT S\'ersic models to create multi-Gaussian expansion \citep[MGEs;][]{Emsellem94a, Emsellem94b, Cappellari04} models.  These MGEs are comprised of a total of 18, 10, 14, and 14 Gaussians to provide a satisfactory fit to the surface mass density profiles of M32, NGC~205, NGC~5102, and NGC~5206 respectively.   We fit our MGEs out to $\sim$$110\farcs0$ for M32 and NGC~5206, $15\farcs0$ for NGC~205, $21\farcs0$ for NGC~5102 and obtain the mass models by straightforward multiplying the \ml~profiles with the 2D light GALFIT MGEs at the corresponding radii.   We choose to parameterize the mass models using GALFIT (as opposed to using the \texttt{fit\_sectors\_mge} code) because (1) it enables us to properly incorporate the complex \hst~PSF, and (2) it enables simple separation of NSC and galaxy components.

%%%%%%%%%%%
\subsection{Color--\emph{\ml}~Relations}\label{ssec:colorm2ls}

To turn our stellar luminosity profiles into stellar mass profiles, we assume a \ml--color relation.    There is a strong correlation between the color and the \ml~of a stellar population: we use two different color--\ml~relations including the \citet{Bell03} (hereafter B03), \citet{Roediger15} (henceforth R15) color--\ml~correlations, as well as models with constant \ml.   This is similar to the method presented by \citet{Nguyen17} for analysis of the NGC~404 nucleus.    For our best-fit models we use the R15 relation, which was found by \citet{Nguyen17} to be within 1$\sigma$ of the color--\ml~relation derived from stellar population fits to STIS data within the nucleus.  The B03 and constant \ml~models are used to assess the systematic uncertainties in our mass models.  The B03 and R15 relations are built based on the $(V-I)_0$ color profiles in the bottom panel of each plot in Figure~\ref{sbp_lmp}, except in NGC~5102. For NGC~5102, we use the $(B-V)_0$ color estimated from the combined \hst~F547M--F656N data at small radii and F450W--F569W data at larger radii as described in the appendix \ref{n5102color}. 

We also create mass models with a constant central \ml.  For these models, we take a reasonable reference value for the \ml, but still allow these values to be scaled in our dynamical models, just as for the varying \ml~models.  We note that these models are not used in any of our final results, but are used to analyze potential systematic errors in our dynamical modeling. The reference values used for the constant \ml~models are:
\begin{itemize}
  \item {\it M32:} We use $M/L_I = 1.41$ (\Msun/\Lsun) based on the Schwarzschild model fits to Sauron data from \citet{Cappellari06}.
  \item {\it NGC~205:} We use a similar nucleus of $M/L_I=1.95$ (\Msun/\Lsun) to that based on Schwarzschild models fit to STIS kinematics by \citep[$M/L_I=1.94$ (\Msun/\Lsun),][]{Valluri05}; note that this is the value found for the nucleus (they find a significantly higher \ml~for the galaxy as a whole).
  \item {\it NGC~5102:} We use \ml$_V = 0.55$ (\Msun/\Lsun) based on the stellar population model fits near the center of the galaxy from Figure~11 of \citet{Mitzkus17} which assume use the MILES libraries with a Salpeter IMF.
  \item {\it NGC~5206:} We use the photometric $M/L_I = 1.98$ (\Msun/\Lsun) based on the color $(V-I)_o=1.1$ and using estimates from \citet{Bell03, Bruzual03}.
\end{itemize}

%%%%%%%%%%%  
\subsection{Mass Models}\label{ssec:massmodels}
We create our final mass models for use in our dynamical modeling by multiplying the MGE luminosity created from the 2D GALFIT fit models of the images with PSF-deconvolution (Section~\ref{ssec:2d}) with the \ml~profiles discussed in Section~\ref{ssec:colorm2ls}.    We fit the MGE luminosity by using the {\tt sectors\_photometry} + {\tt mge\_fit\_sectors} \texttt{IDL}\footnote{available at {\tt http://purl.org/cappellari/software}} package.   During these fits we set their PAs and axial ratios of the Gaussians to be constants as their values obtained in GALFIT.   More specifically, in NGC~5102 we calculate the $V$--band \ml~and apply this to the F547M data, while in the others, we calculate $I$--band \ml s and apply these to the F814W data.

The top-panel in each plot of Figure~\ref{massden} shows the mass surface density and the bottom-panel shows the relative error of the \ml~profile relative to the R15 color--\ml~correlation for each galaxy.     It is clear that the R15 color--\ml~correlation (red line) predicts less mass at the center of each galaxy than the prediction of B03 color--\ml~correlation (purple line).    However, the assumption of constant \ml s (blue line) predict more mass at the centers but less mass at larger radii than the predictions of both B03 and R15 color--\ml~correlation.  This is as expected based on the bluer centers and consequently lower \ml s at the centers.   The MGEs of mass surface densities of M32, NGC 205, NGC~5102, and NGC 5206 are presented in column (2) of Table~\ref{tab_mges} (Appendix~\ref{tabs}).

%%%%%%%%%%%
\subsection{$K$--Band Luminosities}\label{ssec:kband}

Together with the stellar mass distribution, the dynamical models also require as input the distribution of the stellar tracer population from which the kinematics is obtained.  Given that the kinematics was derived in the $K$--band, which covers the wavelengths from 2.29 to 2.34$\mu$m, we create high-resolution synthetic images in that band \citep{Ahn17}.  We do not use the $K$--band images obtained from the IFU observations directly as they have lower resolution and too limited FOV.  To make these we employ Padova SSP model \citep{Bressan12} to fit color--color correlations.  The purpose of this is to transform our \hst~imaging of F555W and F814W (of M32, NGC~205, and NGC~5206) into $K$--band images.   Specifically, we fit the linear correlations of \hst~F555W--F814W vs. F814W--$K$ based on the specific nucleus stellar populations and metallicities for M32 \citep[$Z=0.019$,][]{Corbin01}, NGC~205 \citep[$Z=0.008$,][]{Butler05}, and NGC~5206 ($Z=0.004-0.008$; N. Kacharov et al., in prep.).   We use these to create $K$--band images based on the reference \hst~images (i.e., we add the color correction to the F814W image).   These $K$ band images are similar to the $I$--band images, with deviations in the SB profile being comparable to differences between mass models shown in Figure~\ref{massden}.

In order to create a $K$--band image for NGC~5102, we use its F450W--F569W colors, which are inferred from the F547M--F656N data as described in the appendix \ref{n5102color}.  Next, we fit a linear correlation of \hst~F450W--F569W vs. F569W--$K$ using metallicities $Z=0.004$ \citep{Davidge15}, then infer for its $K$--band images.    We then fit the MGEs using the {\tt mge\_fit\_sectors} \texttt{IDL}\footnote{available at {\tt http://purl.org/cappellari/software}} package.  The MGEs $K$--band luminosity surface densities of M32, NGC~205, NGC~5102, and NGC~5206 are given in column (1) Table~\ref{tab_mges} (Appendix~\ref{tabs}). 

%%%%%%%%%%%%%%%%%%%%%%%%%%%%%%%%%%%%%%%%%%%%%%%%%%%%%%%%%%%%%%%%%%%%%%%%%%%%%%
%% THE SECTION KINEMATICS OF NSC -- Dispersion, velocity, Corrected for FOV Stars!!! 
%%%%%%%%%%%%%%%%%%%%%%%%%%%%%%%%%%%%%%%%%%%%%%%%%%%%%%%%%%%%%%%%%%%%%%%%%%%%%%
\section{Stellar Kinematics Results}\label{sec:kine}

 We use adaptive optics NIFS and SINFONI  spectroscopy to determine the nuclear stellar kinematics in all four galaxies.   We first re-bin spatial pixels within each wavelength of the data cube using the Voronoi binning method \citep{Cappellari03} to obtain S/N $\gtrsim$ 25 per spectral pixel.   Next, we use the pPXF method \citep{Cappellari04, Cappellari17} to derive the stellar kinematics from the CO band-heads of the NIFS and SINFONI  spectroscopy in the wavelength range of 2.280--2.395~$\mu$m and determine the line-of-sight velocity distribution (LOSVD).   We fit only a Gaussian LOSVD, measuring the radial velocity ($V$) and dispersion ($\sigma$) using high spectral resolution of stellar templates of eight supergiant, giant, and main-sequence stars with spectral types between G and M with all luminosity classes \citep{Wallace96}.  These templates are matched to the resolution of the observations by convolving them by a Gaussian with dispersion equal to that of the line-spread function (LSF) of the observed spectra at every wavelength.   These LSFs are determined from sky lines.  For NIFS, the LSF is quite Gaussian with a FWHM $\sim$4.2~\AA, but the width varies across the FOV \citep[e.g.,][]{Seth10b}.  
 However, the SINFONI LSF appears to be significantly non-Gaussian as seen from the shape of the OH sky lines.     To characterize the shape of the LSF and its potential variation across the FOV, we reduced the sky frames in the same way as the science frames, with the difference that we did not subtract the sky.     We then combined the reduced sky cubes using the same dither pattern as the science frames. This ensured that the measured LSF on the resulting sky lines fully resembles the one on the object lines.    From these dithered sky lines we measured the LSF.    We used 6 isolated, strong sky lines, all having close doublets, except one (the 21,995 \AA \ line), to measure the spectral resolution across the detector.    Since the LSF appears to be constant along rows (constant y values), the sky cubes were collapsed along the y-direction.
 
Sky emissions OH-lines can be described as a delta function, $\delta_{\lambda}$. Once they reach the spectrograph they will be dispersed so that the intensity pattern is redistributed with wavelength according to the LSF of the instrument. Since we know the central wavelength of the sky emission lines we assume that their shapes represent the LSF. Once located, the peak values in the spectra and a region around the peaks is defined, the continuum is subtracted, the line flux is normalized to the peak flux and finally the lines are summed up.

The spectral resolution across the detector has a median value of 6.32 \AA \ FWHM ($\frac{\lambda}{\Delta\lambda} = 3820$) with values ranging from 5.46 to 6.7 \AA \ FWHM (R $\sim$ 3440 -- 4300).  The last step before the kinematic extraction is to perform the same binning on the LSF cube as for the science cube. This varying LSF is then used in the kinematic extraction with pPXF.  
    
To obtain optimal kinematics from the SINFONI data we had to correct for several effects: (1) a scattered light component and (2) velocity differences between individual cubes.   For the first issue, the SINFONI data cubes have imperfect sky subtraction, with clear additive residuals remaining despite subtraction of sky cubes.     These residuals appear to have a uniform spectrum that is spatially constant across the field, however the residuals are not clearly identified as sky or stellar spectra.  To remove these residuals, we create a median residual spectrum for each individual data cube using pixels beyond $1\farcs3$ radius and subtract it from all spaxels in the cube.  This subtraction greatly improved the quality of our kinematic fits to our final combined data cube, but we note that this is likely also subtracting some galaxy light from each pixel.  Second, fits to sky lines revealed variations in the wavelength solutions corresponding to velocity offsets of up to 20~\kms~between individual data cubes.  Therefore before combining the cubes, we applied a velocity shift to correct these shifts; 4/12 cubes for NGC~5102 and 3/12 cubes for NGC~5206 were shifted before combining bringing the radial velocity errors $\lesssim$2~\kms~among the cubes to their means for both galaxies.   We note however that applying these velocity shifts had minimal impact ($<$0.5~\kms) on the derived dispersions.  The systemic velocity in each galaxy was estimated by taking a median of pixels with radii $<$$0\farcs15$, and is listed in Table~\ref{host_tab}.

 To calculate the errors on the LOSVD, we add Gaussian random errors to each spectral pixel and apply Monte Carlo simulations to re-run the pPXF code.     The errors used differ in NIFS and SINFONI; for NIFS we have an error spectrum available, and these are used to run the Monte Carlo, while with SINFONI, no error spectrum is available, and we therefore use the standard deviation of the pPXF fit residuals as a uniform error on each pixel.    We test further the robustness of our kinematic results by (1) fitting the spectra toward the short wavelength range 2.280--2.338~$\mu$m of the CO band-heads (the highest S/N portion of the CO bandhead) and (2) using the PHOENIX model spectra \citep{Husser13}, which have higher resolution ($R=500,000$ or 0.6 \kms) than the \citet{Wallace96} ($R=45,000$ or 6.67 \kms) templates.    We find consistent kinematic results within the errors with (1) returning dispersions ~1--2~\kms~higher than the full spectral range, while (2) yields fully consistent results.  We also verified our SINFONI kinematic errors by combining independent subsets of the data and identical binning and found that their distribution had a scatter similar to the error; a similar verification of the NIFS errors was done in \citet{Seth14}.  This analysis indicates that the radial velocities and dispersion errors range from 0.5--20~\kms.  These kinematic data are presented in Table~\ref{kine_tab_n205}, \ref{kine_tab_n5102}, and \ref{kine_tab_n5206} for NGC~205, NGC~5102, and NGC~5206 respectively in the appendix~\ref{tabs}, except for M32, which was published in Table~1 of \citet{Seth10b}.   
 
 Due to increased systematic errors affecting our kinematic measurements at low surface brightness (where many pixels are binned together), we eliminate the outermost bins beyond an ellipse with semi-major axes of 1$\farcs$3, 1$\farcs$2, 0$\farcs$7, and 0$\farcs$7 in M32, NGC~5102, NGC~5206, and NGC~205.  In general, we find that in the noisy data in outer regions we overestimate our $V_{\rm RMS}$ values.  For instance in the M32 data published in \citet{Seth10b}, we find a rise in $V_{\rm RMS}$ beyond 1$\farcs$3, however comparison with data from \citet{Verolme02} suggests this rise may be due to systematic error, and is not used in the remainder of our analysis.  In addition, we only plot and fit bins with dispersion and velocity errors $<$15~\kms.   We correct the systemic velocities (then radial velocities) for the barycentric correction (37~\kms, 20~\kms, $-$20~\kms, and $-$37~\kms~for M32, NGC~205, NGC~5102, and NGC~5206).      The systemic velocities are determined from the central velocity as references and double checked with using the  {\tt fit\_kinematic\_pa} \texttt{IDL} package\footnote{available at {\tt http://purl.org/cappellari/software}}.

 The final kinematic maps for NGC~205, NGC~5102 and NGC~5206 are shown in Figure~\ref{kine_map}.  The left column shows the radial velocity maps, the middle column shows the velocity dispersion maps.  The right most column shows the $V_{\rm RMS}$ measurements (open gray diamonds with error bars, top panel) and the bi-weighted in the circular annuli (fill black dots with error bars).  The bottom panel of the right column shows the results of KINEMETRY fits \citep{Krajnovic06}, which determines the best-fit rotation and dispersion profiles along ellipses.  We discuss the kinematics for each galaxy below.

%%%%%%%%%%%
\subsection{M32}\label{ssec:Kinem32}
The kinematics of M32 were derived with pPXF using a four-order of Gauss--Hermite series including $V$, $\sigma$, skewness ($h_3$), and kurtosis ($h_4$) \citep{Seth10b}.  These measurements are the highest quality kinematic data available that resolve the sphere-of-influence (SOI) of the BH.     These kinematics show a very clear signature of a rotating and disky elliptical galaxy with strong rotation at a PA of $-25$\deg~(E of N) with an amplitude of $\sim$55~\kms~beyond the radius of $0\farcs3$ with the maximum of $V/\sigma$ reaching 0.76 at $r=0\farcs5$ (2~pc).    The dispersion has a peak of $\sim$120~\kms, which we will show is due to the influence of the BH, and flattens out at $\sim$76 \kms~beyond $0\farcs5$.   
   
%%%%%%%%%%%
\subsection{NGC~205}\label{ssec:Kinn205}
Of the nuclei considered here, NGC~205 is by far the one most dominated by individual stars.  These are clearly visible in the \hst~images (Figure~\ref{hst_gal_res}), and their effects can be seen in our kinematics maps.  These maps show bright individual stars that often have decreased dispersion and larger offsets from the systemic velocity, and therefore we have attempted to remove them before deriving the kinematics as done in \citet{Kamann13}.   The spectrum of the star weighted by the PSF was subtracted from all adjacent spectra of the cube.     We summarize this process briefly here.   First, we identify bright stars in the FOV of NIFS data manually by-eye and note down their approximate spaxel coordinates in all layers of the data cube; there are 32 of these identified bright stars in total.  Second, we estimate PSFs for these stars that are well-described analytically either by Moffat or double-Gaussian; the differences between these two PSF representations are small, but we use the Moffat profile in our final analysis.    Next, we perform a combined PSF fit for all identified stars; in this analysis, we model the contribution from the galaxy core as an additional component that is smoothly varying spatially.   Third, we  obtain spectra for all identified stars so that their information were used to subtract off from each layer of the cube.    Finally, we use this star-subtracted cube to determine the kinematics of underlying background light of the galaxy.    

 We show our original map of relative velocity in the left panel of Figure~\ref{n205_restars}.  One of the subtracted stellar spectra is shown in the middle panel, while velocity map after the stellar subtraction is shown in the right panel. Subtraction of stars yields a much smoother velocity map and isophotal contours than our original map. The complete kinematic maps determined after star subtraction are shown in the top row plots of Figure~\ref{kine_map}.      We use this map for all further analysis, but also include the original map in tests on the robustness of our results.    

We examine the stellar rotational velocity ($V$) and dispersion velocity ($\sigma_{\star}$) of NGC~205.    The dispersion velocity drops to $\sim$15~\kms~within the $0\farcs2$ radius and reaches $\sim$23~\kms~in the outer annulus of $0\farcs2$--$0\farcs8$.     This dispersion velocity map is consistent with the radial dispersion profile obtained from \hst/STIS data of the NGC~205 nucleus \citep{Valluri05} at the radius \textgreater$0\farcs2$,  although we find a slightly lower dispersion at the very nucleus than \citet{Valluri05} ($\Delta$$\sigma\sim3$~\kms).  This difference in dispersions is $\sim$1~\kms~higher the velocity dispersion errors of the central spectral bins (Table~\ref{kine_tab_n205} in Appendix~\ref{tabs}).  However, we caution that both sets of kinematics are close to the spectral resolution limits of the data.  However, with the high S/N of the central  bins, and using our careful LSF determination, we believe we can reliably recover dispersions even at $\sim$15~\kms \citep{Cappellari17}.

The $V_{\rm RMS}=\sqrt{V^2+\sigma^2}$ profile has a value of $20\pm5$~\kms~out to $0\farcs7$. The maximum $V/\sigma\sim0.33$ occurs at $0\farcs6$ (2.6 pc), and this low level of rotation is seen throughout the cluster.

%%%%%%%%%%%
\subsection{NGC~5102}\label{ssec:Kinn5102}
The radial velocity map of the NGC~5102 nucleus shows clear rotation.     This rotation reaches an amplitude of $\sim$30~\kms at $0\farcs6$ from the nucleus.    Beyond this radius, the rotation curve is flat out to the edge of our data.  The dispersion is fairly flat at $\sim$42~\kms~from $r = 0\farcs3$--$1\farcs2$ with a peak at the center reaching $59\pm1$~\kms.  The maximum $V/\sigma$ is $\sim$0.7 at $0\farcs6$. This suggests the second S\'ersic component, which is identified as part of the NSC, and is the most flattened component, may be strongly rotating.  This components position in the $V/\sigma$ vs. $\epsilon$  ($\epsilon=0.4$ and $V/\sigma=0.7$) is consistent with a rotationally flattened system, especially given the best-fit inclination of $\sim$71$^\circ$ we derive below. 

Our kinematics are consistent with those from \citet{Mitzkus17}, who use MUSE spectroscopy to measure kinematics out to radii of $\sim$30$\arcsec$.  On larger scales they find the galaxy has two counter-rotating stellar disks, with maximum rotation amplitudes of $\sim$20~\kms and dispersions very similar to our measured dispersion throughout the central $\sim$10$\arcsec$.   
    
%%%%%%%%%%%
 \subsection{NGC~5206}\label{ssec:Kinn5102}
 The radial velocity map of NGC520 shows a small but significant rotation signal in the nucleus; this rotation appears to gradually rise outwards with a maximum amplitude of 10--15~\kms~at $0\farcs5$. Our decomposition of the nucleus shows a nearly round system ($b/a = 0.96$) for both NSC components, however, the maximum $V/\sigma_{\star}\sim 0.3$ at $0\farcs5$ suggests the outer of these two components may be more strongly rotating. Similar to NGC~5102,  the NGC~5206 dispersion map shows a fairly flat dispersion of 33~\kms~from $r=0\farcs4$--$0\farcs7$, increasing to 46~\kms at the center.  

%%%%%%%%%%%%%%%%%%%%%%%%%%%%%%%%%%%%%%%%%%%%%%%%%%%%%%%%%%%%%%%%%%%%%%%%%%%%
%%%%%%%%%%%%%%%%%%%%%%%%%%%%% STELLAR KINEMATICS DYNAMICAL MODELS SECTION %%%%%%%%%%%%%%%%%
%%%%%%%%%%%%%%%%%%%%%%%%%%%%%%%%%%%%%%%%%%%%%%%%%%%%%%%%%%%%%%%%%%%%%%%%%%%%
\section{Stellar Dynamical Modeling and BH Mass Estimates}\label{sec:jeans}  

In this section, we fit the stellar kinematics using JAM models and present estimates of the BH masses.  

%%%%%% Table 6 of dynamical JAM model results for best-fit BH models for 5 galaxies %%%%%%%%
\begin{table*}[ht]
\caption{Gemini/NIFS and VLT/SINFONI Jeans modeling best-fit results}
\hspace{-4mm}
\begin{tabular}{cccccccccccc}
\hline \hline
Object &Filter&Color--\ml&  \Mbh  & \betaz &$\gamma$&  $i$  & No. of Bins  &$\chi^2_{\rm r}$&$\chi^2_{\rm r,\; no BH}$&    $r_{\rm g,\; dyn.}$    &   $r_{\rm g,\; dyn.}$  \\ 
           &        &                &(\Msun)&            &                &(\deg)&          &             &            &($\arcsec$)&   (pc)     \\
                      
  (1) &               (2)           &                    (3)            &  (4) &       (5)     &(6)   &  (7)      &      (8)        & (9)   & (10)          &             (11)      &   (12)   \\[1mm]        
\hline 
%%%% M32 %%%   
M32   &F814W$^{\rm G}$&  R15 &$2.5^{+0.6}_{-1.0}\times10^6$&$-0.20^{+0.30}_{-0.35}$&$1.08^{+0.20}_{-0.30}$&$70.0^{+6.0}_{-6.0}$&1354 &1.12&6.72&0.404&1.61\\[1mm]
%%%%%%% unsubtraction nucleus stars    NGC 205 %%%%%
NGC~205 &F814W$^{\rm G}$& R15   &$2.5^{+4.7}_{-2.5}\times10^4$&$0.27^{+0.07}_{-0.10}$&$1.10^{+0.10}_{-0.18}$&$59.0^{+5.0}_{-8.0}$&256 &1.24&1.20&0.033&0.14\\[1mm]                                                                                     
%%%% NGC 5102 %%%
NGC~5102&F547M$^{\rm G}$& R15 &$8.8^{+4.2}_{-6.6}\times10^5$&$0.15^{+0.16}_{-0.16}$&$1.15^{+0.06}_{-0.03}$&$71.5^{+9.0}_{-4.0}$&1017&1.12&3.57&0.070&1.20\\[1mm]
%%%% NGC 5206 %%%
NGC~5206&F814W$^{\rm G}$& R15 &$4.7^{+2.3}_{-3.4}\times10^5$&$0.25^{+0.08}_{-0.12}$&$1.06^{+0.05}_{-0.05}$& $44.0^{+8.5}_{-5.0}$&240 &1.20&2.69&0.058&1.00\\[1mm]                                                                                                                                                   
\hline
\end{tabular}
\tablenotemark{}
\tablecomments{\normalsize   Column 1: galaxy name.    Column 2: the filter in which the luminosity and mass models are constructed; the superscript D or G means the stellar mass profile that are constructed from the \hst~imaging or its GALFIT model (see Table~B6 for full listing of modeling results).   Column 3: the color--\ml~relation used to construct mass models.    Columns 4--7: the best-fit parameters and their $3\sigma$ error intervals of JAM models including dynamical BH masses ($M_{BH}$), anisotropies ($\beta_z$), mass scaling factors ($\gamma$), and the inclination angles ($i$), respectively.    Column 8: number of stellar kinematic measurement bins.    Columns 9 and 10: the reduced $\chi^2_{\rm r}$ and $\chi^2_{\rm r,\; no\;BH}$ of the best-fit JAM models and no BH cases, respectively.  Columns 11 and 12: the best-fit BH sphere-of-influence (SOI) in arcsec and parsecs.}
\label{kine_jeanmodel_tab}
\end{table*}
%%%%%%%%%%%%%%%%%%%%%%%%%%%%%%%%%%%%%%%%%%%%%%%%%%%%%%%% 
%%%%%%%%%%%
\subsection{Jeans Anisotropic Models}\label{ssec:jam}   

We create dynamical models using the Jeans anisotropic models (JAM) of \citet{Cappellari08}\footnote{Specifically, we use the IDL version of the code, available at {\tt http://purl.org/cappellari/software}}.   Given the strong gradients in the stellar population of some of the galaxies, the mass-follows-light assumption is not generally acceptable. For this reason we distinguish the parametrization of the stellar mass, from that of the tracer population. This approach was also used for the same reason in the JAM models by \citet{Ahn17, Li17, Mitzkus17, Nguyen17, Poci17, Thater17}.   Specifically, we used the MGEs derived using the default R15 color--relation to parametrize the stellar mass, but we adopted the $K$--band luminosity MGEs to parametrize the tracer population in JAM.    We use JAM to predict the second velocity moment, $V_{\rm RMS}=\sqrt{V^2+\sigma^2}$ -- where $V$ is the radial velocity relative to the systemic velocity and $\sigma$ is the line-of-sight velocity dispersion (Section~\ref{sec:kine}).  The JAM models have four free parameters: BH mass (\Mbh), mass scaling factor ($\gamma=(M/L_{\rm dyn.})/(M/L_{\rm pop.}$)) assuming Salpeter IMF, anisotropy ($\beta_z$), and inclination angle ($i$), which relate the gravitational potential to the second velocity moments of the stellar velocity distribution.     These second velocity moment predictions are projected into the observational space to predict the $V_{\rm RMS}$ in each kinematic bin using the luminosity model (synthetic $K$--band luminosity) and kinematic PSF.    The anisotropy parameter ($\beta_z$) relates the velocity dispersion in the radial direction ($\sigma_R$) and z-direction ($\sigma_z$): $\beta_z=1-\sigma_z^2/\sigma_R^2$ assuming the velocity ellipsoid is aligned with cylindrical coordinates.  The predicted $V_{\rm RMS}$ is compared to the observations, and a $\chi^2$ is determined for each model. The number of kinematic bins for the four galaxies are shown in Column 8 of Table~\ref{kine_jeanmodel_tab}.  To find the best-fit parameters, we construct a grid of values for the four parameters ($\beta_z$, $\gamma$, \Mbh, $i$).     For each triplet parameters  of ($\beta_z$, \Mbh, $i$), we linearly scale $\gamma$ parameter to match the data in a $\chi^2$ evaluation.  We run coarse grids in our parameters to isolate the regions with acceptable models, and then run finer grids over those regions as shown in Figure~\ref{jam_plots}.

Figure~\ref{jam_plots} shows $\chi^2$ contours as a function of \Mbh~vs. $\beta_z$ (left), \Mbh~vs. $\gamma$ (middle), and \Mbh~vs. $i$ (right).   The best-fit BH masses are shown by red dots with three contours showing the 1$\sigma$, 2$\sigma$ (thin red lines), and 3$\sigma$ (thick red line) levels or $\Delta\chi^2$ = 2.30 ($\sim$68\%), 6.18 ($\sim$97\%), and 11.81 ($\sim$99.7\%) after marginalizing over the other two parameters.  The best-fit \Mbh, $\beta_z$, $\gamma$, and $i$, and minimum reduced $\chi^2$ are shown Table~\ref{kine_jeanmodel_tab}.    We quote our uncertainties below on the BH mass and other parameters at the 3$\sigma$ level. Given the restrictions JAM models place on the orbital freedom of the system (relative to e.g., Schwarzschild models), it is common to use 3$\sigma$ limits/detections in quoting BH masses (see \citet{Seth14} and Section 4.3.1 of \citet{Nguyen17} for additional discussion).    

\subsubsection{M32}\label{sssec:jamm32}  
The best-fit JAM model of M32 gives \Mbh~$=2.5^{+0.6}_{-1.0}\times10^6$\Msun, $\beta_z=-0.20^{+0.30}_{-0.35}$, $\gamma=1.08^{+0.20}_{-0.30}$, and $i=70$\deg$^{+6}_{-6}$.  We note the wide range in uncertainty in $\beta_z$ is likely due to the bulk of our kinematic data lying within the SOI of the BH.  Our BH mass,~\ml~and $i$ estimates are fully consistent with values presented in \citet{Verolme02}, \citet{Cappellari06}, and \citet{vandenBosch10} after correcting to common distance of 0.79~Mpc and correct for extinction $A_B=0.177$ mag \citep{Schlafly11} and assuming A$_I$ = A$_B$/2.22.  The consistency of the BH masses in M32 shows the good agreement between JAM and Schwarzschild modeling techniques (see Section~\ref{ssec:uncer_dynmodel}).  

\subsubsection{NGC~205}\label{sssec:jamn205} 
The JAM models of NGC~205 are compared to the kinematics derived after subtracting off bright stars in the nucleus (see Section~\ref{ssec:Kinn205}).   The data is consistent with no BH even at the 1$\sigma$ level, and we find a 3$\sigma$ upper limits would be \Mbh~$=7\times10^4$\Msun. The $\beta_z$, $\gamma$, and $i$ have a range of values of (0.17--0.34), (0.97--1.22), and (51--65)\deg, respectively.   We note that if we use the original kinematics (without star subtraction), we get a similar BH mass, inclination, anisotropy, and gamma values, with a slightly higher 3$\sigma$ BH mass upper limit of $10^5$\Msun.  Our best-fit BH mass for NGC~205 is twice the $3\sigma$ upper mass limit of \citet{Valluri05}, although we find the same \ml$_I$.  

\subsubsection{NGC~5102 $\&$ NGC~5206}\label{sssec:jamn5102and5206} 
A BH of NGC~5102 is detected with a zero-mass black hole excluded at more than the 3$\sigma$ level.  The BH of NGC~5206 is also detected here at the 3$\sigma$ level in the default R15 models, but the B03 and constant~\ml~models are consistent with zero mass at the $\gtrsim$4$\sigma$ level.   We emphasize here this is the first time these BHs are clearly detected and these masses are measured in the sub-million solar mass regime with the properties listed in Table~\ref{kine_jeanmodel_tab}.

To illustrate the best-fit JAM models with and without a BH, we plot 1D and 2D $V_{\rm RMS}$ predicted by JAM and compare them with their corresponding data in Figure~\ref{jam1d} and Figure~\ref{jam2d}, respectively.    The 1D $V_{\rm RMS}$ profiles show the bi-weighted average over circular annuli as shown in the top-right plots in Figure~\ref{kine_map}.    Overplotted on this data are the best fit models, those with the maximum and minimum BH mass in the 3$\sigma$ contours, and the best-fit no BH model.   These show that models with \Mbh~=~0\Msun~do not provide a good fit to the data except for NGC~205.  We quote both the corresponding $\chi^2_{\rm r}$ and $\chi^2_{\rm r,\; no\;BH}$ for the best-fit with and without BH in Table~\ref{kine_jeanmodel_tab}. 

The 1D profiles show that the radial data agree quite well with the models, with the only siginficant discrepancy being at the center of NGC~205.  This may be due to mass model issues in NGC~205 due both to the partially resolved kinematic data and the strong color gradient near the center.  However, we also note the dispersion measurements are very low, falling below the instrumental resolution of the telescope ($\sigma_{\rm LSF} \sim 23$~\kms).  

With the best-fit BH mass value, the SOI is calculated based on the dynamics: $r_{\rm g,\;dyn.}=GM_{\rm BH}/\sigma^2$, where $\sigma_{\star}$ is the stellar velocity dispersion of the bulge, $G$ is the gravitational constant.   The calculations of $\sigma$ will be discussed in Section~\ref{ssec:possible_BHs}.   Alternatively, we also estimate these BH SOI via our mass profile with $r_{\rm g, \;\star}$ is the radius at which the enclosed stellar mass is equal to \Mbh~\citep{Merritt13}.    The results of $r_{\rm g,\;dyn.}$ of these BH SOI are presented in Table~\ref{kine_jeanmodel_tab} in both arc-second (Column 10) and pc (Column 11) scales.    Both methods give consistent values of BH SOI $(r_{\rm g,\;dyn.}~\simeq~r_{\rm g,\;\star})$.  The $r_{\rm g,\;dyn.}$ of  $\sim$$0\farcs07$ and $\sim$$0\farcs06$ in NGC~5102 and NGC~5206 are comparable to the FWHMs of the data, suggesting that the observational signature of these BHs are just marginally resolved, while in M32, its BH's signature is easily resolved by our PSF.  For NGC~205, its $3\sigma$ upper limit SOI ($\lesssim$$0\farcs03$) is comparable to the resolution of ACS HRC imaging. 

\subsection{Sources of BH Mass Uncertainty}\label{ssec:uncer}
In this section, we discuss possible sources of uncertainties in our dynamical models for BH mass measurements.   The confidence intervals in our analysis are based on the kinematic measurement errors but do not include any systematic uncertainties in mass models, uncertainties in the PSF (particularly for NGC~5206), and uncertainties due to the JAM dynamical model that we used.  We focus on these uncertainties here, and show that for our \Mbh~estimates, these systematics are smaller than the quoted 3$\sigma$ confidence intervals.

%%%%%%%%%%%
 \subsubsection{Uncertainties due to Mass Models}\label{ssec:uncer_massmodel}
 We examine the mass model uncertainties by analyzing JAM model results from using (1)  models with constant \ml~or constructed using the B03 color--\ml~correlation, (2) models based on different \hst~filters, and (3) models fit directly to the HST imaging rather than based on the GALFIT model results.    Our tests show that our BH mass results do not strongly depend on the mass models we use for M32, NGC~205, and NGC~5102, although in some cases the best-fit $\gamma$ vary significantly because the centers of the galaxies are bluer than their surroundings.  For NGC~5206, the constant~\ml~or B03 models are basically consistent at the $3\sigma$ level with no BH, although the best-fit BH from this model is still within our 3$\sigma$ uncertainties.  We note that the color gradient in the central 1'' is quite modest, but the R15 model predicts 10\% lower mass at the center than the other two mass models.   The complete results of these mass model test are given in detail in Table~\ref{kine_jeanmodel_tabfull}. 

{\it  (a) Existing Color--\ml~Models:}  We use the R15 color--\ml~relation for our default mass models (Section~\ref{ssec:jam}).  Here we examine how much our JAM modeling results would change if we use the shallower B03 color--\ml~relation, or a constant~\ml~to construct our mass models.  Because our target galaxies are mostly bluer at their centers (except for M32), we expect that the use of a constant~\ml~will reduce the significance of any BH component due to the increased stellar mass at the center relative to our reference R15 model.  We show the best-fit JAM model parameters using these alternative mass models in Figure~\ref{jam_plots}, with the purple lines showing the B03 color--\ml~relation model, and the blue lines showing a constant \ml~model with dashed lines for $1\sigma$ and $2\sigma$ and solid lines for $3\sigma$.  As expected these alternative models typically have lower BH masses than our default model, but in M32, NGC~205, and NGC~5102, the shifts are within 1$\sigma$, and do not significantly affect our interpretation.  However, for NGC~5206 the shift is larger; the best fits are still within the R15 models 3$\sigma$ uncertainties, but no BH models are permitted within $\sim$3$\sigma$ for both the constant~\ml~and B03 based mass models.  For NGC~5102, the mass scaling ($\gamma$) values differ significantly, with the best-fit constant~\ml~model being more than 3$\sigma$ below the R15 model confidence intervals.  Overall, these results are in agreement with those determined for NGC~404 \citep{Nguyen17}.   

{\it (b) Mass Models from Other Filters:}  We recreate our mass models based on \hst~images of other filters.  Specifically, we use the R15 and B03 color--\ml~relations, as well as  the constant \ml~on the F555W images for M32, NGC~205 and NGC~5206.   We find that these mass models do not significantly change our results. 

{\it (c) Mass models from direct fits to the \hst~images:}  Our mass models are constructed based on GALFIT model fits.
While this has the strength of properly incorporating a \texttt{Tiny Tim} synthetic image of the \hst~PSF, it also makes parametric assumptions about the NSC SB profiles (that they are well-described by S\'ersic profiles).  We test for systematic errors in this approach by directly fitting the \hst~images using the {\tt mge\_fit\_sectors} \texttt{IDL} package \footnote{available at {\tt http://purl.org/cappellari/software}}, after approximating the PSF as a circularly-symmetric MGE as well.   These MGEs result in the best-fit of BH masses (and upper limits for NGC~205) that change by 7--15\%, with similar $\beta_z$, $\gamma$, and $i$.  These best-fit results with multiple existing color--\ml~relations are presented in Table~\ref{kine_jeanmodel_tabfull} with the superscript D and G stand for fitting their \hst~images and GALFIT models.

%%%%%%%%%%%
\subsubsection{Uncertainties due to Kinematic PSF in the Spectroscopic Data of NGC~5206}\label{ssec:sinfoni_psf}
As discussed in Section~\ref{ssec:psf},   the PSF of our kinematic data of NGC~5206 has a larger uncertainty than our \hst~PSFs and the other kinematic PSFs due to a significant scattered light component.   The JAM model for \Mbh, $\beta_z$, and $\gamma$ for NGC~5206 with its using the PSF derived from the half-scattered light subtraction gives \Mbh~$=4.1^{+2.6}_{-4.1}\times10^5$\Msun, $\beta_z=0.30^{+0.15}_{-0.35}$, $\gamma=1.09^{+0.15}_{-0.09}$ (assuming fixed $i=44$\deg).   So, along with the mass models, the PSF uncertainty also reduces the detection detection significance of this BH.  

%%%%%%%%%%%
  \subsubsection{Dynamical Modeling Uncertainties}\label{ssec:uncer_dynmodel}
 
 We have used JAM modelling \citep{Cappellari08} to dynamically model our galaxies.  The agreement we find between our JAM modeling and previous measurements with the the more flexible Schwarzschild modeling technique in M32 and NGC~205 (see above) is consistent with previous studies  \citep[e.g.,][]{Cappellari09, Seth14, Feldmeier-Krause17,Leung18}, and JAM models have also been shown to give BH mass estimates consistent with maser estimates \citep{Drehmer15}.
  
    The two dynamical modeling techniques differ quite significantly.  In both cases, deprojected light/mass models are used as inputs, but in the Schwarzschild models, these models are used to create a library of all possible orbits, and these orbits are then weighted to fit the mass model and full line-of-sight velocity distribution \citep[e.g.,][]{vandenBosch08}. This enables determination of the orbital structure as a function of radius (including anisotropy).  In the JAM models, only the $V_{\rm RMS}$ is fit, and the Jeans anisotropic equations used make a number of simplifying assumptions; most notably, the anisotropy is fit as the parameter $\beta_z = 1 - (\sigma_z^2/\sigma_R^2)$, and is aligned with the cylindrical coordinate system \citep{Cappellari08}. We use constant anisotropy models here.  In practice this assumption likely does not lead to significant inaccuracies, as the anisotropy has been found to be small and fairly constant in the Milky Way \citep{Schodel09}, M32 \citep{Verolme02},  NGC~404 \citep{Nguyen17}, and Cen~A \citep{Cappellari09}, as well as the more distant M60-UCD1 \citep{Seth14} and NGC~5102 and NGC~5206 in this work.

    The lack of orbital freedom in the JAM models relative to the Schwarzschild models means that the error bars are typically smaller in JAM;  \citet{Seth14} find errors in BH mass from M60-UCD1 from JAM models that are 2--3 times smaller than the errors from the Schwarzschild models, for this reason we quote conservative, 3$\sigma$ errors in this work.

%%%%%%%%%%%%%%% Table 7 of NSC properties of 4 galaxies %%%%%%%%%%%%%%%%%%
\begin{table*}[ht]
\centering
\caption{Dynamical Nuclear Star Cluster Properties}
\begin{tabular}{cccccccccc}
\hline \hline
      Object  &  Filter   & $r_{\rm eff.}$&$r_{\rm eff.}$ &$\mu_{\rm NSC}$&    $L_{\star}$          &         $M_{\star, \rm dyn.}$     &\ml$_{\rm pop., NSC}$&$\rho_{r_{\rm eff.}}$&$t_{\rm relaxation}$\\ 
                   &            & ($\arcsec$)   &       (pc)        &         (mag)         &($\times10^7$\Lsun)&($\times10^7$\Msun)&         (\Msun/\Lsun)    & (\Msun~pc$^{-3}$) &            (yrs)            \\                     
        (1)      &    (2)    &        (3)         &       (4)          &               (5)         &              (6)              &            (7)                 &                  (8)            &             (9)               &               (10)        \\[1mm]        
\hline  
M32           & F814W &$1.1\pm0.1$&$4.4\pm0.4$  &  $11.0\pm0.1$    &   $1.10\pm0.10$     &    $1.65\pm0.31$      &             1.35               &  $1.1\times10^5$   &    $6.8\times10^9$ \\
NGC~205  & F814W &$0.3\pm0.1$&$1.3\pm0.4$  &  $13.6\pm0.4$   &   $0.10\pm0.04$      &    $0.20\pm0.10$      &	            1.80  	         &  $2.4\times10^4$   &    $5.8\times10^8$ \\                                                                                    
NGC~5102& F547M &$1.6\pm0.1$&$26.3\pm1.6$&  $12.1\pm0.2$    &   $12.51\pm2.49$   &    $7.30\pm2.34$      &		    0.50               &  $2.0\times10^3$   & $3.8\times10^{10}$\\
NGC~5206& F814W &$0.5\pm0.1$&$8.1\pm1.7$  &  $14.4\pm0.2$   &   $0.73\pm0.16$     &    $1.54\pm0.51$       &             1.98               &  $3.8\times10^3$   & $2.0\times10^{10}$\\                                                                                                                                                 
\hline
\end{tabular}
\tablenotemark{}
\tablecomments{\normalsize  Columns 1: host galaxy name. Column 2: the filter used for the lumiosity modeling.  Columns 3 and 4: the effective radii of NSCs in arcseconds and parsecs.  Column 5: the total apparent magnitude of each NSC. Note that the total apparent magnitude of M32 and NGC~205's NSCs are taken from Table~\ref{seric_nsc_tab}, while those of NGC~5102 and NGC~5206's NSCs are estimated from their two inner S\'ersic components in Table~\ref{seric_nsc_tab}.   Columns 6 and 7: the luminosity (in filter from Column 2) and dynamical masses of NSCs.  Here we assume the mass-scaling factor $\gamma$ are calculated from the JAM dynamical models using the R15 color--\ml~mass models.  Column 8: the color-based population \ml~in the band given in Column 2 (see Table~\ref{seric_nsc_tab}).     Column 9: the mass densities at the effective radii of the NSCs.  Column 10:  the relaxation timescales of the nuclei evaluated at the effective radii of NSCs.}
\label{nsc_tab}
\end{table*}
%%%%%%%%%%%%%%%%%%%%%%%%%%%%%%%%%%%%%%%%%%%%%%%%%%%%%%%% 

%%%%%%%%%%%%%%%%%%%%%%%%%%%%%%%%%%%%%%%%%%%%%%%%%%%%%%%%%%%%%%%%%%%
% THE SECTION OF NUCLEAR STAR CLUSTERS!!! 
%%%%%%%%%%%%%%%%%%%%%%%%%%%%%%%%%%%%%%%%%%%%%%%%%%%%%%%%%%%%%%%%%%%
\section{Nuclear Star Clusters}\label{sec:nscs}
%%%%%%%%%%%
\subsection{Dynamical NSCs Masses}\label{ssec:dyn_nscs}
We assume the NSCs are described by the innermost (M32 and NGC~205) or the inner + middle (NGC~5102 and NGC~5206) components of our GALFIT light profile models (see Table~\ref{seric_nsc_tab}).   To estimate the masses of the NSCs, we scale the population estimates for these components from our best-fit R15 MGEs using the best-fit dynamical models.  The 1$\sigma$ errors in the $\gamma$ values are combined with the errors on the luminosities of the components to create the errors given in Table~\ref{nsc_tab}.

Our mass estimate for M32 is quite consistent with that of \citet{Graham09}; this however appears to be a coincidence as our NSC luminosity is 3$\times$ lower, while our dynamical $M/L$ is 3$\times$ higher than the stellar population value they assume; however we note that the stellar synthesis estimates of \citet{Coelho09} appear to match our color-based estimates.  In NGC~205, our mass estimate is somewhat higher than the dynamical estimate of $1.4 \times 10^6$~\Msun~in \citet{DeRijcke06}; their model assumed a constant $M/L$ with radius, and assumed a King profile for the NSC.  Our dynamical mass of NGC~5102's NSC is an order of magnitude higher than the estimate of \citet{Pritchet79} in $V$--band, in part because we use \ml$_{\rm F547M, \; dyn.}\sim\ml_{V}=0.55$ (\Msun/\Lsun) \citep{Mitzkus17} or $\ml_{\rm F547M, \; dyn.}=\ml_{\rm F547M,\; pop.}\times\gamma_{\rm F547M}$ with \ml$_{\rm F547M, \; pop.}\sim0.5$ (\Msun/\Lsun) estimated from R15 color--\ml~relation within 0$\farcs$5, while \citet{Pritchet79} adopt a very low \ml$_V=0.11$ (\Msun/\Lsun).

%%%%%%%%%%%%%%%%%%%%%%%%%%%%%% Figure 8 of Mass vs. size for NSCs %%%%%%%%%%%%%%%%%%%%%%
\begin{figure}[h]
     \centering
      \epsscale{1.15}
        \plotone{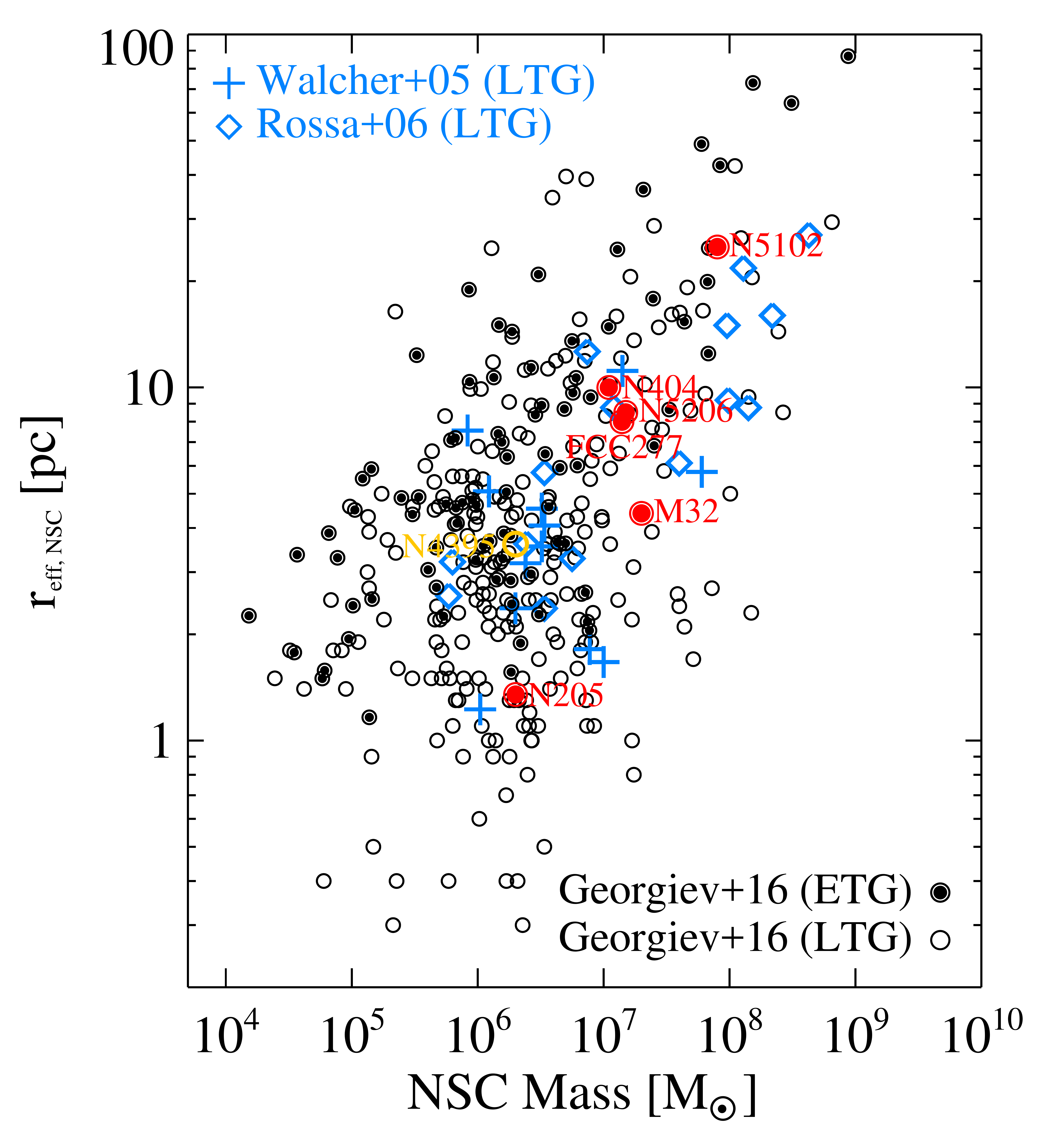}
            \caption{\normalsize   Mass vs. size for NSCs.  Our dynamical mass determinations plus that of NGC~404 \citet{Nguyen17} are shown in red (dots + outer circles); these are the only dynamically determined NSC masses in early-type galaxies (ETGs).  These are compared to photometric estimates from \citet{Georgiev16} plotted both in black circles with central dots for ETGs and in black open circles for late-type galaxies (LTGs).   LTG NSCs with the dynamical \ml~estimates from \citet{Walcher05} (late-type spirals) and the spectroscopic \ml~estimates \citet{Rossa06} (Sa-Sbc) are plotted in blue pluses and diamonds, respectively.  We also add the dynamical \ml~estimate for LTG NSC of NGC~4395 \citep{denBrok15} in yellow open circle and ETG NSC of FCC~277 \citep{Lyubenova13} in the same notation of this work NSCs.}
\label{reff_mnsc}
\end{figure}   
%%%%%%%%%%%%%%%%%%%%%%%%%%%%%%%%%%%%%%%%%%%%%%%%%%%%%%%%%%%%%%%%%%%%%%%%%%

To calculate the effective radii ($r_{\rm eff}$) of the multiple components NSCs in NGC~5102 and NGC~5206, we integrate the two components together to get an effective radius of $0\farcs5$ (8 pc) for NGC~5206 and $1\farcs6$ (26 pc) for NGC~5102.  In Figure~\ref{reff_mnsc}, we compare the sizes and dynamical masses of our ETG NSCs to previous measurements in other galaxies.  Most of the previous measurements are photometric mass estimates taken from \citet{Georgiev16}.  The dynamical \ml~estimates are taken from \citet{Walcher05} for LTGs,  \citet{Lyubenova13} for FCC~277,  \citet{denBrok15} for NGC~4395,  \citet{Nguyen17} for NGC~404, and the spectroscopic \ml~estimates are taken from \citet{Rossa06} for LTGs; we note here that our data are the first sample of dynamical NSC masses measured in ETGs.  We find that all four of our NSCs fall within the existing distribution in the size-mass plane, with M32 and NGC~205 being on the compact side of the distribution.

%%%%%%%%%%%
\subsection{Uncertainties on NSC Mass Estimates}\label{ssec:nsc_errors}

For the quoted errors in the NSC masses shown in Table~\ref{nsc_tab}, we take the 3$\sigma$ errors on the mass scaling factors, and combine these in quadrature with uncertainties in the total luminosity of the NSCs to provide a conservative error estimate.  The uncertainty in the total luminosity of the NSC (which comes from how much of the center is considered to be part of the NSC) ends up being the dominant error.   The total fractional errors range between 11\% and 21\%.  In addition to these errors, we consider the systematic error on the mass estimates from using different color--\ml~relations.  We find that the maximum differences are $<$12\%, given that these are below the quoted uncertainties in all cases, we don't quote separate systematic errors.

 %%%%%%%%%%%     
\subsection{Structural Complexity and NSC formation}\label{ssec:nscform}

The NSCs of NGC~5102 and NGC~5206 have complex morphologies, made up of multiple components, while NGC~205 also shows clear color gradients within the nucleus.  A similarly complex structure has also been seen in the NSC of nearby ETG NGC~404, where a central light excess in the central 0$\farcs$2 (3~pc) appears to be counter-rotating relative to the rest of the NSC \citep{Seth10a} and has an age of $\sim$1~Gyr \citep{Nguyen17}.  The sizes of NSCs in spirals also change with wavelength, with bluer bands having smaller sizes and being more flattened, suggesting young populations are concentrated near the center \citep{Seth06, Georgiev14, Carson15}.   In NGC~5102, the central component dominates the light within the central 0$\farcs$2 (3~pc), while in NGC~5206, the central component dominates the light out to 0$\farcs$5 (8~pc).  However, it is challenging to determine if these in fact are distinct components.  In both NSCs the color gets bluer near the center, suggesting a morphology similar to that seen in the late-type NSCs.  In NGC~5102, the outer component is somewhat more flattened, and has an elevated $V/\sigma$ value suggesting that it may be more strongly rotating than the inner component.  In NGC~5206, both components have similar flattening and position angles, and no clear rotation kinematic differences are seen.

As NSCs have complex stellar populations, it is not surprising that their morphology may also be complex.  These multiple components are likely linked to their formation history.  In particular, the central bluer/younger components seen in 3/4 of our clusters strongly argue for {\em in situ} star formation, as in-spiraling clusters are expect to be tidally disrupted in the outskirts of any pre-existing cluster  \citep[e.g.,][]{Antonini13}.    The gas required for in-situ star formation could be funneled into the nucleus by dynamical resonances in the galaxy disk \citep[e.g.,][]{Schinnerer03}, or be produced within the NSC by stellar winds \citep{Bailey80} or collisions \citep{Leigh16}.   The large $r_{\rm eff} = 32$~pc, high-mass fraction (1\% of total galaxy mass) outer component in NGC~5102 may also be akin to the extra-light components expected to be formed from gas inflows during mergers \citep{Hopkins09a}.  Regardless of the source of the gas, it is clear that NSCs in ETGs frequently host younger stellar populations within their centers, similar to late-type NSCs.   

%%%%%%%%%%%%%%%%%%%%%%%%%%%%%%%%%%%%%%%%%%%%%%%%%%%%%%%%%%%%%%%%%%%%%
% THE SECTION OF DISCUSSIONS!!! 
%%%%%%%%%%%%%%%%%%%%%%%%%%%%%%%%%%%%%%%%%%%%%%%%%%%%%%%%%%%%%%%%%%%%%
\section{Discussion}\label{sec:disc}

%%%%%%%%%%% Figure 10 of BH and hosts scaling relationships including this work%%%%%%%%%%%%
\begin{figure*}[ht]
     \centering
      \epsscale{1.2}
          \plotone{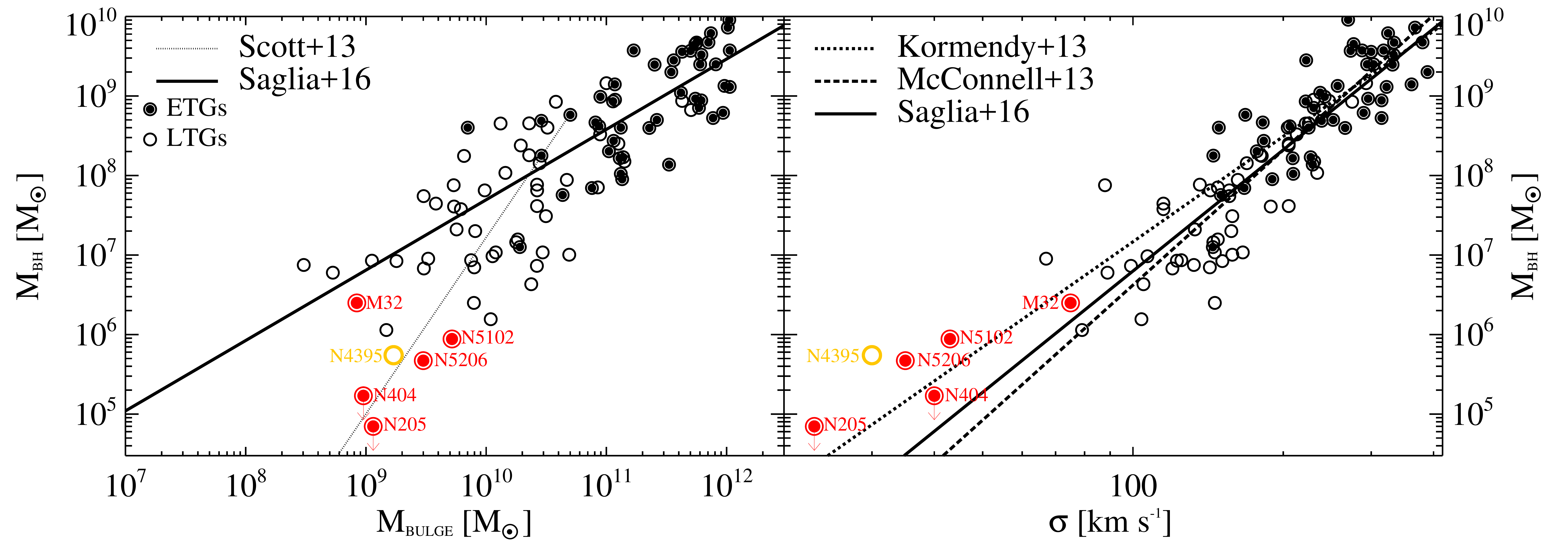}
          \caption[SCP06C1]{\normalsize  Our four low-mass early-type galaxies  in context of the \Mbh--\Mbulge~(left) and \Mbh--$\sigma$ (right) scaling relations.    The data of both early-type galaxies (ETGs, black dots with open circles) and late-type galaxies (black open circles) are taken from the compilation of \citet{Saglia16}.  The dotted-, dashed-, long dashed-, and solid-lines indicate their linear best-fit in $\log$ scale of the relations from \citet{Scott13, McConnell13, Kormendy13}, and \citet{Saglia16} for early-type galaxies (LTGs), respectively.     The BH masses of M32, NGC~5102, and NGC~5206 and 3$\sigma$ upper limit mass of NGC~205 with the downward arrow are plotted in red dots + open circles as in the legend.   We also add BHs with dynamical masses below $10^6$\Msun, including the dwarf AGN late-type galaxy NGC~4395 \citep[yellow open circle;][]{denBrok15} and early-type galaxy NGC~404 \citep{Nguyen17}.}
\label{scale_rela_bh}
\end{figure*}
%%%%%%%%%%%%%%%%%%%%%%%%%%%%%%%%%%%%%%%%%%%%%%%%%%%%%%%%%%%%%%%%%%%%
%%%%%%%%%%% Figure 11 of BH and hosts scaling relationships including this work%%%%%%%%%%%%
\begin{figure*}[ht]
     \centering
      \epsscale{1.2}
          \plotone{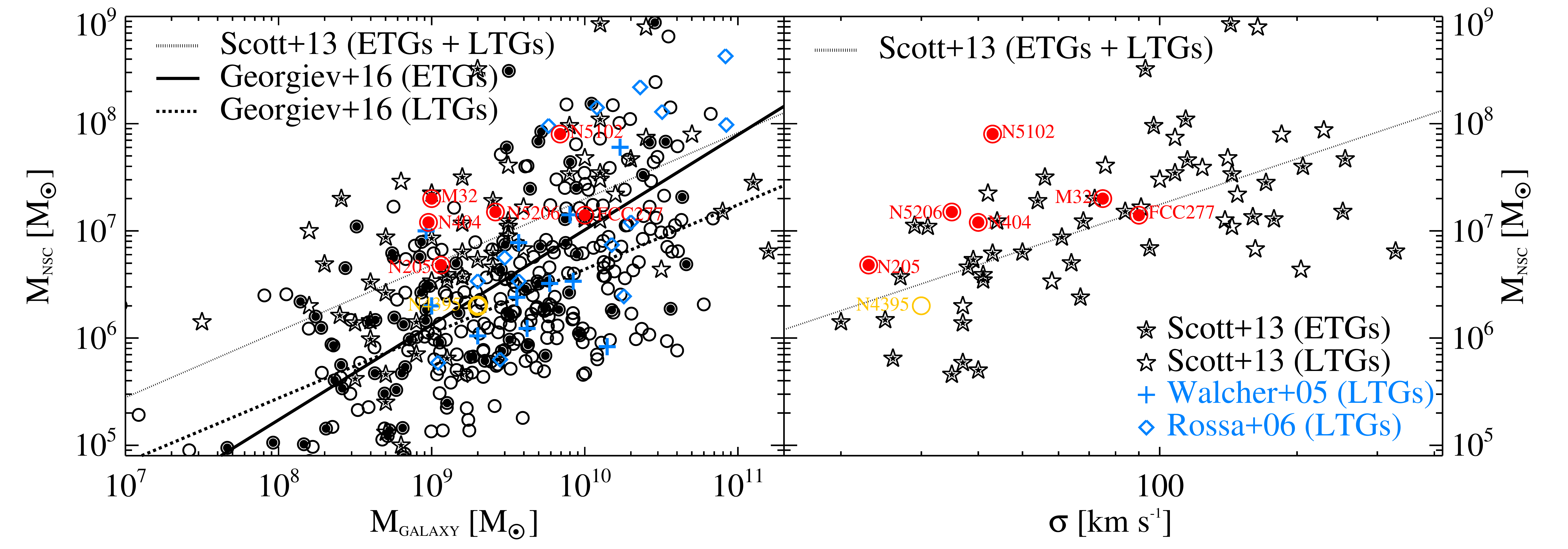}
          \caption[SCP06C1]{\normalsize  Our four galaxies in context of the \Mnsc--galaxy mass (left) and \Mnsc--$\sigma$ (right) scaling relations.     The data are taken from early-type galaxies (double stars) and late-type galaxies (open stars) \citet{Scott13},  and \citet{Georgiev16} with early-type galaxies (ETGs) are plotted in black dots with circles, while late-type galaxies (LTGs) are plotted in open circles; and their corresponding linear best-fit in $\log$ scale.    We note that the \citet{Scott13} linear relation has been fit to the sample of both early-type (solid line) galaxies and late-type galaxies (dashed line).  The M32, NGC~205, NGC~5102, and NGC~5206 NSCs are shown in color dots with circles as in the legend.   We also add two more NSCs including of late-type galaxy NGC~4395 \citep{denBrok15} and early-type galaxy NGC~404 \citep{Nguyen17}.   NSCs have dynamical or spectroscopic \ml~estimates from \citet{Walcher05} (late-type spirals) and \citet{Rossa06} (Sa-Sbc) are plotted in gray pluses and diamonds, respectively.   We also add the dynamical estimate of the  ETG NSC of FCC~277 \citep{Lyubenova13} in the same notation of this work NSCs.  The total stellar of FCC~277 was estimated assuming its $B$--band absolute magnitude of $-17.78$ (\href{HypeLEDA}{http://leda.univ-lyon1.fr/search.html}) and \ml$_B\sim5.0$ (\Msun/\Lsun) at the highest $3\sigma$ of its Schwarzschild dynamical estimate in \hst/ACS z--band \citep{Lyubenova13}. } 
\label{scale_rela_nsc}
\end{figure*}
%%%%%%%%%%%%%%%%%%%%%%%%%%%%%%%%%%%%%%%%%%%%%%%%%%%%%%%%%%%%%%%%%%%%%

%%%%%%%%%%%%
\subsection{BH demographics}\label{ssec:BH_demographic}

We have dynamically constrained the BH masses of M32, NGC~205, NGC~5102, and NGC~5206.  While we find no evidence of a BH in NGC~205 (our 3$\sigma$ upper limit is $7\times10^4$\Msun), we detect BHs in the other three systems.  Two of these best-fit BHs are below 10$^6$\Msun, in NGC~5206 ($4.7\times10^5$\Msun)~and in NGC~5102 ($8.8\times10^5$\Msun).  This doubles the previous sample of BHs with dynamical constraints placing them below 10$^6$~\Msun; the previous two were found in late-type spiral NGC~4395 \citep{denBrok15} and the accreting BH in NGC~404 \citep{Nguyen17}.  This adds significantly to the evidence that $\lesssim$$10^6$\Msun~BHs do inhabit the nuclei of low-mass galaxies \citep[see also][]{Reines15}. 

Combining our results with the accretion evidence for a BH in NGC~404 \citep{Nguyen17}, we can make a measurement of the fraction of the nearest ETGs that host BHs.  These five galaxies have total stellar masses between $5\times10^8$\Msun--$10\times10^9$\Msun.  This sample represents a complete, volume limited sample of ETGs in this mass range, and thus forms a small but unbiased sample of ETGs (Section~\ref{sec:hosts}).  Most of the galaxies are satellite galaxies, with only NGC~404 being isolated, thus these represent only group and isolated environments.  With four of five of these galaxies now having strong evidence for central BHs, we measure an occupation fraction of 80\% over this stellar mass range.

We can directly compare this to the occupation fraction estimates from the AMUSE survey of X-ray nuclei in ETGs \citep{Gallo08, Miller15}.  Of the 57 galaxies observed in the mass range of our ETGs, only 9 (16\%) have detected X-rays sources, however, modeling of the X-ray luminosity function suggests occupation fraction of 30--100\% for these galaxies \citep{Miller15}.  Thus both dynamical and accretion evidences points to high mass occupation fraction for ETGs down to $\sim$10$^9$~M$_\odot$.  

The high occupation fraction we find here can be used to constrain the seed formation mechanism of BHs in the early Universe.  Our result appears to be more consistent with the scenario where BH seeds are formed from the remnants of the first generation of massive stars \citep[e.g., Population III stars (light seeds),][]{Volonteri08, Volonteri10, Greene12, Reines16}, rather than heavier seeds.   More specifically, the 80\% observed occupation fraction  is consistent with the semi-analytic model predictions at $z=0$ for light seeds as a function of velocity dispersion \citep{Volonteri08, vanWassenhove10}.  A more recent semi-analytic prediction predicts a significantly lower BH occupation of $\sim$20\% in our mass range for the low-mass seed scenario \citet{Antonini15a}.  This study also breaks down the prediction by galaxy type, with even lower occupation fraction expected in similar mass LTGs.  We note that our measurements apply only to ETGs, which make up a minority of galaxies at the stellar masses probed \citep[e.g.,][]{Blanton05}.

Our BH masses are also consistent with their formation from runaway tidal capture of NSC stars, a scenario outlined by \citet{Stone17}.   \citet{Stone17} find that this runaway tidal capture processes requires dense NSCs that are typically found only in galaxies with $\sigma \gtrsim 35$~\kms.   For sufficiently dense NSCs (in higher dispersion galaxies), the runaway tidal encounters supplies mass for the BH seed to grow, and this growth is larger for higher dispersion galaxies.  Over a Hubble time, this BH growth saturates at a specific mass.  Using their equation~37, we find saturation masses of  $\sim$$6.5\times10^5$\Msun, $\sim$$5\times10^5$\Msun, $\sim$$1\times10^6$\Msun~for NGC~5102, NGC~5206, and M32, respectively using the velocity dispersions are estimated in Section~\ref{ssec:possible_BHs}.  These BH masses are within a factor of $\sim$2 of the BH masses we dynamically measure in Section~\ref{ssec:jam}.   NGC~205 and NGC~404 both fall below the threshold dispersion where runaway tidal growth is expected, and our upper mass limits in these galaxies are therefore also consistent with this scenario.

Additional constraints on the occupation fraction in a wider range of galaxy types are possible in the local volume, and in future work we plan to provide additional constraints using existing data on LTGs with both stellar kinematics using Gemini/NIFS or VLT/SINFONI and upcoming gas kinematic data from ALMA.

%%%%%%%%%%%%
\subsection{BH Mass Scaling Relations}\label{ssec:possible_BHs}
We examine our best-fit BH masses in the context of scaling relations including \Mbh--\Mbulge~and \Mbh--$\sigma$ \citep{Kormendy13, McConnell13, Saglia16} for ETGs.    Since our dynamical constraints suggest $\gamma\sim1$, and we do not have direct dynamical measurements of the bulge masses, we assume population based masses for the bulge components here for all four galaxies; these are listed in Column 12 of Table~\ref{seric_nsc_tab}.

We note that we assume the galactic bulges are comprised of all two (NGC~205 and M32) and three (NGC~5102 and NGC~5206) S\'ersic components.  \citet{Kormendy13}, \citet{McConnell13}, and \citet{Saglia16} report the bulge mass and luminosity of M32 are $\sim$$4.5\times10^8$\Msun~and $\sim$$4\times10^8$\Lsun~($M_V=-16.64$), while our estimates of mass and luminosity from F814W are \Mbulge~$\sim8.2\times10^8$\Msun~and \Lbulge~$\sim4.5\times10^8$\Lsun;  these are consistent with \citet{Graham09, Kormendy09} estimated in $I$--band.      For NGC~205, \citet{McConnachie12} reports a total luminosity of $M_V=-16.5$ ($3\times10^8$\Lsun) and \citet{Angus16} simulation gives a mass of $4.0\times10^8$\Msun.  We find a consistent luminosity but higher mass for the bulge of $6.5\times10^8$\Msun~from F814W \hst~data.   

Figure~\ref{scale_rela_bh} shows the best-fit BH mass of M32 is within the scatter of the \Mbh--\Mbulge~relations of \citet{Kormendy13}, \citet{McConnell13}, and \citet{Saglia16}.   However, the best-fit BHs masses of NGC~205, NGC~5102, and NGC~5206 clearly fall below these \Mbh--\Mbulge~relations by 1-2 orders of magnitude.  Combined with other recent work on BH mass measurements in low-mass systems, e.g.,  NGC~404 \citep{Nguyen17}, NGC~4395 \citep{denBrok15}, RGG118 \citep{Baldassare15}, Pox52 \citep{Barth04, Thornton08}, an increasing number of galaxies appear to fall below the BH mass vs. bulge mass relationship extrapolated from massive ETGs to the low-mass regime.  This suggests that the \Mbh--\Mbulge~relationship may steepen as suggested by \citet{Scott13} for low-mass systems with bulge mass $<$$2\times10^{10}$\Msun.  Alternatively, the scaling relations may just break down at these masses, with significant scatter to lower masses as also seen in the megamaser BH measurements in spiral galaxies \citep{Laesker16, Greene16}.  We note that this increased scatter may occur at bulge masses where the occupation fraction also drops below unity.  
 
Next, we explore these BHs in the context of the \Mbh--$\sigma$ scaling correlation.  While typically the dispersion used is that estimated within the effective radius, we do not have this information for all of our galaxies.  For M32 and NGC~205, the $\sigma_e$ have been measured by \citet{Ho09} who measure the dispersions of $72.1\pm1.9$~\kms and $23.3\pm3.7$~\kms~in a $2\arcsec\times4\arcsec$ aperture, respectively. To estimate the central velocity dispersions of NGC~5102 and NGC~5206, we co-add all the spectra in the FOV of 1$\farcs$5 and fit the resulting single spectrum with pPXF for velocity dispersions. This gives dispersions of $42.3\pm1.1$~\kms~for NGC~5102 and $35.4\pm1.0$ \kms~for NGC~5206.

%%%%%%%%%%%%%%%%%
\subsection{Possibility of Collections of Dark Remnants}\label{ssec:darkremnants}
Recent work has suggested that mass segregation of stellar mass BHs can occur on timescales well below the relaxation time \citep{Bianchini17}, significantly enhancing the central \ml~of globular clusters.  An enhancement of \ml~is in fact seen in the globular cluster M15 \citep{denBrok14b}.     This leads to the possibility that a significant number of stellar mass BHs could collect at the center of NSCs, leading to a false detection of a massive BH.  We note that the current simulations are of much lower mass systems than the massive NSCs we discuss here, and that significant uncertainties exist on how many BHs would segregate and be retained during this process.  Nonetheless, we can get a sense of the maximum dark mass that can plausibly be accumulated by stellar remnants.  For a Solar metallicity population with a standard Kroupa IMF, $\sim$4\% of the mass of the cluster will turn into stellar mass BHs \citep{Shanahan15}.  This represents a maximum mass fraction of the NSCs that could be concentrated at the center with the retention fraction of BHs likely being much lower than this.  The fraction of the NSC mass found in the BH in M32, NGC~205, NGC~5102, and NGC~5206 are 13\%, $<$3.5\%, 1.1\%, and 3.3\% of their NSC masses, respectively. Therefore, based on the mass fractions, apart from M32, a collection of dark remnants is a possibility.

We can also evaluate stellar relaxation times ($t_{\rm relaxation}$) of these nuclei at the effective radii of their NSCs using the following equation \citep{Bahcall77, Valluri05}:
\begin{equation}
t_{\rm relaxation}\simeq(1.4\times10^8)\sigma_{20}^3\rho_5^{-1}(\ln \Lambda_{10})^{-1}~({\rm yrs})
\end{equation}
where $\sigma_{20}=\sigma_{\star}/20$~\kms, $\rho_5=(\rho_{\rm g,\; dyn.})/10^5$~\Msun~pc$^{-3}$, $\ln \Lambda_{10}=\ln \Lambda/10$, and $\ln \Lambda=10$.  $\sigma_{\star}$ is the stellar dispersion estimated in Section~\ref{ssec:possible_BHs}, and the mass densities $\rho_{\rm g,\; dyn}$ are estimated at the effective radii of NSCs.   These mass densities and relaxation times are provided in Table~\ref{nsc_tab}.   Our estimated relaxation times for the BH detections are long (6.8-38~Gyr), while in NGC~205 (the one source without a BH detection), the relaxation time is quite short (0.58~Gyr).  The fact that we detect no BH in this shortest relaxation time system \citep[as well as the tight upper limit on the mass of the M33 BH][]{Gebhardt01} argues against the detected BHs  being collections of remnants.  However, more realistic simulations are required to evaluate whether significant fractions of the NSC BHs can really be mass segregated and retained.  %

%%%%%%%%%%%%
\subsection{NSC Masses and Scaling Relations}\label{ssec:scaling_nscs}
Our dynamical mass measurements of the four NSCs in this work along with the measurement of NGC~404 NSC \citep{Seth10a, Nguyen17} and FCC~277 \citep{Lyubenova13} are the first six dynamical mass measurements of NSCs in ETGs.  These NSC masses are plotted against the mass and dispersion of galaxies in Figure~\ref{scale_rela_nsc}.

Comparing our dynamical NSC masses to the galaxy stellar masses and dispersions, we find that they are within the scatter of the previous (mostly photometric) NSC mass measurements.  However, almost all of the dynamical measurements lie on or above previous relations, with NGC~5102 the highest mass outlier. The closest previously published relations are those presented in \citep{Scott13}.  The \citep{Georgiev16} NSC--galaxy mass relation for ETGs lies below all 6 dynamical measurements.

The ratios between the NSCs masses and their host bulges/galaxies masses are  $\sim$1.7\%, 0.2\%, 1.1\%, 0.7\%, and $<$1.5\% for M32, NGC~205, NGC~5102, NGC~5206, and NGC~404 \citep{Nguyen17} respectively.   Apart from NGC~205, these NSCs are more massive than the typical $\sim$0.3\% mass fraction for NSCs seen in ETGs with similar stellar mass in the Virgo cluster \citep{Cote06, Spengler17}.  A trend towards higher mass NSCs in lower mass galaxies was suggested by \citet{Graham09} and is verified for ETGs in recent work \citep{Spengler17}; based on the \citet{Graham09} relation a mass fraction of $\sim$1\% is expected for our sample galaxies.  The high masses and mass fractions inferred here might be a sign that the relation of NSC to galaxy mass changes outside cluster environments, as might be expected based on the increased availability of gas to accrete to the present day.  

%%%%%%%%%%%
\subsection{Relative Formation Between NSCs and BHs}\label{ssec:nscbh_rela}
\begin{figure*}[ht]  
      \epsscale{1.16}
            \plottwo{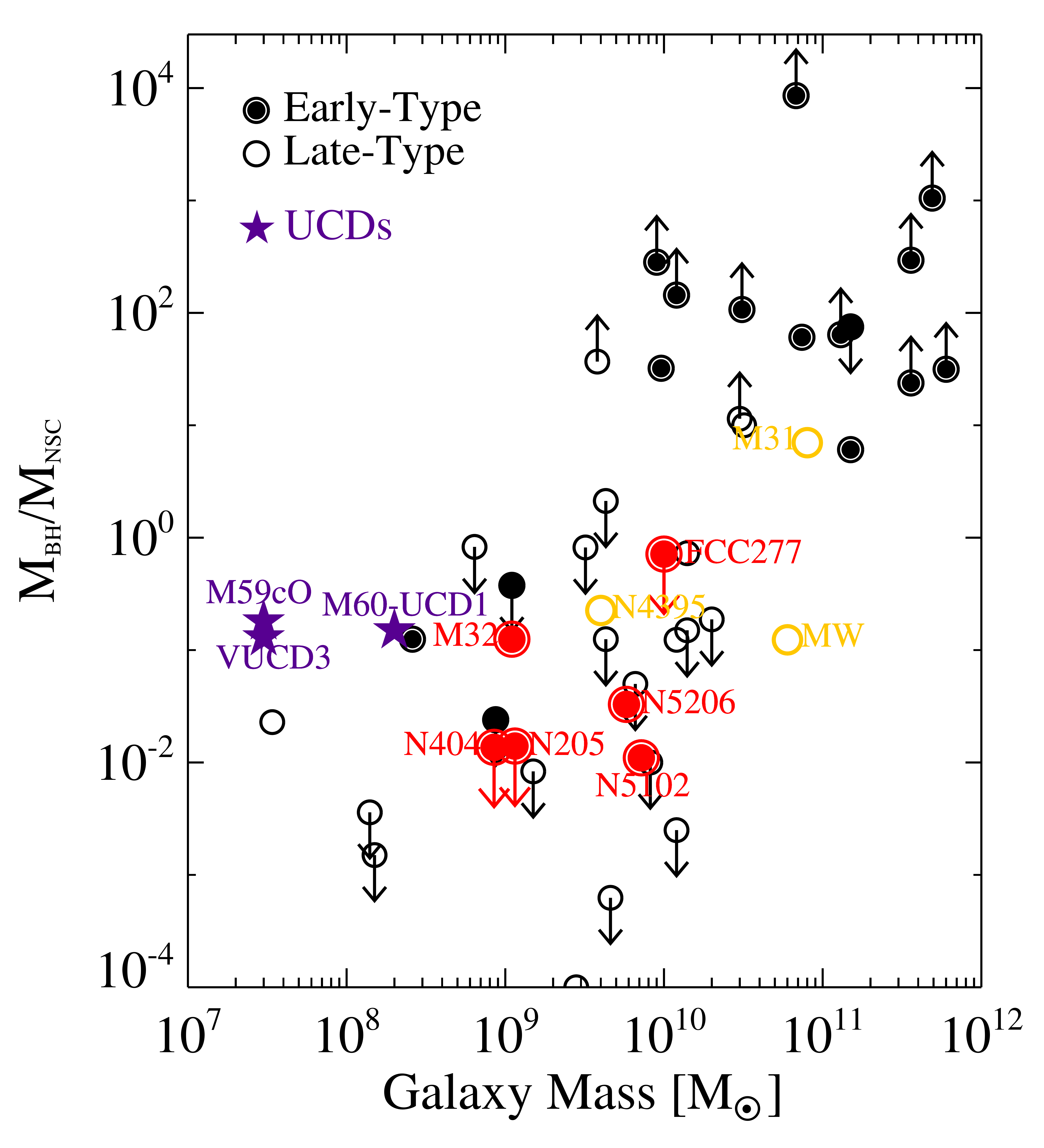}{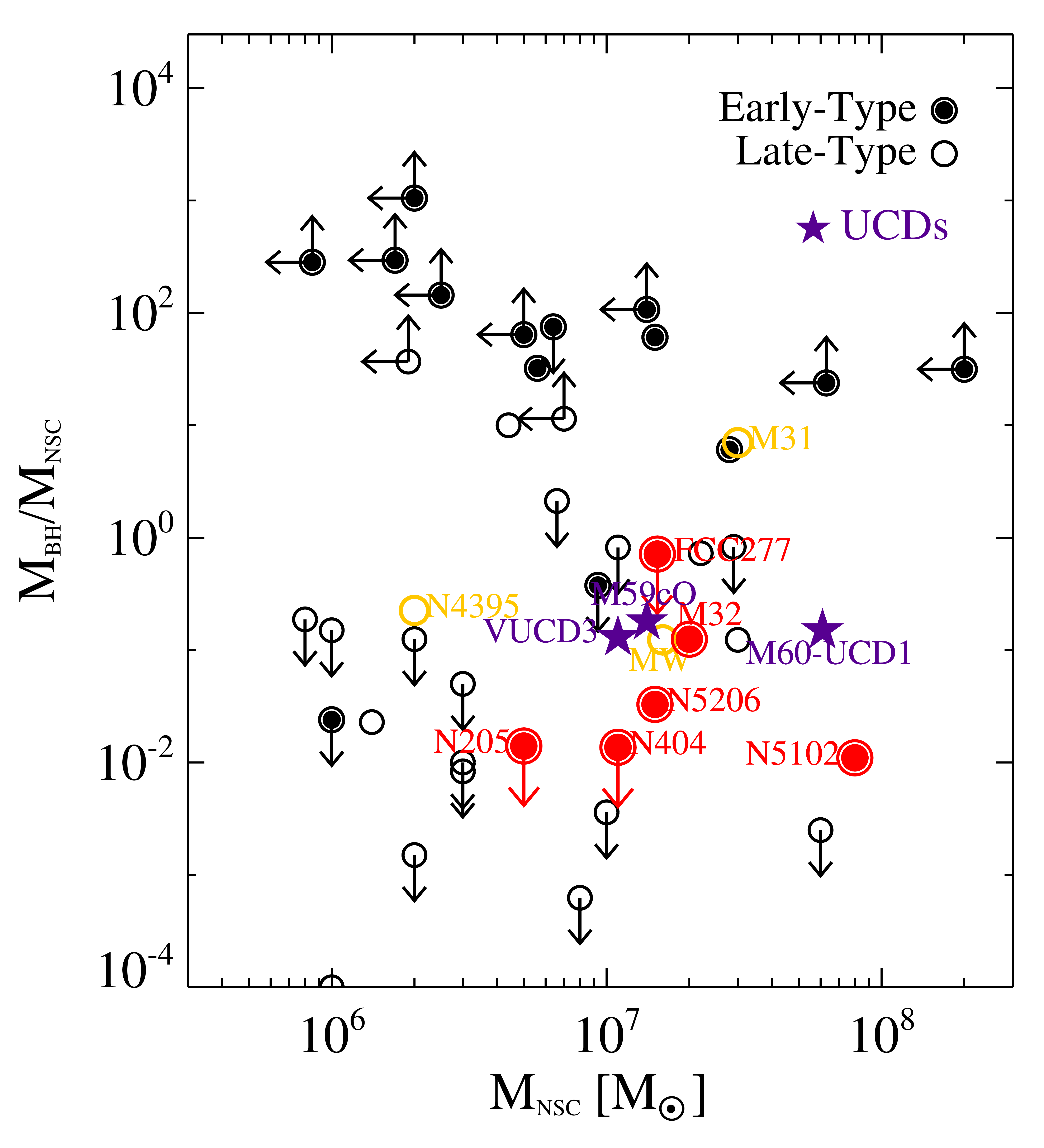}
            \caption{\normalsize   Ratio of the BH and NSC mass against their host-galaxy stellar mass (left) and their NSC dynamical masses based on dynamical measurements of \ml$_{\rm dyn.}$ (right).  Original data with dynamical measurements are shown with early- (black dots), late-type (gray dots) galaxies, and current BH mass measurements of UCDs (black stars) are taken from \citet{Neumayer12},  \citet{Graham09}, \citet[FCC~277;][]{Lyubenova13}, \citet[M60-UCD1;][]{Seth14}, and \citet[VUCD3 and M59cO;][]{Ahn17}.  The upward/downward arrows are used for galaxies with lower/upper limits of BH and NSC masses.  We note that the for the UCD NSC masses, we plot only the mass of their inner components.}
\label{mbh_mnsc}
\end{figure*}

As discussed in the introduction, the relationship between NSCs and BHs is not yet clear.  Our galaxies add significantly to the sample of galaxies where we have mass measurements for both the BH and NSC.  Our three BH detections have a \Mbh/\Mnsc~ratio between $10^{-2}$ and $0.3$.  We plot these against the  galaxy mass and NSC mass in Figure~\ref{mbh_mnsc}. The \Mbh/\Mnsc~ratio in our galaxies is consistent with galaxies like the MW \citep[e.g.,][]{Schodel14}, and previous measurements of Ultracompact Dwarfs (UCDs) \citep{Seth14, Ahn17}, where just the inner component of the UCD is assumed to be the remnant NSC.  However, the lack of NSCs in massive ETGs clearly separates these from our sample \citep{Neumayer12}.  On the other hand, for spiral galaxies with upper limits on the BHs \citep{Neumayer12}, our NGC~205 limit joins previous measurements constraining the BH mass to be at most $10^{-2}$ to $10^{-3}$ that of the NSC.  Even amongst the sample where both BH and NSC are found, there is a scatter of $\sim$3 orders of magnitude at similar NSC mass.  This wide range of \Mbh/\Mnsc~ratios suggests that NSC and BH formation must not always be tightly linked.

%%%%%%%%%%%%%%%%%%%%%%%%%%%%%%%% Figure 11 of Mass vs. size for NSCs %%%%%%%%%%%%%%%%%%%%%%
%%%%%%%%%%%%%%%%%%%%%%%%%%%%%%%%%%%%%%%%%%%%%%%%%%%%%%%%%%%%%%%%%%%%%%

%%%%%%%%%%%
\subsection{Relation of NSCs to UCDs}\label{ssec:nsc_ucd}

The infall and ongoing stripping of the Sagittarius dwarf spheroidal galaxy shows that nuclei similar to those we sample here can be stripped of their host galaxies \citep[e.g.,][]{Monaco05}.  Our NSCs clearly overlap with the radii and mass distribution of UCDs \citep{Norris14}, and some of these UCDs may be stripped galaxy nuclei \citep[e.g.,][]{Pfeffer14}.  UCDs typically have higher than expected integrated dispersion \citep[e.g.,][]{Hasegan05}.  Specifically, virial mass estimates based on these dispersions give  $M/L$s higher than expected from their stellar populations -- these can be interpreted as indirect evidence for massive BHs \citep{Mieske13}, as would be expected if UCDs were tidally stripped NSCs.  The presence of these BHs with mass fractions above 10\% has been verified in a few massive UCDs ($>$$10^7$\Msun) using resolved kinematic measurements  \citep{Seth14, Ahn17}.  We can examine the apparent increase in $M/L$ of our galaxies if we were to ignore the BHs by looking at the no BH model $M/L$s; these are inflated by 4--10.0\% in our three galaxies with detected BHs; this is smaller than the effect observed in known UCDs due to the smaller mass fractions of the BHs (1--13\%) in these systems.  We note that the UCDs are often observed to have multiple components, thus the NSCs may make up only a fraction their total mass, suggesting even lower BH mass fractions if these galaxies were to be stripped.

%%%%%%%%%%%%%%%%%%%%%%%%%%%%%%%%%%%%%%%%%%%%%%%%%%%%%%%%%%%%%%%%%%%%%%%
% THE SECTION OF CONCLUSIONS AND FUTURE PROSPECTS!!! 
%%%%%%%%%%%%%%%%%%%%%%%%%%%%%%%%%%%%%%%%%%%%%%%%%%%%%%%%%%%%%%%%%%%%%%%
\section{Conclusions $\&$ Future Prospects}\label{sec:concl_future}
%%%%%%%%%%%%%%%%
\subsection{Conclusions}\label{ssec:concl}

We have examined the nuclear morphology and kinematics of a sample of four ETGs with stellar masses of $10^{9}-10^{10}$\Msun~using adaptive optics Gemini/NIFS and VLT/SINFONI spectroscopic data and \hst~imaging. We use \hst~data to estimate the sizes, masses, and luminosities of their NSCs.  Using dynamical modeling of the spectroscopic data, we constrain the mass of their central BHs and NSCs. Our primary findings are as follows.

 \begin{enumerate}
 \setlength\itemsep{0em}

\item Our dynamical models yield significant detection of central BHs in M32, NGC~5102, and NGC~5206 and an upper limit in NGC~205.  The NGC~205 upper limit and M32 BH masses are consistent with previous measurements.  

\item We present the first dynamical measurements of the central BH masses in NGC~5102 and NGC~5206.  Both BHs have masses below 1 million \Msun, thus doubling the number of dynamical measurements of sub-million solar-mass BHs \citep{denBrok15, Nguyen17}.

\item Including our recent work on the BH in NGC~404 \citep{Nguyen17}, we find that the fraction of $10^{9}-10^{10}$\Msun~($\sigma_{\star}=$20~\kms~to 70~\kms) ETGs that host central BHs is at least 80\% (4/5 with detected BHs).  These galaxies represent a volume limited sample of ETGs in this mass range. This rather high occupation fraction suggests that the early universe produced an abundance of BH seeds, favoring a scenario where BH seeds formed from the population III stars \citep[e.g.,][]{Volonteri10}.  Our BH masses are also consistent with growth by runaway tidal encounters \citep{Stone17}.

\item The effective radii of the NSCs are $\sim$2--26~pc.  The NSC SB profiles of NGC~205 and M32 are both fit well by a single S\'ersic function, but the NSCs of NGC~5102 and NGC~5206 appear to have more complicated nuclear structures requiring two S\'ersic components.  

\item The high-quality ($\sim$0$\farcs$1 resolution) kinematics of M32 and NGC~5102 show strong rotation ($V/\sigma \gtrsim 0.7$), while NGC~205 and NGC~5206 have less significant rotation ($V/\sigma \sim 0.3$).  

\item We measure the masses of the NSCs in all four galaxies dynamically, some of the first dynamical mass estimates of ETG NSCs.  Their dynamical masses range from $0.2\times10^7$\Msun~to $7.3\times10^7$\Msun.  These NSC masses are higher than the typical ETGs of similar galaxy mass in the Virgo and Fornax clusters, suggesting a possible environmental dependence on the NSC mass.  

\item The BHs and NSCs in these galaxies appear to follow the \Mbh--$\sigma$ and \Mnsc--$\sigma$ relations.   Their BHs fall below the bulge mass/total stellar mass--BH mass relations, while their NSCs are located above similar relations.

\end{enumerate}

%%%%%%%%%%%%%%%%
\subsection{Future Prospects}\label{ssec:future}
Our dynamical modeling using high resolution stellar kinematics from NIFS and SINFONI spectroscopy confirm the existence of an BH in M32 and put an upper limit on the NGC~205's BH mass.  We also measure the masses of BHs in NGC~5102 and NGC~5206 for the first time.    However, we model these BH masses assuming a constant \ml~or the B03 and R15 color--\ml~relations.     Although the assumption of a constant \ml~is good for M32, the nuclei of NGC~205, NGC~5102, and NGC~5206 show the clear spatial population variations and hence their nuclear \ml s.   We will model these \ml~variations using new STIS data obtained this year using the techniques we developed in \citet{Nguyen17}.    This data will enable stellar population synthesis modeling on the same 0$\farcs$05 scale as our kinematic observations.    With this information we can significantly reduce the uncertainties in the stellar mass profiles and the resulting BH mass estimates.     Moreover, our upcoming ALMA data will also be able to allow us to measure the central BH masses using cold gas kinematic models of the circumnuclear gas disk (CND) within the nuclei of a variety Hubble type, nearby, and low-mass galaxy sample.  This larger and high quality data sample will provide a more precise, complete, and unbiased occupation fraction dynamically.   

%%%%%%%%%%%%%%%%%%%%%%%%%%%%%%%%%%%%%%%%%%%%%%%%%%%%%%%%%%%%%%%%%%%%%
% THE SECTION OF ACKNOWLEDGMENTS!!! 
%%%%%%%%%%%%%%%%%%%%%%%%%%%%%%%%%%%%%%%%%%%%%%%%%%%%%%%%%%%%%%%%%%%%%
\acknowledgments

The authors are grateful to Professor Elena Gallo of The University of Michigan, Department of Astronomy for her refereeing role and provided comments that significantly improved the presentation and accuracy of the paper.    We also would like to thank Aaron Barth of the Department of Physics and Astronomy, University of California, Irvine, Luis Ho and Li Zhao-Yu of the Department of Astronomy, School of Physics, Peking University for generously sharing us with their NGC~5102 and NGC~5206 CGS SB profiles in four Bessel--Johnson filters $BVRI$; Iskren Y. Georgiev of Max-Planck Instiut f\"ur Astronomie for providing his NSC data; and the Physics and Astronomy Department, University of Utah for supporting this work.    D. D. N. and A. C. S. acknowledge financial support from NSF grant AST-1350389.  M. C. acknowledges support from a Royal Society University Research Fellowship.\\

{\it Facilities:} Gemini: Gemini/NIFS/ALTAIR, ESO--VLT/SINFONI, and \hst~(WFPC2, ACS HRC)

%%%%%%%%%%%%%%%%%%%%%%%%%%%%%%%%%%%%%%%%%%%%%%%%%%%%%%%%%%%%%%%%%%
% THE SECTION OF APPENDIX
%%%%%%%%%%%%%%%%%%%%%%%%%%%%%%%%%%%%%%%%%%%%%%%%%%%%%%%%%%%%%%%%%%
\appendix
\section{Measurement of the NGC~5102 Nucleus $(B-V)_0$ Color Profile}\label{n5102color}
%%%%%%%%%%%%%%%%%% Figure 3 of central saturated color-color correction %%%%%%%%%%%%%%%%%%%%
\begin{figure}[!htb]
\minipage{0.49\textwidth}
  \includegraphics[width=\linewidth]{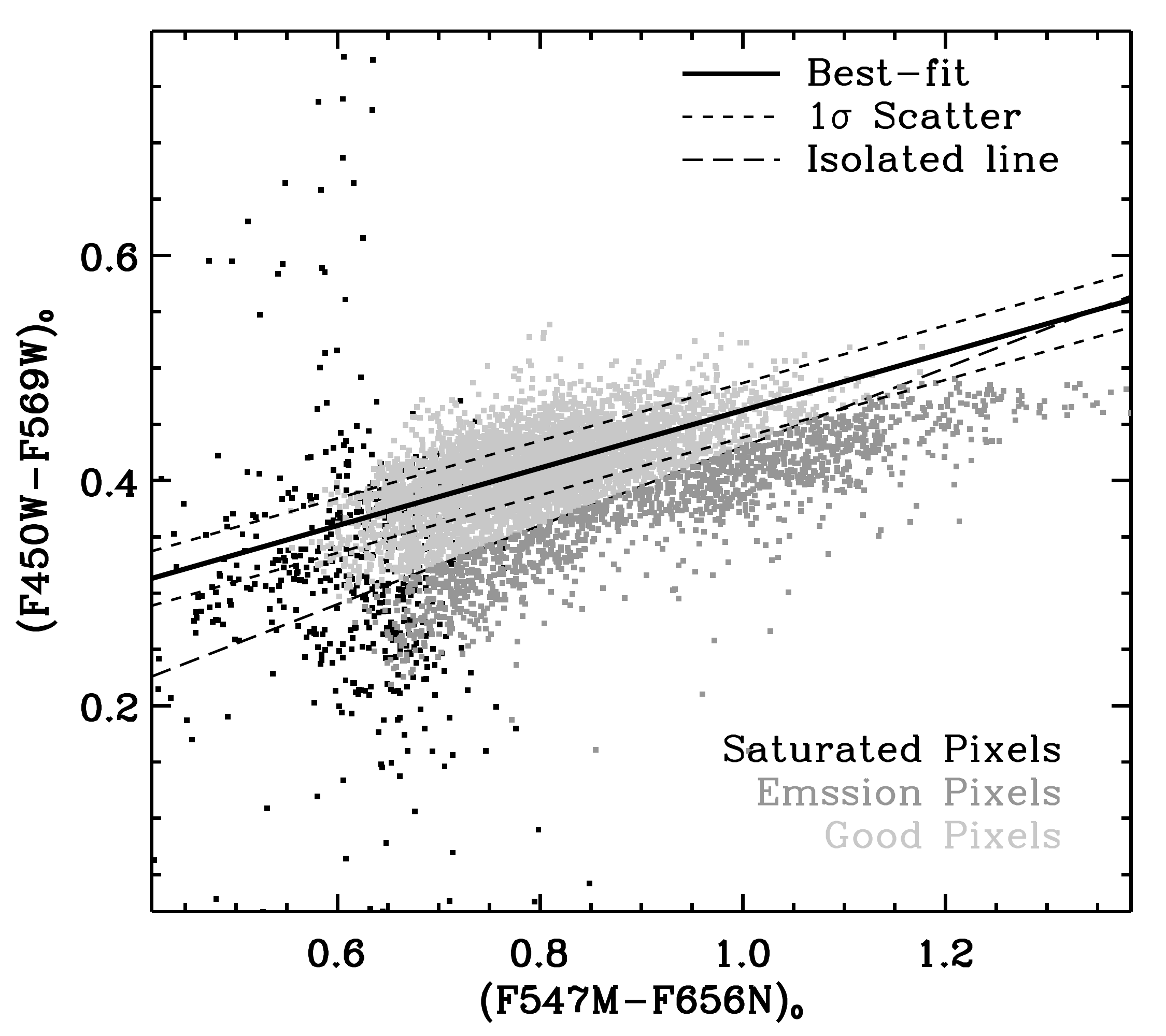}
\endminipage\hfill
\minipage{0.49\textwidth}\vspace{5mm}
  \includegraphics[width=\linewidth]{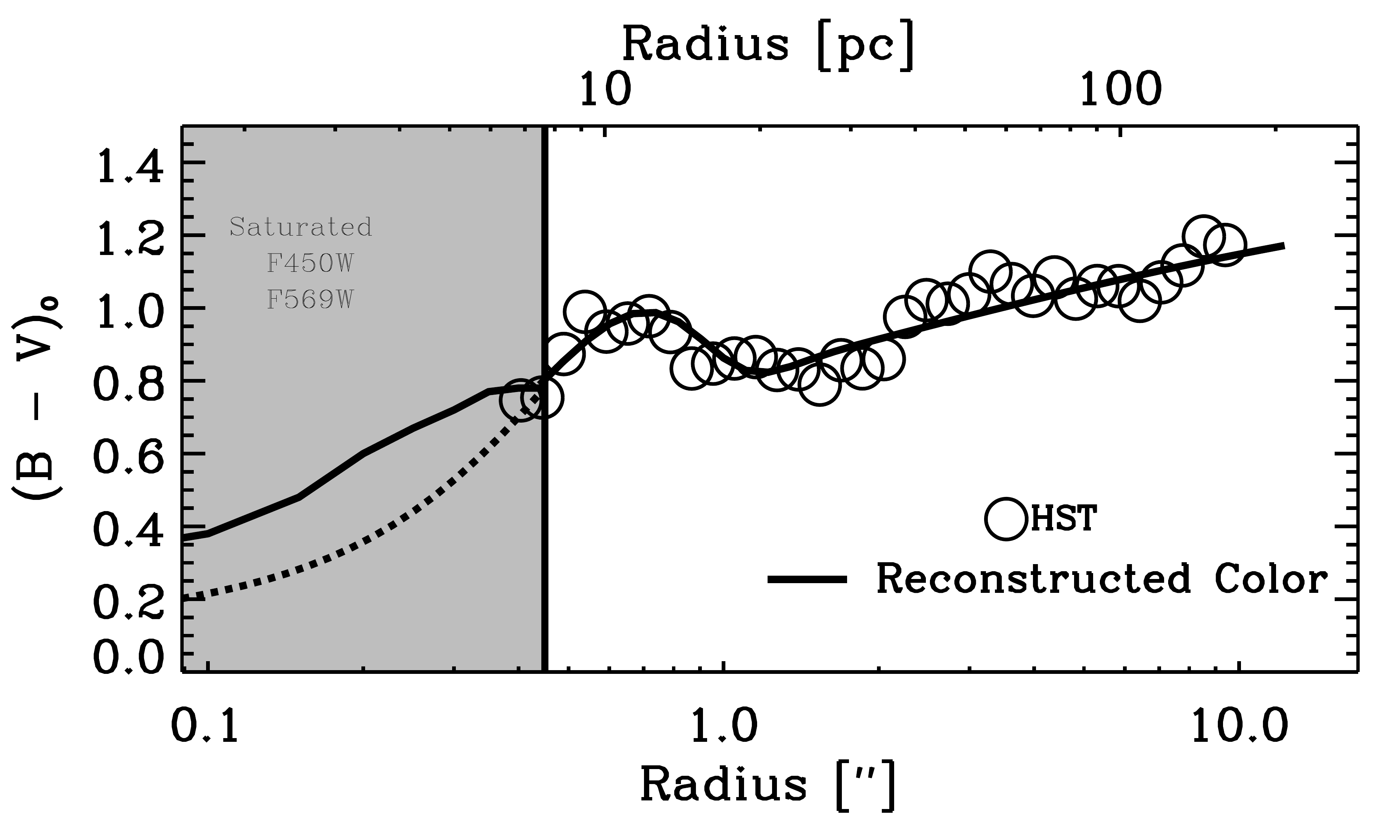}
\endminipage
\caption[SCP06C1]{\normalsize   Top panel: F450W--F569W vs. F547M--F656N color--color correlation for NGC~5102 nucleus.  Black dots are the saturated pixels within $0\farcs4$ in F450W and F569W.  Gray dots are the pixels that are possibly contaminated by H${\alpha}$ emission in F656N band.   White-gray dots are pixels outside $0\farcs4$, which are used for the fit.  The solid line is the best-fit of (F450W--F569W)$_{0}$ vs. (F547M--F656N)$_{0}$ relation, while the dashed lines represent 1$\sigma$ scatter of the best-fit.  The long-dashed line shows the isolation criterion of H$\alpha$, (F547M--F656N)$_{0}\gtrsim2.5\times$(F450W--F569W)$_{0}$.   Bottom panel:   the $(B-V)_0$ color profile, which is inferred from the \hst~photometric transformation from WFPC2 $\rightarrow$~$BVRI$ \citep{Sirianni05}.    Open circles are the \hst~data outside $0\farcs4$ and the gray region is the saturated \hst~data within $0\farcs4$.   The solid line is the best-fit model of $(B-V)_{\rm 0}$ colors, while the dashed line is the original saturated color.}
\label{n5102_nucleus_ml}
\vspace{3mm}
\end{figure}
%%%%%%%%%%%%%%%%%%%%%%%%%%%%%%%%%%%%%%%%%%%%%%%%%%%%%%%%%%%%%%%%%

We determine $(B-V)_0$ colors of NGC~5102 inside the radii of $0\farcs4$ from F450W--F569W WFPC2 data.   However, these data are saturated at the center. We can use the F547M--F656N data to predict the central $(B-V)_0$ colors where does not appear to be strong H${\alpha}$ emission.   The upper panel of Figure~\ref{n5102_nucleus_ml} shows the scheme that we use to explore the $\lesssim$$0\farcs4$ $(B-V)_0$ colors.      First, we create the (F450W--F569W)$_{0}$ and (F547M--F656N)$_{0}$ color maps with PSF cross convolution and Galactic extinction correction in each band.    Second, we get rid of all the colors with saturated pixels from F450W and F569W in both (F450W--F569W)$_{0}$ and (F547M--F656N)$_{0}$ maps (balck dots).    The pixels with possible contaminated by H${\alpha}$ emission in F656N band were also eliminated based on the criterion of (F547M--F656N)$_{0}\gtrsim2.5\times$(F450W--F569W)$_{0}$ (gray dots under the long-dashed line).    All the good pixels are presented in white-gray dots.   Third, we use these good pixels of colors to fit the linear correlation between (F450W--F569W)$_{0}$ and (F547M--F656N)$_{0}$.   The best-fit is presented by the solid black line, while the scatter by 1$\sigma$ of the fit is shown by the dashed black lines.     The central (F450W--F569W)$_{0}$ color profile is inferred from this (F450W--F569W)$_{0}$--(F547M--F656N)$_{0}$ correlation. 

We transform the (F450W--F569W)$_{0}$ colors into the  $(B-V)_{\rm 0}$ colors using the \hst~photometric transformation from WFPC2 $\rightarrow$ $BVRI$ \citep{Sirianni05}.    The  $(B-V)_{\rm 0}$ colors include the central region ($\lesssim$$0\farcs4$) that is shown in the lower panel of Figure~\ref{n5102_nucleus_ml}.    We plot the $(B-V)_{\rm 0}$ color profile of the outer part with radii $>$$0\farcs4$ from $B$-- (transformed from \hst~F450W) and $V$--band (transformed from \hst~F569W) (open circles for data, and black solid line for the best-fit 1D model), while the saturated $(B-V)_{\rm 0}$ within the radii $r\lesssim0\farcs4$ is plotted in gray region with 1D profile is shown black dotted line.    The best-fit reconstruction of $(B-V)_{\rm 0}$ is presented in black solid line.   This new color profile results a smaller mass of the NGC~5102 NSC of \Mnsc~$=(8.0\pm1.3)\times10^7$\Msun, $\sim$23\% smaller than its mass with the assumption of constant $(V-I)_{\rm 0}\sim1$~in the $\lesssim$$1\arcsec$ region inferred from the ground-based observation (CGS), but totally consistent to the dynamical estimate in Table~\ref{nsc_tab}.    Also, we note that we will use this $(B-V)_{\rm 0}$ color profile to model the mass model for NGC~5102 throughout this paper.   \\

%%%%%%%%%%%%%%%%%% Tables in Appendix %%%%%%%%%%%%%%%%%%%%%%%%%%
\section{Kinematics Tables, MGE Models, and JAM's Results }\label{tabs}
%%%%%%%%%%%%%%%%%%%%%%%%%%%%%%%%%%%%%%%%%%%%%%%%%%%%%%%
 We make the full kinematic data set available in Table~\ref{kine_tab_n205}, Table~\ref{kine_tab_n5102}, and Table~\ref{kine_tab_n5206} of NGC~205, NGC~5102, and NGC~5206 for use in future studies respectively.  Table~\ref{tab_mges} of MGE models of four galaxies used in JAM models (Section~\ref{sec:jeans}) are presented in this section as well.  Also, the complete results of JAM models indicate for its systemic uncertainties are shown in Table~\ref{kine_jeanmodel_tabfull}.   

%%%%%%%%%%%%%%% Table 8 of  nucleus NGC~205 kinematics %%%%%%%%%%%%%%%%%%
\begin{table}[ht]
\centering
\caption{Gemini/NIFS Kinematic Data of the NGC~205 Nucleus}
\begin{tabular}{cccc}
\hline \hline
$\Delta$R.A. ($\arcsec$)&$\Delta$Decl. ($\arcsec$) &$V_{\rm r}$ (\kms)& $\sigma$ (\kms)\\                  
\hline  
   ...             &      ...    &                 ...                &            ...             \\ 
$-$0.10      &$-$0.09  &$-$263.33 $\pm$ 3.06& 15.26 $\pm$ 2.20\\
$-$0.10      &0.01       &$-$269.34 $\pm$ 1.10& 15.63 $\pm$ 0.98\\
$-$0.15      &$-$0.04  &$-$267.32 $\pm$ 2.69& 16.34 $\pm$ 2.65\\
   ...             &      ...    &                 ...                &            ...             \\                                                                                                    
\hline
\end{tabular}
\tablenotemark{}
\tablecomments{\normalsize  Only a portion of this table is shown here to demonstrate its form and content. A machine-readable version of the full table is available.}
\label{kine_tab_n205}
\end{table}
%%%%%%%%%%%%%%%%%%%%%%%%%%%%%%%%%%%%%%%%%%%%%%%%%%%%%%%% 

%%%%%%%%%%%%%%% Table 9 of  nucleus NGC~5102 kinematics %%%%%%%%%%%%%%%%%%
\begin{table}[ht]
\centering
\caption{VLT/SINFONI Kinematic Data of the NGC~5102 Nucleus}
\begin{tabular}{cccc}
\hline \hline
$\Delta$R.A. ($\arcsec$)&$\Delta$Decl. ($\arcsec$) &$V_{\rm r}$ (\kms)& $\sigma$ (\kms)\\                  
\hline  
   ...             &      ...    &                 ...            &            ...             \\ 
0.02            & $-$0.02 & 492.93 $\pm$ 1.29 & 56.53 $\pm$ 2.84\\
0.00            &      0.03 & 490.75 $\pm$ 1.42 & 56.04 $\pm$ 3.26\\
$-$0.03       &$-$0.04 & 492.97 $\pm$ 1.28 &52.16 $\pm$ 2.36\\   
   ...             &      ...    &                 ...            &            ...             \\                                                                                                    
\hline
\end{tabular}
\tablenotemark{}
\tablecomments{\normalsize  Only a portion of this table is shown here to demonstrate its form and content. A machine-readable version of the full table is available.}
\label{kine_tab_n5102}
\end{table}
%%%%%%%%%%%%%%%%%%%%%%%%%%%%%%%%%%%%%%%%%%%%%%%%%%%%%%%% 

%%%%%%%%%%%%%%% Table 10 of  nucleus NGC~5206 kinematics %%%%%%%%%%%%%%%%%%
\begin{table}[ht]
\centering
\caption{VLT/SINFONI Kinematic Data of the NGC~5206 Nucleus}
\begin{tabular}{cccc}
\hline \hline
$\Delta$R.A. ($\arcsec$)&$\Delta$Decl. ($\arcsec$) &$V_{\rm r}$ (\kms)& $\sigma$ (\kms)\\                  
\hline  
   ...             &      ...    &                 ...           &            ...             \\ 
0.01            &   0.02   &610.18 $\pm$ 2.55 & 44.85 $\pm$ 3.46\\
0.01            &$-$0.03 &608.95 $\pm$ 2.39 & 44.57 $\pm$ 2.76\\
0.01            &0.07      &607.90 $\pm$ 2.07 & 44.24  $\pm$ 2.66\\ 
   ...             &      ...    &                 ...            &            ...             \\                                                                                                    
\hline
\end{tabular}
\tablenotemark{}
\tablecomments{\normalsize  Only a portion of this table is shown here to demonstrate its form and content. A machine-readable version of the full table is available.}
\label{kine_tab_n5206}
\end{table}
%%%%%%%%%%%%%%%%%%%%%%%%%%%%%%%%%%%%%%%%%%%%%%%%%%%%%%%% 

%%%%% Table 11 of MGE of GalFit Mass Map Profile in HST/WFC3 F814W with DRZSKY adding %%%%% 
\begin{table}[!th]
\centering
\caption{Luminosity and Mass MGE Models Using in Dynamical Modeling}
\begin{tabular}{ccccc}
\hline\hline   
     &                &M32&        &                     \\
\hline 
Mass &  Light   &  $\sigma$  &  $a/b$ &  P.A. \\
\hline
(\Mppc) & (\Lppc) & ($^{\prime\prime}$) &    &   (\deg)  \\[1mm]
(1) &    (2)        &    (3)      &  (4)  &   (5)     \\
\hline
5.01   & 4.74 & $-$2.76 & 0.75 & $-$25.0  \\
5.38   & 5.11 & $-$2.25 & 0.75 & $-$25.0  \\ 
5.64   & 5.37 & $-$1.78 & 0.75 & $-$25.0  \\
5.80   & 5.53 & $-$1.34 & 0.75 & $-$25.0  \\
5.85   & 5.58 & $-$0.94 & 0.75 & $-$25.0  \\
5.79   & 5.52 & $-$0.57 & 0.75 & $-$25.0  \\
5.62   & 5.35 & $-$0.23 & 0.75 & $-$25.0  \\
5.34   & 5.07 &   0.095  & 0.79 & $-$25.0  \\
4.93   & 4.66 &   0.398  & 0.79 & $-$25.0  \\
4.41   & 4.14 &   0.680  & 0.79 & $-$25.0 \\
3.77   & 3.50 &   0.946  & 0.79 & $-$25.0 \\ 
3.32   & 3.06 &   1.030  & 0.79 & $-$25.0 \\
3.00   & 2.73 &   1.195  & 0.79 & $-$25.0 \\
3.54   & 3.28 &   1.382  & 0.79 & $-$25.0 \\
3.19   & 2.92 &   1.611  & 0.79 & $-$25.0 \\
2.87   & 2.60 &   1.775  & 0.79 & $-$25.0 \\
2.71   & 2.44 &   1.907  & 0.79 & $-$25.0 \\
2.46   & 2.19 &   2.028  & 0.79 & $-$25.0 \\
\hline \hline
     &              &NGC~205&        &          \\ 
 \hline   
 Mass &  Light  &$\sigma$&$(a/b)$&  P.A. \\   
 \hline
(\Mppc) & (\Lppc) & ($^{\prime\prime}$) &    &   (\deg)  \\[1mm]
(1) &    (2)        &    (3)      &  (4)  &   (5)  \\
\hline 
4.87 & 4.58 & $-$1.75 & 0.95 & $-$40.4 \\
4.73 & 4.44 & $-$1.50 & 0.95 & $-$40.4 \\ 
5.23 & 4.94 & $-$1.15 & 0.95 & $-$40.4 \\
4.38 & 4.09 & $-$0.92 & 0.95 & $-$40.4 \\
4.93 & 4.64 & $-$0.68 & 0.95 & $-$40.4 \\
4.84 & 4.55 & $-$0.23 & 0.95 & $-$40.4 \\
4.36 & 4.07 & $-$0.00 & 0.95 & $-$40.4 \\
3.64 & 3.35 &   0.201  & 0.95 & $-$40.4 \\
3.62 & 3.33 &   0.871  & 0.90 & $-$40.4 \\
3.02 & 2.74 &   1.165  & 0.90 & $-$40.4 \\
\hline  \hline
\end{tabular}
\tablenotemark{}  
\tablecomments{\normalsize  MGE models using in JAM model fits in Section~\ref{ssec:jam} .  Columns 1:   the MGE models which represented for galaxies' mass models.  Columns 2: the  MGE models of $K$--band luminosity densities, created from the Padova SSP linear fit color-color correlation models of the nuclei color \hst/WFPC2 maps (Section~\ref{ssec:kband}).  Columns 3: the Gaussian width along the major axis.  Column 4: the axial ratios ($a/b$).  Column 5: the position angles.   Colums 1, 2, and 3 are shown in $\log$ scale.} 
\label{tab_mges}
\end{table}
%%%%%%%%%%%%%%%%%%%%%%%%%%%%%%%%%%%%%%%%%%%%%%%%%%%%%%% 
\begin{table}[!th]%\ContinuedFloat
\caption{Luminosity and Mass MGE Models Using in Dynamical Modeling (continue of Table~\ref{tab_mges}) }
\centering
\begin{tabular}{ccccc}
\hline\hline  
     &              &NGC~5102&     &         \\ 
 \hline  
 Mass &  Light  &$\sigma$&$(a/b)$&  P.A. \\
 \hline
(\Mppc) & (\Lppc) & ($^{\prime\prime}$) &    &   (\deg)  \\[1mm]
(1) &    (2)        &    (3)      &  (4)  &   (5)  \\
\hline    
6.95  & 5.81 & $-$1.28 & 0.68 & 50.5  \\
6.70  & 5.26 & $-$1.11 & 0.68 & 50.5  \\ 
5.94  & 4.50 & $-$0.77 & 0.68 & 50.5  \\
4.95  & 3.51 & $-$0.40 & 0.68 & 50.5  \\
4.89  & 4.05 & $-$0.26 & 0.68 & 50.5  \\
4.77  & 3.93 &      0.25 & 0.60 & 50.5  \\
4.65  & 3.63 &      0.37 & 0.60 & 50.5  \\
3.47  & 2.46 &      0.82 & 0.60 & 50.5  \\
3.11  & 2.09 &      0.88 & 0.60 & 50.5  \\
2.88  & 2.04 &      1.38 & 0.60 & 50.5  \\
2.82  & 1.98 &      1.64 & 0.60 & 50.5  \\ 
2.48  & 1.63 &      1.75 & 0.60 & 50.5  \\
2.76  & 1.92 &      1.77 & 0.60 & 50.5  \\
2.47  & 1.62 &      1.32 & 0.60 & 50.5  \\ 
\hline\hline
     &              &NGC~5206&    &         \\ 
\hline 
Mass &  Light  &$\sigma$&$(a/b)$&  P.A. \\
\hline
(\Mppc) & (\Lppc) & ($^{\prime\prime}$) &    &   (\deg)  \\[1mm]
(1)     &    (2)   &    (3)      &  (4)  &   (5)  \\
\hline 
5.60   & 5.49  & $-$1.74 & 0.96 & 38.3 \\
5.58   & 5.48  & $-$1.20 & 0.96 & 38.3 \\ 
5.43   & 5.33  & $-$1.11 & 0.96 & 38.3 \\
5.34   & 5.23  & $-$0.78 & 0.96 & 38.3 \\
4.16   & 4.06  & $-$0.66 & 0.96 & 38.3 \\
3.93   & 3.82  & $-$0.41 & 0.96 & 38.3 \\
3.83   & 3.72  & $-$0.27 & 0.96 & 38.3 \\
2.88   & 2.79  & $-$0.16 & 0.96 & 38.3 \\
2.98   & 2.88  & $-$0.08 & 0.98 & 38.3 \\
2.88   & 2.80  &   0.760  & 0.98 & 38.3 \\
2.61   & 2.73  &   0.890  & 0.98 & 38.3 \\ 
2.58   & 2.64  &   0.988  & 0.98 & 38.3 \\
2.40   & 2.33  &   1.316  & 0.98 & 38.3 \\
2.29   & 2.19  &   2.169  & 0.98 & 38.3 \\
\hline  \hline
\end{tabular}
\tablenotemark{}  
\end{table}
 
%%%%% Table 12 of full bets-fit JAM models for BHs of 4 galaxies %%%%%%%%
\begin{table*}[ht]
\centering
\caption{Jeans modeling results (full table)}
\begin{tabular}{ccccccccccccc}
\hline \hline
Object &Filter&Color--\ml&  \Mbh  & \betaz & $\gamma$ & $i$   &Number of Bins&$\chi^2_{\rm r}$&    $r_g$    &   $r_g$  \\ 
            &        &         &(\Msun)&        &       &(\deg)&       &      &($\arcsec$)&   (pc)     \\
                      
  (1)    &      (2)    &      (3)   &  (4) &       (5)     &(6)   &  (7)      &      (8)        & (9)   & (10)      &    (11)    \\
            
\hline 
%%% M32 %%% 
M32     &F555W$^{\rm G}$&  R15 &$2.3^{+1.3}_{-1.5}\times10^6$&$-0.25^{+0.35}_{-0.25}$&$1.05^{+0.20}_{-0.15}$&$74^{+12}_{-9}$&1354&1.35&0.455&1.81\\[1mm]
            &F555W$^{\rm D}$&  R15 &$2.5\times10^6$&$-$0.25&1.05&74&        &1.39&0.403&1.60\\[2mm]
%%%%%%%%%%%%%%                        
            &F814W$^{\rm G}$& B03  &$2.0^{+0.5}_{-0.8}\times10^6$&$-0.15^{+0.35}_{-0.35}$&$1.07^{+0.14}_{-0.10}$&$72^{+6}_{-6}$    &        &1.31&0.337&1.35\\[1mm]         
            &F814W$^{\rm G}$&Conts&$2.0^{+0.3}_{-1.0}\times10^6$&$-0.20^{+0.40}_{-0.40}$&$1.03^{+0.17}_{-0.08}$&$70^{+4}_{-5}$    &        &1.97&0.337&1.35\\[2mm]
%%%%%%%%%%%%%%             
            &F814W$^{\rm D}$&  R15 &$2.4\times10^6$&$-$0.20&1.08&70&        &1.75&0.387&1.61\\[1mm]          
            &F814W$^{\rm D}$& B03  &$2.2\times10^6$&$-$0.15&1.05&72&        &1.83&0.319&1.35\\[1mm]          
            &F814W$^{\rm D}$&Conts&$2.1\times10^6$&$-$0.20&1.03&70&        &2.04&0.303&1.35\\[1mm]                                                             
\hline
%%% NGC 205 %%%
% unsubtraction nucleus stars   
NGC~205&F555W$^{\rm D}$& R15  &$2.7\times10^4$&0.27&1.07&60& 256 &1.33&0.070&0.30\\[2mm]
% subtraction nucleus stars   
            &F555W$^{\rm G*}$&R15   &$2.1\times10^4$&0.25&1.07&60&    &1.37&0.051&0.17\\[2mm]                                               
 % subtraction nucleus stars    
            &F555W$^{\rm D*}$&R15   &$3.0^{+3.2}_{-3.0}\times10^4$&$0.27^{+0.08}_{-0.18}$&$1.07^{+0.18}_{-0.20}$&$60^{+5}_{-8}$&197   &1.37&0.074&0.35\\[1mm]               
            &F814W$^{\rm D*}$&B03   &$2.0^{+4.0}_{-2.0}\times10^4$&$0.25^{+0.06}_{-0.04}$&$1.09^{+0.07}_{-0.10}$&$66^{+4}_{-8}$&          &1.40&0.063&0.22\\[1mm]
            &F814W$^{\rm D*}$&Const&$1.8^{+3.2}_{-1.8}\times10^4$&$0.27^{+0.04}_{-0.06}$&$1.07^{+0.08}_{-0.07}$&$67^{+4}_{-7}$&          &2.51&0.060&0.20\\[1mm]          
\hline
%%% NGC 5102 %%%
NGC~5102&F547M$^{\rm G}$& B03 &$7.0^{+10.0}_{-2.0}\times10^5$&$0.25^{+0.08}_{-0.14}$&$1.13^{+0.03}_{-0.04}$&$76.5^{+8.5}_{-6.0}$& 1017  &1.23&0.056&0.89\\[1mm]
            &F547M$^{\rm G}$&Const&$6.0^{+9.0}_{-3.5}\times10^5$&$0.09^{+0.08}_{-0.12}$&$1.08^{+0.03}_{-0.05}$&$72.5^{+5.5}_{-4.5}$&          &2.32&0.048&0.76\\[2mm]     
%%%%%%%%%%%%%%  
           &F547M$^{\rm D}$& R15 &$8.0\times10^5$&0.15&1.15&71.5&          &1.57&0.070&1.20\\[1mm]
           &F547M$^{\rm D}$& B03 &$7.5\times10^5$&0.18&1.13&76.0&          &1.73&0.056&0.89\\[1mm]
           &F547M$^{\rm D}$&Const&$7.2\times10^5$&0.13&1.08&72.5&         &2.92&0.048&0.82\\[1mm]
\hline
%%% NGC 5206 %%%
NGC~5206&F555W$^{\rm G}$& R15 &$5.1^{+2.3}_{-1.2}\times10^5$&$0.28^{+0.07}_{-0.14}$&$1.08^{+0.20}_{-0.40}$& $43.5^{+4.0}_{-4.5}$&  240 &1.33&0.061&1.04\\[1mm]  
           &F555W$^{\rm D}$& R15 &$4.8\times10^5$&0.28&1.08& 43.5&         &1.36&0.071&1.21\\[2mm] 
%%%%%%%%%%%%%%                                          
           &F814W$^{\rm G}$& B03 &$2.4^{+1.1}_{-1.3}\times10^5$&$0.27^{+0.04}_{-0.15}$&$1.10^{+0.05}_{-0.10}$& $41.0^{+2.5}_{-2.5}$&    &1.57&0.045&0.75\\[1mm]  
           &F814W$^{\rm G}$&Const&$2.1^{+0.8}_{-1.4}\times10^5$&$0.29^{+0.06}_{-0.04}$&$1.06^{+0.10}_{-0.05}$& $45.0^{+2.5}_{-1.5}$&   &2.41&0.038&0.65\\[2mm]                   
%%%%%%%%%%%%%%  
           &F814W$^{\rm D}$& R15 &$4.5\times10^5$&0.25&1.06& 44.0&         &1.70&0.058&1.00\\[1mm]                   
           &F814W$^{\rm D}$& B03 &$2.3\times10^5$&0.27&1.10& 41.0&         &1.66&0.045&0.75\\ [1mm]                  
           &F814W$^{\rm D}$&Const&$2.0\times10^5$&0.29&1.06& 45.0&        &2.78&0.038&0.65\\[1mm]                                               
\hline
\end{tabular}
\tablenotemark{}
\tablecomments{\normalsize  This table's notations are the same in Table~\ref{kine_jeanmodel_tab}. Numbers without errors were fixed during JAM models fitting.   For NGC~205, the filters with/without * means the best-fit models of after/before contaminated stars subtraction of the NIFS kinematic spectroscopy.}
\label{kine_jeanmodel_tabfull}
\end{table*}
%%%%%%%%%%%%%%%%%%%%%%%%%%%%%%%%%%%%%%%%%%%%%%%%%%%%%%% 
%%%%%%%%%%%%%%%%%%%%%%%%%%%%%%%%%%%%%%%%%%%%%%%%%%%%%%%%%%%%%%%%%%
% THE SECTION OF BIBLIOGRAPHY SECTION!!! 
%%%%%%%%%%%%%%%%%%%%%%%%%%%%%%%%%%%%%%%%%%%%%%%%%%%%%%%%%%%%%%%%%%
\bibliographystyle{apj}
\bibliography{apjref}

\end{document}